%% file: Throughput_Optimal_and_Fast_Near-Optimal_Scheduling_with_Heterogeneously_Delayed_Network-State_Information__Extended_Version__v2.tex
\pdfoutput=1

\documentclass[journal]{IEEEtran}
\ifCLASSINFOpdf
\else
\fi
\usepackage{graphicx}
\usepackage[shortlabels]{enumitem}
\usepackage[cmex10]{amsmath}
\usepackage{amssymb}
\usepackage{url}
\usepackage{mathtools} 
\usepackage{cite} 
\usepackage{makecell} 

\usepackage{amsthm}

\newtheorem{theorem}{Theorem}
\newtheorem{lemma}{Lemma}
\newtheorem{proposition}{Proposition}
\newtheorem{corollary}{Corollary}
\usepackage[flushleft]{threeparttable}
\usepackage{xspace} 

\usepackage{algorithm}
\usepackage{algpseudocode}
\makeatletter
\def\BState{\State\hskip-\ALG@thistlm}
\makeatother

\newcommand{\Ex}{\mathbb{E}}
\DeclareMathOperator*{\argmax}{arg\,max}

\DeclareMathOperator*{\argminwithsuperscript}{arg\,min^{\S}}
\newcommand{\idoperator}{\mathbf{1}}
\newcommand{\BigO}{\ensuremath{\mathcal{O}}}

\newcolumntype{C}[1]{>{\centering\let\newline\\\arraybackslash\hspace{0pt}}m{#1}}
\mathchardef\mhyphen="2D 
\newcommand{\LCVARIANTONE}{$LC\mhyphen ELDR$\xspace}
\newcommand{\LCVARIANTTWO}{$LC\mhyphen ERDMC$\xspace}

\usepackage{mathtools}
\usepackage{amsmath}
\usepackage{amssymb}
\usepackage{multicol}
\usepackage{lipsum}
\usepackage{xcolor}
\usepackage{dsfont}
\usepackage{setspace}
\usepackage[font={footnotesize}]{caption}
\usepackage{subcaption}
\usepackage[normalem]{ulem} 
\usepackage{bm} 

\newsavebox\myVerb
\newenvironment{verbbox}{\lrbox\myVerb}{\endlrbox}
\newcommand*{\verbBox}{\usebox\myVerb}

\newcommand{\OnePlus}{\mathbf{1}_{\textbf{+}}}
\newcommand{\One}{\mathbf{1}}
\newcommand{\OneInfty}{\mathbf{1}_{\mathbf{\infty}}}

\newcommand*{\bfrac}[2]{\genfrac{}{}{0pt}{}{#1}{#2}}

\def\nespace{\hskip\fontdimen2\font\relax}
\newcommand{\Expr}{Expr.\nespace }

\newif\ifExtendedVersion
\newif\ifSubmitVersion
\newif\ifSupplementaryMaterial
\newif\ifIncludeAppendices
\newif\ifMainContents
\newif\ifReferences
\newif\ifAffiliations

\ExtendedVersiontrue 

\newcommand{\ifelseEV}[2]{\ifExtendedVersion#1\else#2 in \cite{SupplementaryMaterial}\fi}

\newcommand{\ifelseEVWithoutCite}[2]{\ifExtendedVersion#1\else#2\fi}

\newcommand{\FigVSVDVSVCTenLinksInExtendedVersion}{5 in \cite{ExtendedVersion}}

\newcommand{\twopartdef}[4]
{
	\left\{
	\begin{array}{lp{0.1cm}l}
		#1 & &\mbox{if } #2 \\
		#3 & & #4
	\end{array}
	\right.
}

\newcounter{ExampleCtr}

\begin{document}

\ifExtendedVersion
	\MainContentstrue
	\IncludeAppendicestrue
	\Referencestrue
	\Affiliationstrue
	\SubmitVersionfalse
\fi

%
\ifExtendedVersion
\title{Throughput Optimal and Fast Near-Optimal Scheduling with Heterogeneously Delayed Network-State Information \\ (Extended Version)}
\fi
%
%
%

\ifAffiliations
\author{Srinath~Narasimha, 
        Joy~Kuri
        and~Albert~Sunny
\thanks{S. Narasimha, J. Kuri and A. Sunny are with the Department
of Electronic Systems Engineering, Indian Institute of Science, Bengaluru
560012 India. e-mail: srinathn, kuri, salbert@dese.iisc.ernet.in} }
\else
\author{Srinath~Narasimha 
        and~Joy~Kuri
}
\fi

\maketitle


\ifMainContents
\begin{abstract}
\input{section-abstract}

\end{abstract}


%
\IEEEpeerreviewmaketitle

\section{Introduction}
\input{section-Introduction}

%
%
%
%
%
%


\section{System Model}
\input{section-System-Model}

\section{Brief Description of the $R$ Policy} 
\label{section-Brief-Description-of-R-Policy}
\input{section-Brief-Description-of-R-Policy}
\section{The $H$ Policy}
\label{section-The-H-Policy}
\input{section-The-H-Policy}

\section{Analytical Characterization of Delay Performance}
\label{section-section-Analytical-Characterization-of-Delay-Performance}
\input{section-Analytical-Characterization-of-Delay-Performance}

\section{Low Complexity Scheduling Policies}
\label{section-low-complexity-scheduling-policies}
\input{section-low-complexity-scheduling-policies}

\section{Numerical Results and Discussion}
\label{section-numerical-results}
\input{section-numerical-results}

\section{Conclusion}
\input{section-conclusion}

\fi


%
\ifSubmitVersion
	\ifReferences

\input{biblio}
	\fi
\begin{figure*}[!th]
{\centering \Huge Throughput Optimal and Fast Near-Optimal Scheduling with Heterogeneously Delayed Network-State Information\\}
{\Huge \centering \vspace{0.2cm} \hspace{3cm} (Appendices\normalsize${}^{{}^{\footnotemark{}}}$\Huge)} \\ 
{\\ \large \centering \vspace{0.75cm}  Srinath Narasimha and Joy Kuri\\} 
\end{figure*}
\newpage
\footnotetext{(a) This supplementary material is cited as Ref. \cite{SupplementaryMaterial} in the submitted manuscript. (b) Section numbers, equation numbers, etc. referenced in this supplementary material (but not found here) refer to the corresponding labels in the submitted manuscript.}
\else
	\ifReferences
	\input{biblio}
	\fi
\begin{figure*}[!th]
\end{figure*}
\fi

\ifIncludeAppendices
\appendices
\section{Proof of Proposition \ref{lemma-characterization-of-general-case-func-eval-complexity-of-R-policy}}
\label{appendix-proof-characterization-of-general-case-func-eval-complexity-of-R-policy}
\input{appendix-proof-characterization-of-general-case-func-eval-complexity-of-R-policy}

\section{Proof of Proposition \ref{lemma-characterization-of-worst-case-delay-values-for-R-policy}}
\label{appendix-proof-characterization-of-worst-case-delay-values-for-R-policy}
\input{appendix-proof-characterization-of-worst-case-delay-values-for-R-policy}

\section{Proof of Proposition \ref{lemma-characterization-of-worst-case-func-eval-complexity-of-R-policy}}
\label{appendix-proof-characterization-of-worst-case-func-eval-complexity-of-R-policy}
\input{appendix-proof-characterization-of-worst-case-func-eval-complexity-of-R-policy}
\section{Proof of Proposition \ref{lemma-characterization-of-sample-path-complexity-of-R-policy}}
\label{appendix-proof-characterization-of-sample-path-complexity-of-R-policy}
\input{appendix-proof-characterization-of-sample-path-complexity-of-R-policy}

\section{Proof of Proposition \ref{lemma-characterization-of-general-case-func-eval-complexity-of-H-policy}}
\label{appendix-proof-characterization-of-general-case-func-eval-complexity-of-H-policy}
\input{appendix-proof-characterization-of-general-case-func-eval-complexity-of-H-policy}

\section{Proof of Proposition \ref{lemma-critical-set-of-link-l-R-policy-equals-critical-set-of-link-l-H-policy}}
\label{appendix-proof-lemma-critical-set-of-link-l-R-policy-equals-critical-set-of-link-l-H-policy}
\input{appendix-proof-lemma-critical-set-of-link-l-R-policy-equals-critical-set-of-link-l-H-policy}

\section{Proof of Proposition \ref{lemma-characterization-of-worst-case-delay-values-for-H-policy}}
\label{appendix-proof-lemma-characterization-of-worst-case-delay-values-for-H-policy}
\input{appendix-proof-lemma-characterization-of-worst-case-delay-values-for-H-policy}

\section{Proof of Proposition \ref{lemma-characterization-of-worst-case-func-eval-complexity-of-H-policy}}
\label{appendix-proof-lemma-characterization-of-worst-case-func-eval-complexity-of-H-policy}
\input{appendix-proof-lemma-characterization-of-worst-case-func-eval-complexity-of-H-policy}

\section{Proof of Lemma \ref{lemma-Lambda-encompasses-all-supportable-arrival-rates}}
\label{appendix-proof-lemma-Lambda-encompasses-all-supportable-arrival-rates}
\input{proof-lemma-Lambda-encompasses-all-supportable-arrival-rates}

\section{Proof of Theorem \ref{theorem-H-policy-is-throughput-optimal}}
\label{appendix-proof-of-theorem-H-policy-is-throughput-optimal}
\input{appendix-proof-of-theorem-H-policy-stabilizes-all-arrival-rate-vectors-in-its-thpt-region}

\section{Proof of Proposition \ref{lemma-computational-complexity-LC1-LC2-policies}}
\label{appendix-proof-run-time-complexity-of-LC1-and-LC2-policies}
\input{appendix-proof-run-time-complexity-of-LC1-and-LC2-policies}

\section{Proof of Proposition \ref{lemma-saturated-system-throughput-of-LC1-policy}}
\label{appendix-proof-saturated-system-throughput-of-LC1-policy}

\input{appendix-proof-Exact-analytical-expression-for-the-saturated-system-throughput-of-LC-ELDR-policy}

\section{Proof of Proposition \ref{lemma-saturated-system-throughput-of-LC2-policy}}
\label{appendix-proof-saturated-system-throughput-of-LC2-policy}
\input{appendix-proof-Exact-analytical-expression-for-the-saturated-system-throughput-of-LC-ERDMC-policy}
\fi




\ifCLASSOPTIONcaptionsoff
  \newpage
\fi

\end{document}

%% file: section-abstract.tex
We consider the problem of distributed scheduling in wireless networks where heterogeneously delayed information about queue lengths and channel states of all links are available at all the transmitters. In an earlier work (by Reddy et al. in \emph{Queueing Systems}, 2012), a throughput optimal scheduling policy (which we refer to henceforth as the $\bm{R}$ policy) for this setting was proposed. We study the $\bm{R}$ policy, and examine its two drawbacks -- (i) its huge computational complexity, and (ii) its non-optimal average per-packet queueing delay. We show that the $\bm{R}$ policy unnecessarily constrains itself to work with information that is more delayed than that afforded by the system. We propose a new policy that fully exploits the commonly available information, thereby greatly improving upon the computational complexity and the delay performance of the $\bm{R}$ policy. We show that our policy is throughput optimal. Our main contribution in this work is the design of two fast and near-throughput-optimal policies for this setting, whose explicit throughput and runtime performances we characterize analytically. While the $\bm{R}$ policy takes a few milliseconds to several tens of seconds to compute the schedule once (for varying number of links in the network), the running times of the proposed near-throughput-optimal algorithms range from a few microseconds to only a few hundred microseconds, and are thus suitable for practical implementation in networks with heterogeneously delayed information.

%% file: section-Introduction.tex
\IEEEPARstart{A}{}central problem in any wireless network is one of scheduling the different users in the network with the objective of optimizing some desired metric -- for example, maximizing the system throughput -- in the presence of challenges that are unique to the wireless medium -- namely, channel fading and interference due to transmissions from other users in the network. This problem has been studied extensively in the literature. A highly influential and often cited work in this area is the work by Tassiulas and Ephremides \cite{Tassiulas_Ephremides_92}, who proposed the \textit{Back-Pressure} scheduling algorithm (a version of the \textit{Max-Weight} algorithm \cite{Moharir_Shakkottai_13, Tassiulas_Ephremides_93}), which is a centralized algorithm that schedules the links in the network based on global knowledge of the instantaneous queue lengths at all the links. Even though this algorithm is provably throughput optimal (see Sec. \ref{section-performance-metric}), it is a centralized algorithm that requires solving a global optimization problem in each time slot, and it also requires knowledge of instantaneous queue lengths at all links in the network to determine the schedule \cite{Reddy_et_al_12} \cite{Tassiulas_Ephremides_92}.

The Max-Weight algorithm, being a centralized policy, involves the computationally costly task of finding a maximal independent (i.e., non-interfering) set of links that can be activated simultaneously and whose summation of link-weights is maximum \cite{Rajagopalan09}. To circumvent this limitation, researchers have considered two broad approaches \cite{Rajagopalan09} -- namely, design of random-access algorithms in which access probabilities are dependent on queue sizes \cite{Rajagopalan09} or on arrival rates \cite{Bordenave08, Marbach207, Gupta06, Liu07, Stolyar08, Jiang08}, and design of distributed implementations of the Max-Weight algorithm \cite{Modiano06, Sanghavi07}. Some of these approaches require knowledge of instantaneous queue lengths and/or instantaneous channel states to attain their objective.




Even though it may be reasonable to assume that each node has knowledge of instantaneous queue lengths and channel states (at all times) for its own links (links emanating from itself), it is less pragmatic to assume that any node possesses instantaneous information about any  link in the network other than its own links (at any time instant). This could be because, for example, these quantities vary quickly with time (e.g., fast fading), or because the propagation delay of the feedback channel is large \cite{Pantelidou07}. In \cite{Ying08}, the authors consider networks where each node possesses knowledge of instantaneous queue lengths and channel states for its own links, but only has these information from other links in the network with some globally fixed delay (commonly referred to as \textit{homogeneous delay}). The assumption of homogeneous delays, however, is not satisfied in many networks where there is often a mismatch in the delays with which each node can acquire information about queue lengths and channel states of other links in the network \cite{Reddy_et_al_12}. An example of such a case is a network where the ``quality'' of information that a node possesses, of queue lengths and channel states of links other than its own direct links, is a decreasing function of their ``distances'' from this node \cite{Reddy_et_al_12}. Non-homogeneous delays are commonly referred to as \textit{heterogeneous delays}.

One serious issue with heterogeneous delays is that the nodes could potentially have different views of the state of the network \cite{Reddy_et_al_12}. In \cite{Reddy_et_al_12}, the authors consider distributed scheduling in a wireless network where information about the queue lengths and channel states of links in the network, available at the nodes, are heterogeneously delayed. They characterize the system throughput region for this setting, propose a scheduling policy, and show that their policy is throughput optimal. We study the limitations of the policy proposed in \cite{Reddy_et_al_12}.

As in \cite{Reddy_et_al_12}, we consider the problem of distributed scheduling in wireless networks where delayed queue length and delayed channel state information of each wireless link in the network, are available at all the transmitters in the network, but with possibly different delays. We refer to these as heterogeneously delayed queue state information (QSI) and heterogeneously delayed channel state information (CSI) respectively, and collectively refer to them as heterogeneously delayed network state information (NSI). We refer the reader to Sec. \ref{section-Structure-of-Heterogeneously-Delayed-NSI} for detailed information about the structure of heterogeneously delayed NSI that we have considered in this work. Our contributions are summarized below.

\subsection{Our Contributions}
\label{section-Our-Contributions}
\input{section-Our-Contributions}

%% file: section-Our-Contributions.tex
\noindent
Our primary contribution in this work is the design, analysis and performance evaluation of two fast and near-throughput-optimal scheduling policies for the case of networks with heterogeneously delayed NSI. Starting with the policy proposed in \cite{Reddy_et_al_12} (which we refer to henceforth as the $R$ policy), we arrive at our fast and near-throughput-optimal policies through the following steps of identifying issues with the $R$ policy and making progressive refinements to address these issues:

\begin{enumerate}[leftmargin=*] 

  \item First, we study the limitations of the $R$ policy. The $R$ policy is formulated as a functional optimization problem that searches for optimal threshold functions (see Sec. \ref{section-Brief-Description-of-R-Policy}) that maximize the queue-length weighted aggregate expected throughput of the system. We study the computational complexity of the $R$ policy and note that the $R$ policy is computationally very costly -- (i) it needs to evaluate a number of functions in its optimization domain that grows as a double exponential in the number of links in the network, and grows as an exponential in the number of levels into which the channel states on the wireless links are quantized, and (ii) it needs to consider, in computing the expected rate, a number of sample paths that is exponential in the number of links in the network and in the maximum of the heterogeneous delay values.

  \item We show that the delay performance of the $R$ policy is non-optimal, and that it can be improved upon significantly.
  
 
  \item Next, we show that the structure of heterogeneously delayed NSI as defined in the system model in \cite{Reddy_et_al_12} affords each node access to NSI that is \emph{less delayed} than that used in \cite{Reddy_et_al_12}, and yet commonly available at all nodes in the network. We propose two modifications to the $R$ policy to make use of this less delayed NSI (we call the resulting policy the $H$ policy). We show that the $H$ policy hugely improves upon the computational complexity of the $R$ policy, and numerically show that the $H$ policy also improves upon the delay performance of the $R$ policy substantially. We establish that, like the $R$ policy, the $H$ policy too is throughput optimal.
 
  \item Finally, we show that despite the huge leap that the $H$ policy takes in reducing the computational complexity in comparison to that of the $R$ policy, the computational complexity of the $H$ policy still remains impracticably large, making the case for low-complexity scheduling policies. Taking design cues from the $H$ policy, we propose two low-complexity near-throughput-optimal algorithms that are several orders of magnitude faster than the $R$ and $H$ policies. We obtain explicit analytical expressions for the expected saturated system throughputs of these policies, evaluate their performance and show that these policies closely approximate the optimal throughputs of the $R$ and $H$ policies. We also show that these policies possess desirable queueing delay characteristics.
  
	
\end{enumerate}

%% file: section-System-Model.tex
Our system model is precisely the same as that used in \cite{Reddy_et_al_12}. We use a slotted-time system and restrict our attention to single-hop transmissions. We borrow heavily in definitions, notations and nomenclature from \cite{Reddy_et_al_12} to keep our presentation easily cross-referenced and compared with \cite{Reddy_et_al_12}.

\subsection{Network Model}
\label{section-Network-Model}
Our model of the wireless network has $L$ transmitter-receiver pairs (or links); the set of links is denoted $\mathcal{L}$. We abstract the channel condition on each of the wireless links by the link's capacity. We model the time-varying capacity of each link $l$ as a separate discrete-time Markov chain (DTMC) denoted $\{C_l[t]\}$ on the same state space $\mathcal{C} = \{c_1, c_2, \ldots c_M\}$, where $c_1 < c_2 < \ldots < c_M$ are non-negative integers, and with the same transition probabilities\footnote{We use $\mathcal{C}$ to denote both the set and its cardinality.}${}^{,}$\footnote{We remark that the above channel model is assumed for making notations simpler and to enhance clarity. Our results hold even for the case of networks where each link is modeled as a separate DTMC (with different state spaces and/or different transition probabilities).}. We assume that the channel conditions on the wireless links are all independent of each other, but identically distributed, with transition probabilities $p_{ij} := \Pr[C_l[t + 1] = c_j \; | \; C_l[t] = c_i]$. We assume that these DTMCs are irreducible and aperiodic, and therefore have a stationary distribution, with the stationary probability of being in state $c_j, j \in \{1, 2, \ldots, M\} $ denoted $\pi(c_j)$.


\subsection{Interference Model}
For each link $l$, we let $I_l$ denote the set of links in the network that interfere with transmissions on link $l$; thus, $I_l \cup \{l\}$ is an \textit{interference set}.\footnote{$I_l$ can capture arbitrary interference constraints.}${}^{,}$\footnote{An \textit{interference set} is a set of wireless links such that if there is transmission on more than one link in the set in the same time slot, then these transmissions interfere with one another, possibly resulting in only a part of the transmission being received successfully at (any of) the intended receiver(s).} We say that a \textit{collision} occurs with a transmission scheduled on link $l$ if, in the same time slot, a transmission is scheduled on at least one link $l' \in I_l$. When a link $l$ successfully transmits (i.e. when the packets transmitted on link $l$ do not encounter a collision), a number of packets equal to $C_l[t]$ in the case of saturated queues, and equal to $\min(C_l[t], Q_l[t])$ in the case of non-saturated queues, are successfully received by the receiver on the other side of the link. However, in the case of a collision, $\gamma_l C_l[t]$ packets in the case of saturated queues, and $\min(\gamma_l C_l[t], Q_l[t])$ in the case of non-saturated queues, where $\gamma_l \in [0, 1]$, are received at the intended receivers.\footnote{We assume that $\{\gamma_l\}_{l \in \mathcal{L}}$ are such that $\{\gamma_l c_1, \ldots, \gamma_l c_M\}_{l \in \mathcal{L}}$ for $\gamma_l \in [0, 1]$ are all integers.}

\subsection{Traffic Model and Queue Dynamics}
\label{subsec:Traffic-Model-and-Queue-Dynamics}
As noted earlier, we use a slotted-time model. Packets arriving at a transmitter, depending on their intended destination, are assigned a wireless link. Prior to transmission on this link, these packets are buffered in the queue associated with this link. We denote by $Q_l[t]$ the length of the queue associated with link $l$, at time $t$. We model the packet arrivals into the queue associated with link $l$, by an arrival process denoted $A_l[t]$, that describes the number of packets that arrive into the queue, at time $t$. For every link $l$, we assume that $A_l[t]$ is an integer-valued process, independent across time slots $t$, with $0 \leq A_l[t] \leq A_{max} < \infty$ almost surely, with $\lambda_l := \Ex[A_l[t]] < \infty$. The queue lengths are governed by the update equation:
  \begin{equation}
    \label{equation-queue-update-equation}
    Q_l[t + 1] = (Q_l[t] + A_l[t] - S_l[t])^+,
  \end{equation} where $S_l[t] \coloneqq C_l[t]$ is the maximum possible service rate of link $l$ at time $t$, and $x^{+} \vcentcolon= \mbox{max}(0, x)$.

\subsection{Information Model and Structure of Heterogeneously Delayed NSI}
\label{section-Structure-of-Heterogeneously-Delayed-NSI}
\input{table_Small_Delays_For_Simple_Wireless_Network}
We use precisely the same structure for heterogeneously delayed NSI that each node has access to, as that used in \cite{Reddy_et_al_12}. In our model, at time $t$, the transmitter node of each link $l$ has information of the current queue length and the current channel state of link $l$, but only has delayed queue length and delayed channel state information of other links in the network, where these delays are possibly heterogeneous but time-invariant. As an example, consider the delay values in Table \ref{tab:Small_Delays_For_Simple_Wireless_Network} for a wireless network with three links $l_1$, $l_2$, $l_3$ with $A$, $B$, $C$ as the transmitter nodes of these links respectively, such that they form a network with \textit{complete interference}.\footnote{A network is said to have the complete interference property if each link in the network is in the interference set of all other links in the network.} From the first column of this table, transmitter $A$ has NSI of link $l_1$ (a link emanating from transmitter $A$) with a delay of $0$ time slots, NSI of links $l_2$ and $l_3$ with delays of $2$ and $1$ time slots respectively. Similarly, transmitter $B$ has NSI of links $l_1$, $l_2$ (a link emanating from transmitter $B$), and $l_3$ with delays of $1$, $0$, and $2$ time slots respectively, and transmitter $C$ has NSI of links $l_1$, $l_2$, and $l_3$ (a link emanating from transmitter $C$) with delays of $3$, $4$, and $0$ time slots respectively. Analogously, transmitters $A$, $B$ and $C$ possess the NSI of link $l_1$ with delays of 0, 1, and 3 time slots respectively, of link $l_2$ with delays of 2, 0, and 4 time slots respectively, and of link $l_3$ with delays of 1, 2, and 0 time slots respectively. Additionally, the entire table of time-invariant delay values is assumed to be available at all transmitters in the network.

We will need some notations. Let $\tau_l(h)$ denote the time-invariant delay with which the NSI of link $h$ is available at the transmitter node of link $l$. For example, referring to the $(2,3)$ entry of Table \ref{tab:Small_Delays_For_Simple_Wireless_Network}, $\tau_{l_3}(l_2)$ is $4$. Let $\tau_{max}$ denote the maximum of the delay values across the whole network, i.e., $\tau_{max} := \mbox{max}_{l,h \in \mathcal{L}, \, l \neq h} \tau_l(h)$. Also, let $\tau_{l, max}$ denote the maximum of the delays with which the NSI of link $l$ are available at the transmitter node of any link in the network, i.e., $\tau_{l, max} := \mbox{max}_{h \in \mathcal{L}, \, l \neq h} \tau_h(l)$. In other words, referring to the table of delay values, $\tau_{max}$ refers to the maximum of all the entries in the table ($\tau_{max} = 4$ in Table \ref{tab:Small_Delays_For_Simple_Wireless_Network}), and $\tau_{l_{i},max} \, \forall l_i \in \mathcal{L}$ refers to the maximum of the entries in row $i$ ($\tau_{l_{1}, max} = 3$, $\tau_{l_{2}, max} = 4$, and $\tau_{l_{3}, max} = 2$ in Table \ref{tab:Small_Delays_For_Simple_Wireless_Network}). We note that $\tau_l(l) = 0 \; \forall l \in \mathcal{L}$ in our model (the diagonal entries in Table \ref{tab:Small_Delays_For_Simple_Wireless_Network}).

Let $C_l[t](0:\tau) := \{C_l[t-\tau], C_l[t-\tau+1], \ldots, C_l[t]\}$,  $\boldsymbol{\mathrm{C}}[t](0:\tau_{max}) := \{ C_l[t](0:\tau_{max}) \}_{l \in \mathcal{L}}$ and $\boldsymbol{\mathrm{C}}[t](0:\tau_{\textbf{.}, max}) := \{ C_l[t](0:\tau_{l, max}) \}_{l \in \mathcal{L}}$. Let $\boldsymbol{\mathrm{Q}}[t](0:\tau_{max})$ and $\boldsymbol{\mathrm{Q}}[t](0:\tau_{\textbf{.}, max})$ be similarly defined. We denote the NSI available at the transmitter node of link $l$ by $\{ \mathcal{P}_l(\textrm{\textbf{Q}}[t](0:\tau_{\textbf{.}, max})), \mathcal{P}_l(\textrm{\textbf{C}}[t](0:\tau_{\textbf{.}, max}))\}$, where
\vspace{-1.5pt}
\begin{equation}
\begin{aligned}
\mathcal{P}_l(\textrm{\textbf{Q}}[t](0:\tau_{\textbf{.}, max})) &:= \{\boldsymbol{\mathcal{P}}_{lm}(\textrm{\textbf{Q}}[t](0 : \tau_{\textbf{.}, max}))\}_{m \in \mathcal{L}}, \\
\boldsymbol{\mathcal{P}}_{lm}(\textrm{\textbf{Q}}[t](0 : \tau_{\textbf{.}, max})) &:= \{Q_m[t-\tau]\}_{\tau=\tau_{m, max}}^{\tau_l(m)} \\
\mathcal{P}_l(\textrm{\textbf{C}}[t](0:\tau_{\textbf{.}, max})) &:= \{\boldsymbol{\mathcal{P}}_{lm}(\textrm{\textbf{C}}[t](0 : \tau_{\textbf{.}, max}))\}_{m \in \mathcal{L}}, \; \mbox{and} \\
\label{definition-Pl-and-Plm}
\boldsymbol{\mathcal{P}}_{lm}(\textrm{\textbf{C}}[t](0 : \tau_{\textbf{.}, max})) &:= \{C_m[t-\tau]\}_{\tau=\tau_{m, max}}^{\tau_l(m)} \end{aligned}
\end{equation}

\noindent
\footnotesize 
\textcolor{darkgray}{\textit{Example \refstepcounter{ExampleCtr}\theExampleCtr\label{Ex:NSI-H-policy}}: With reference to the delay values in Table \ref{tab:Small_Delays_For_Simple_Wireless_Network}, the NSI available at transmitter $A$ (the transmitter node of link $l_1$), $\{ \mathcal{P}_{l_1}(\textrm{\textbf{Q}}[t](0:\tau_{\textbf{.}, max})), \mathcal{P}_{l_1}(\textrm{\textbf{C}}[t](0:\tau_{\textbf{.}, max}))\}$ is the set: $\big\{ \{ Q_{l_1}[t-3], \ldots, Q_{l_1}[t], Q_{l_2}[t-4], \ldots, Q_{l_2}[t-2], Q_{l_3}[t-2], Q_{l_3}[t-1] \}, \{ C_{l_1}[t-3], \ldots, C_{l_1}[t], C_{l_2}[t-4], \ldots, C_{l_2}[t-2], C_{l_3}[t-2], C_{l_3}[t-1] \} \big\}$.} 
\normalsize
	
Note that, in \cite{Reddy_et_al_12}, the NSI available at link $l$ is defined as $\{ \mathcal{P}_l(\textrm{\textbf{Q}}[t](0:\tau_{max})), \mathcal{P}_l(\textrm{\textbf{C}}[t](0:\tau_{max}))\}$, and as a consequence, even though the structure of NSI that we use in our work is the same as that in \cite{Reddy_et_al_12}, the NSI at link $l$ in our model is a subset of the corresponding NSI used in \cite{Reddy_et_al_12} since for each $l$, $\tau_{l, max} \leq \tau_{max}$.\\
\indent
It is crucial to note that $\{Q_{l}[t-\tau_{l, max}], C_{l}[t-\tau_{l, max}]\}_{l \in \mathcal{L}}$ are all common information available at the transmitter node of each link in the network, as are $\{Q_{l}[t-\tau_{max}], C_{l}[t-\tau_{max}]\}_{l \in \mathcal{L}}$. It is intuitively appealing that $\{Q_{l}[t-\tau_{l, max}], C_{l}[t-\tau_{l, max}]\}_{l \in \mathcal{L}}$, being less delayed (i.e., more recent) compared to $\{Q_{l}[t-\tau_{max}], C_{l}[t-\tau_{max}]\}_{l \in \mathcal{L}}$, and also being commonly available at all transmitters in the network, it is preferable that each transmitter in the network base its transmit/no-transmit decision on this information rather than on $\{Q_{l}[t-\tau_{max}], C_{l}[t-\tau_{max}]\}_{l \in \mathcal{L}}$. In Sec. \ref{section-The-H-Policy} we show that this intuition is indeed correct.

\subsection{Performance Metric}
\label{section-performance-metric}
The metric of interest to us is the \textit{throughput region}\footnote{Throughput region (also called \textit{stability region}) is the set of all \textit{supportable} (defined in the next paragraph) mean arrival rate vectors.} (except in Sec. \ref{section-numerical-results} where we are interested in the saturated system throughput\footnote{Saturated system throughput is the sum of the throughputs on each link in the network when the queues at the transmitter node of each link have an infinite number of packets backlogged.}). For this, we define the \textit{state} of the network at time $t$ as the process $\boldsymbol{\mathrm{Y}}[t] := \{ Q_{l}[t](0 : \tau_{l,max}), C_{l}[t](0 : \tau_{l, max}) \}_{ l \in \mathcal{L} }$. We denote  $\boldsymbol{\mathrm{Y}}[t]$ under the scheduling policy\footnote{For each link $l$, a scheduling policy is a map from the NSI available at the transmitter node of link $l$, $\{ \mathcal{P}_l(\textrm{\textbf{Q}}[t](0:\tau_{\textbf{.}, max})), \mathcal{P}_l(\textrm{\textbf{C}}[t](0:\tau_{\textbf{.}, max}))\}$ to a transmit/no-transmit scheduling decision for (the transmitter associated with) link $l$.} $\mathcal{F}$ by $\boldsymbol{\mathrm{Y}}^{\mathcal{F}}[t]$. It is easily seen that this process is a DTMC.
	
Given an arrival rate vector $\{ \lambda_l \}_{l \in \mathcal{L}}$, we say that the network is \textit{stochastically stable} under the scheduling policy $\mathcal{F}$ if the DTMC $\boldsymbol{\mathrm{Y}}^{\mathcal{F}}[t]$ is positive recurrent. We say that an arrival rate vector $\{ \lambda_l \}_{l \in \mathcal{L}}$ is \textit{supportable} if some scheduling policy makes the network stochastically stable for this arrival rate vector. We say that a scheduling policy is \textit{throughput optimal} if it stabilizes the network for any arrival rate vector that is supportable.\footnote{It is inherent in the definition of throughput optimality that a policy that is throughput optimal need only stabilize the network for any arrival rate vector that is supportable by any policy that uses the same information structure. For example, it is possible that a policy that has access to only delayed NSI (with a given structure) to be throughput optimal and yet not support an arrival rate vector that is supportable by a policy that has access to instantaneous NSI for all links in the network.}

\subsection{Characterization of System Throughput Region}
\label{section-Characterization-of-Throughput-Region}

Consider the collection of functions $\{ f_l \}_{l \in \mathcal{L}}$, where $f_l : \mathcal{P}_l(\boldsymbol{\mathrm{C[t]}}(0 : \tau_{\textbf{.}, max})) \rightarrow \{ 0, 1 \}$ has the following semantics -- in each time slot $t$, the transmitter associated with each link $l$, computes the binary value $f_l(\mathcal{P}_l(\boldsymbol{\mathrm{C}}[t](0 : \tau_{\textbf{.}, max})))$ and attempts to transmit on link $l$ if and only if the outcome is $1$. Note that the outcome of $f_l$ is independent of all queue-length information.


Let $S_l(\boldsymbol{\mathrm{c}}, \boldsymbol{\mathrm{f}})$ be the expected data transmission rate (i.e. the expected number of bits or packets transmitted per time-slot) at time $t$ that the transmitter associated with link $l$ would receive if it applies the scheduling policy $f_l$ (where $\boldsymbol{\mathrm{f}} \coloneqq \{f_l\}_{l \in \mathcal{L}}$) when the delayed CSI at time $t$ is $\boldsymbol{\mathrm{C}}[t - \tau_{\textbf{.}, max}] = \boldsymbol{\mathrm{c}}$. That is,	
\ifdefined\ONECOLUMN
  \begin{flalign}
  	 \label{eqn:S-l}
     \textstyle S_l(\boldsymbol{\mathrm{c}}, \boldsymbol{\mathrm{f}}) = \textstyle \; \Ex[ \, C_l[t] \, f_l( \mathcal{P}_l(.)) ( \, \gamma_l + (1 - \gamma_l) \nonumber \displaystyle \prod_{m \in I_l}{ ( 1 - f_m( \mathcal{P}_m(.)) ) } \, ) \; | \;  \boldsymbol{\mathrm{C}}[t - \tau_{., max}] = \boldsymbol{\mathrm{c}} \, ],
  \end{flalign}
\else
  \begin{flalign}
  	 \label{eqn:S-l}
     \textstyle S_l(\boldsymbol{\mathrm{c}}, \boldsymbol{\mathrm{f}}) & = \textstyle \; \Ex\Bigg[ \, C_l[t] \; f_l\big( \mathcal{P}_l(.) \big) \bigg( \gamma_l + (1 - \gamma_l) \nonumber & \\ 
     &\displaystyle \prod_{m \in I_l}{ \Big( 1 - f_m \big( \mathcal{P}_m(.) \big) \Big) } \bigg) \; \bigg\rvert \;  \boldsymbol{\mathrm{C}}[t - \tau_{., max}] = \boldsymbol{\mathrm{c}} \, \Bigg], &
  \end{flalign}
\fi
where $\boldsymbol{\mathrm{C}}[t - \tau_{\textbf{.}, max}] \coloneqq \{C_l[t - \tau_{l, max}]\}_{l \in \mathcal{L}}, \text{ and } \mathcal{P}_l(.) \coloneqq \mathcal{P}_l( \boldsymbol{\mathrm{C}[t]}(0 : \tau_{\textbf{.}, max}) )$. Let $\boldsymbol{\mathrm{S}}(\boldsymbol{\mathrm{c}}, \boldsymbol{\mathrm{f}}) = \{ S_l(\boldsymbol{\mathrm{c}}, \boldsymbol{\mathrm{f}}) \}_{l \in \mathcal{L}}$. Also let 
  \begin{equation}
    \label{equation-eta-c}
	 \eta{(\boldsymbol{\mathrm{c}})} = \mathcal{CH}_{\boldsymbol{\mathrm{f}}}(\, \boldsymbol{\mathrm{S}}(\boldsymbol{\mathrm{c}}, \boldsymbol{\mathrm{f}})  ),
  \end{equation}
where $\mathcal{CH}_{\boldsymbol{\mathrm{f}}}(.)$ is the convex hull taken over all the threshold function vectors $\boldsymbol{\mathrm{f}}$. We note that $\eta{(\boldsymbol{\mathrm{c}})} \subset \mathbb{R}^{L}$ is the convex hull of the expected data transmission rates at time $t$ of all links in the network when $\boldsymbol{\mathrm{C}}[t - \tau_{\textbf{.}, max}] = \boldsymbol{\mathrm{c}}$, taken over all the threshold function vectors $\boldsymbol{\mathrm{f}}$. Let $\Lambda \subset \mathbb{R}^{L}$, as defined in Equation (\ref{eqn:Throughput_Region}), be our candidate for the throughput region of the system.
  \begin{equation}
  \label{eqn:Throughput_Region}
	\displaystyle \Lambda = \{ \lambda : \lambda = \sum_{\boldsymbol{\mathrm{c}} \in \mathcal{C}^{L}} { \pi(\boldsymbol{\mathrm{c}}) x(\boldsymbol{\mathrm{c}}), \, x(\boldsymbol{\mathrm{c}}) \in \eta(\boldsymbol{\mathrm{c}})  } \}
  \end{equation}

%% file: table_Small_Delays_For_Simple_Wireless_Network.tex
\begin{table}
\centering
\ifdefined\ONECOLUMN
\else
\scriptsize
\fi
\caption{{\small An instance of (small) delay values for a wireless network with three links}}
\label{tab:Small_Delays_For_Simple_Wireless_Network}       
\begin{threeparttable}
\ifdefined\ONECOLUMN
\begin{tabular}{p{4cm}C{1.5cm}C{1.5cm}C{1.5cm}}
\else
\begin{tabular}{p{3.5cm}C{1.1cm}C{1.1cm}C{1.1cm}}
\fi
\hline \hline \noalign{\smallskip}
& At TX $A^{\dagger}$ & At TX $B^{\dagger}$ & At TX $C^{\dagger}$  \\
\noalign{\smallskip}\Xhline{2.5\arrayrulewidth}\noalign{\smallskip}
Delay in obtaining NSI of link $l_1$ & $0$ & $1$ & $3$  \\
Delay in obtaining NSI of link $l_2$ & $2$ & $0$ & $4$  \\
Delay in obtaining NSI of link $l_3$ & $1$ & $2$ & $0$  \\
\noalign{\smallskip}\hline \hline
\end{tabular}
\begin{tablenotes}
  \scriptsize
  \item ${}^{\dagger} A, B, C$ are transmitter nodes of links $l_1, l_2,$ and $l_3$ respectively.
\end{tablenotes}
\end{threeparttable}
\end{table}

%% file: section-Brief-Description-of-R-Policy.tex
To put our work (which we develop starting in Sec. \ref{section-The-H-Policy}) in context, we briefly describe the $R$ policy proposed in \cite{Reddy_et_al_12} and investigate its two drawbacks in this section. For the $R$ policy, given $\boldsymbol{\mathrm{C}}[t](0 : \tau_{max})$, the critical NSI of the network at time $t$, is defined as\footnote{We remark that all the quantities that pertain to the $R$ policy that we show with a superscript $R$ (e.g., $CS^{R}(.), \mathcal{P}_{l}^{R}(.)$) appear without superscripts in \cite{Reddy_et_al_12}.}
\begin{equation}
\label{equation-critical-set}
\mathcal{CS}^R(\boldsymbol{\mathrm{C}}[t](0 : \tau_{max})) := \{ \{ C_l[t - \tau_k(l)] \}_{k \in \mathcal{L}, k \neq l} \}_{l \in \mathcal{L}}, \\
\end{equation}
and the critical NSI available at the transmitter node of link $l$ as
\ifdefined\ONECOLUMN
\begin{equation}
\begin{split}
\label{equation-critical-set-of-link-l}
\mathcal{CS}_l^R(\boldsymbol{\mathrm{C}}[t](0 : \tau_{max})) :=  \mathcal{CS^R}(\boldsymbol{\mathrm{C}}[t](0 : \tau_{max})) \cap \mathcal{P}_l^R(\boldsymbol{\mathrm{C}}[t](0 : \tau_{max})), \mbox{where }
\end{split}
\end{equation}
\else
\begin{equation}
\begin{split}
\label{equation-critical-set-of-link-l}
\mathcal{CS}_l^R(\boldsymbol{\mathrm{C}}[t](0 : \tau_{max})) & :=  \mathcal{CS^R}(\boldsymbol{\mathrm{C}}[t](0 : \tau_{max})) \\ 
&\quad\quad \cap \mathcal{P}_l^R(\boldsymbol{\mathrm{C}}[t](0 : \tau_{max})), \mbox{where }
\end{split}
\end{equation}
\fi
\begin{equation*}
\begin{split}
 & \mathcal{P}_l^R(\boldsymbol{\mathrm{C}}[t](0 : \tau_{max})) := \{\mathcal{P}_{lm}^R(\textrm{\textbf{C}}[t](0 : \tau_{max}))\}_{m \in \mathcal{L}}, \mbox{and } \\
 & \mathcal{P}_{lm}^R(\textrm{\textbf{C}}[t](0 : \tau_{max})) := \{C_m[t-\tau]\}_{\tau=\tau_{max}}^{\tau_l(m)} \text{are analogous to}
\end{split}
\end{equation*} the corresponding definitions in Eq. (\ref{definition-Pl-and-Plm}). 

In \cite{Reddy_et_al_12}, the authors proposed a distributed scheduling policy for networks with heterogeneously delayed NSI. We reproduce their algorithm verbatim here. The algorithm has two steps. At each time slot,

\begin{itemize}[leftmargin=*]
	\item \textbf{Step 1:}	all the transmitters compute threshold functions based on common NSI available at all transmitters. These threshold functions, one for each transmitter, map the respective transmitter's critical NSI to a corresponding threshold value, and are computed by solving the following optimization problem:
		\begin{equation}
			\label{eqn:Opt_Step1}
			{\displaystyle \argmax_{\textbf{T}} \sum_{l \in \mathcal{L}} Q_l[t - \tau_{max}] \, R_{l, \tau_{max}}(\textbf{T}) },
		\end{equation}
\ifdefined\ONECOLUMN
	where
		\begin{equation}
		\label{eqn:R_l_tau_max}
		\begin{split}
			\textstyle R_{l, \tau}\left(\textbf{T}\right) := \;\textstyle \Ex[C_l\left[t\right] \idoperator_{\{C_l[t] \geq T_l(.)\}} (\gamma_l + \left(1 - \gamma_l\right) \textstyle \prod_{m \in I_l} \idoperator_{\{C_m[t] < T_m(.)\}} ) \; | \; \boldsymbol{\mathrm{C}}\left[t - \tau \right] \; ], 
		\end{split}
		\end{equation}
	$\mbox{and } T_l(.) := T_l(CS_l^R(\boldsymbol{\mathrm{C}}[t](0 : \tau_{max})))$. \vspace{0.1in}
\else
	where
		\begin{equation}
		\label{eqn:R_l_tau_max}
		\begin{split}
			\textstyle R_{l, \tau}\left(\textbf{T}\right) := \;& \;\, \textstyle \Ex\Bigg[C_l\left[t\right] \One_{\{C_l[t] \geq T_l(.)\}} \Big(\gamma_l + \left(1 - \gamma_l\right) \\
			&\quad \prod_{m \in I_l}
			\One_{\{C_m[t] < T_m(.)\}} \Big) \; \bigg\rvert \; 			 \boldsymbol{\mathrm{C}}\left[t - \tau \right] \; \Bigg], 
		\end{split}
		\end{equation}
	$\quad\, \mbox{and } T_l(.) := T_l(CS_l^R(\boldsymbol{\mathrm{C}}[t](0 : \tau_{max})))$. \vspace{0.1in}
\fi

	\item \textbf{Step 2:} each transmitter observes its current critical NSI, evaluates its threshold function (found in Step 1) at this critical NSI, and attempts to transmit if and only if its current channel rate exceeds the threshold value, i.e., when 
	$C_l[t] \geq T_l(CS_l^R(\boldsymbol{\mathrm{C}}[t](0 : \tau_{max})))$.

\end{itemize}
\vspace{0.1in}

\ifExtendedVersion
To briefly illustrate the key idea behind the $R$ policy and the difficulty that the problem of scheduling with heterogeneously delayed NSI entails, consider a network with three links -- $l_1$, $l_2$ and $l_3$ with perfect collision and complete interference (illustration is easier for this setting). First, let us consider the case where the transmitter node of each link has the $\tau$-unit delayed NSI of each link (the case of homogeneous delays), and no link possesses the NSI of any link (including that of itself) with a delay lesser than $\tau$ units (this of course does not conform to the heterogeneously delayed NSI setting). In this case, the $R$ policy can be thought of as optimally partitioning the space of all sample paths (so as to maximize the achieved system throughput) into $L$ partitions (where some partitions are possibly empty) with the connotation that the $i$th partition contains sample paths for which link $l_i$ will carry transmission when any of these sample paths is realized. To see this, consider the 3-tuple ($C_{l_1}[t - \tau], C_{l_2}[t - \tau], C_{l_3}[t - \tau]$) which give rise to eight possibilities when $\mathcal{C} = \{1, 2\}$. We can think of all the sample paths as grouped into eight classes corresponding to these eight possibilities of ($C_{l_1}[t - \tau], C_{l_2}[t - \tau], C_{l_3}[t - \tau]$). These eight classes are then partitioned optimally into three partitions such that when any sample path in the first partition is realized, link $l_1$ will carry transmission (and links $l_2$ and $l_3$ won't), and so on. When a particular sample path is realized, the transmitter nodes of links $l_1$, $l_2$ and $l_3$ look at the tuple ($C_{l_1}[t - \tau], C_{l_2}[t - \tau], C_{l_3}[t - \tau]$) and determine the partition to which it belongs and thereby decide which of the three links will carry transmission. Now, this is true when the delays are homogeneous. In the case of heterogeneous delays, the problem of partitioning the sample paths is slightly more complicated by the fact that the different transmitters in the network have disparate views of the state of the network. We continue this illustration for the heterogeneously delayed NSI setting after introducing the example in Sec. \ref{section-A-Note-on-the-Computation-Complexity-of-R-Policy}
\fi


\subsection{Computational Complexity}
\label{section-A-Note-on-the-Computation-Complexity-of-R-Policy}
From \Expr (\ref{eqn:Opt_Step1}), we see that the domain of optimization is the set of threshold function vectors \textbf{T}, where $\textbf{T} = \{T_l(.)\}_{l \in \mathcal{L}}$. From \Expr (\ref{eqn:R_l_tau_max}) we see that, while computing the conditional expectation, given a threshold function vector \textbf{T}, each sample path in the evaluation of the conditional expectation is subjected to evaluation by the threshold functions $\{T_l(.)\}_{l \in \mathcal{L}}$. As a consequence, there are two aspects to the computational complexity of the $R$ policy -- namely, (i) \textit{functional evaluation complexity} -- the number of threshold function vectors \textbf{T} over which the optimization is to be done, and (ii) \textit{sample-path complexity} -- the number of sample paths that are to be considered in evaluating the conditional expectation. We first consider the complexity of functional evaluation and obtain an expression for the number of threshold function vectors required in the domain of optimization in the general case. Next, we characterize the structure on the delay values (in the table of delay values) that produces the worst-case functional evaluation complexity in the $R$ policy. We then consider the complexity of functional evaluation in this worst-case setting, and obtain an expression for the number of threshold function vectors required in the domain of optimization in the worst case.
\begin{proposition}
	\label{lemma-characterization-of-general-case-func-eval-complexity-of-R-policy}
	For the $R$ policy, the total number of threshold functions that are needed to be considered in the domain of optimization in \Expr (\ref{eqn:Opt_Step1}), in general, is $(\mathcal{C}+1)^{\sum_{l_i \in \mathcal{L}}\mathcal{C}^{|\mathcal{V}_i|}}$, where $\mathcal{C}$ is the number of channel states, $\mathcal{L}$ is the set of links in the network, and $\mathcal{V}_i \coloneqq CS_{l_i}^R(\boldsymbol{\mathrm{C}}[t](0 : \tau_{max})) \setminus \{C_{l_i}[t - \tau_{max}]\}_{l_i \in \mathcal{L}}$ is the set of input parameters to the threshold function $T_{l_i}(.)$.
\end{proposition}
We present a proof of this result in Appendix \ifelseEV{\ref{appendix-proof-characterization-of-general-case-func-eval-complexity-of-R-policy}}{A}. We illustrate this result in Example \ref{Ex:func-eval-and-sample-path-compl}. Next, we characterize the structure on the delay values that produces the worst-case functional evaluation complexity in the $R$ policy.
\begin{proposition}
\label{lemma-characterization-of-worst-case-delay-values-for-R-policy}
The worst-case complexity of functional evaluation in the $R$ policy is realized when the delays in the table of delay values are all distinct, and the delay values at positions $(i,j)$ in row $i$ of the table of delay values, for $j = 1, 2, \dots, i-1, i+1, \ldots, L$ appear in descending order, for all $i$.
\end{proposition}
\indent
We relegate the proof of this result to Appendix \ifelseEV{\ref{appendix-proof-characterization-of-worst-case-delay-values-for-R-policy}}{A}. We now characterize the functional evaluation complexity of the $R$ policy for the worst-case setting noted in Proposition \ref{lemma-characterization-of-worst-case-delay-values-for-R-policy}.
\begin{proposition}
\label{lemma-characterization-of-worst-case-func-eval-complexity-of-R-policy}
For the $R$ policy, the total number of threshold functions that are needed to be considered in the domain of optimization in \Expr (\ref{eqn:Opt_Step1}), for the worst-case scenario noted in Proposition \ref{lemma-characterization-of-worst-case-delay-values-for-R-policy}, is $(\mathcal{C}+1)^{\sum_{l_i \in \mathcal{L}}\mathcal{C}^{i(L-2)+(L-1)}}$, where $\mathcal{C}$ is the number of channel states and $L$ is the number of links in the network.
\end{proposition}
\indent
From this result, the number of threshold functions required in the domain of optimization in \Expr (\ref{eqn:Opt_Step1}) is doubly exponential in the number of links in the network, and exponential in the number of levels into which the channel states on the wireless links are quantized. We present a proof of this proposition in Appendix \ifelseEV{\ref{appendix-proof-characterization-of-worst-case-func-eval-complexity-of-R-policy}}{B}. Next, we characterize the sample-path complexity of the $R$ policy.
\begin{proposition}
\label{lemma-characterization-of-sample-path-complexity-of-R-policy}
For the $R$ policy, the number of sample paths that are needed to be considered in the evaluation of the conditional expectation in \Expr (\ref{eqn:R_l_tau_max}) is given by $\mathcal{C}^{L \tau_{max}}$, where $\mathcal{C}$ is the number of channel states and $L$ is the number of links in the network.
\end{proposition}
From this result, the number of sample paths involved in the evaluation of the conditional expectation in \Expr (\ref{eqn:R_l_tau_max}) is exponential in the number of links in the network and in the maximum of the heterogeneous delay values. We present a proof of this proposition in Appendix \ifelseEV{\ref{appendix-proof-characterization-of-sample-path-complexity-of-R-policy}}{C}. As noted earlier, given a threshold function vector $\textbf{T} = \{T_l(.)\}_{l  \in \mathcal{L}}$, each of these $\mathcal{C}^{L \tau_{max}}$ sample paths is subjected to evaluation by the $L$ threshold functions in \textbf{T}. We now illustrate the functional evaluation and sample-path complexities with an example.

\vspace{0.2cm}
\noindent
\footnotesize
\textcolor{darkgray}{\textit{Example \refstepcounter{ExampleCtr}\theExampleCtr\label{Ex:func-eval-and-sample-path-compl}}:\footnote{\label{footnote:review-proof-of-func-eval-complexity-of-R-policy} Reviewing the proof of Proposition \ref{lemma-characterization-of-worst-case-func-eval-complexity-of-R-policy} in Appendix \ifelseEV{\ref{appendix-proof-characterization-of-worst-case-func-eval-complexity-of-R-policy}}{B} will be helpful in understanding this example fully.} Consider a three node network with three links -- $l_1$, $l_2$, and $l_3$ with heterogeneous delays {\small $\scriptstyle \tau_{l_2}(l_1) = 11, \tau_{l_3}(l_1) = 7, \tau_{l_1}(l_2) = 9, \tau_{l_3}(l_2) = 8, \tau_{l_1}(l_3) = 12, \tau_{l_2}(l_3) = 6$}. Note that these delay values are already in the worst-case form of Proposition \ref{lemma-characterization-of-worst-case-delay-values-for-R-policy}. From these delays, we get {\small $\scriptstyle CS_{l_1}(.) = \{C_{l_1}[t - 11], C_{l_1}[t - 7], C_{l_2}[t - 9], C_{l_3}[t - 12]\}$}, {\small $\scriptstyle CS_{l_2}(.) = \{C_{l_1}[t - 11], C_{l_2}[t - 9], C_{l_2}[t - 8], C_{l_3}[t - 12], C_{l_3}[t - 6]\}$}, and {\small $\scriptstyle CS_{l_3}(.) = \{C_{l_1}[t - 11], C_{l_1}[t - 7], C_{l_2}[t - 9], C_{l_2}[t - 8], C_{l_3}[t - 12], C_{l_3}[t - 6]\}$}. Therefore, given $\scriptstyle \{C_l[t - \tau_{max}]\}_{l \in \mathcal{L}}$, $\scriptstyle T_{l_1}(.), T_{l_2}(.), T_{l_3}(.)$ are functions of $3, 4$ and $5$ variables respectively (i.e., in the language of Proposition \ref{lemma-characterization-of-general-case-func-eval-complexity-of-R-policy}, $\mathcal{V}_1 = 3, \mathcal{V}_2 = 4$, and $\mathcal{V}_3 = 5$). Taking $\mathcal{C} = \{1, 2\}$, $T_{l_1}(.)$ maps the $2^3$ values in its domain to the real numbers say $0.5$, $1.5$ and $2.5$, independently.\footnote{{See footnote \ref{footnote:review-proof-of-func-eval-complexity-of-R-policy}}} Thus, the number of choices of threshold functions for $\scriptstyle T_{l_1}(.)$ is $3^{8}$. Similarly, the number of choices for $\scriptstyle T_{l_2}(.)$ and $\scriptstyle T_{l_3}(.)$ are $3^{16}$ and $3^{32}$ respectively. Therefore, the total number of threshold function vectors \textbf{T} in the domain of optimization\footnote{{Incidentally, in this example, considering the delay values before rearranging as noted in Proposition \ref{lemma-characterization-of-worst-case-delay-values-for-R-policy}, the total number of threshold function vectors \textbf{T} in the domain of optimization is also $3^{56}$.}} is $3^{56}$.  The number of sample paths to be considered in evaluating the conditional expectation is $2^{36}$ since $\tau_{max} = 12$ and $L=3$. \hfill$\Box$}\\
\normalsize
	
\ifExtendedVersion
Continuing the discussion on the key idea behind the $R$ policy and the difficulty that the problem of scheduling with heterogeneously delayed NSI entails from the last paragraph in Sec. \ref{section-Brief-Description-of-R-Policy}, for the case of heterogeneous delays in the above example, when a particular sample path is realized, the transmitter node of link $l_1$ looks at $T(CS_{l_1}^R(.))$ (i.e., at $T(C_{l_1}[t - 11], C_{l_1}[t - 7], C_{l_2}[t - 9], C_{l_3}[t - 12])$) and decides whether link $l_1$ will carry transmission depending on whether $C_{l_1}[t] \geq T(CS_{l_1}^R(.))$, the transmitter node of link $l_2$ looks at $T(CS_{l_2}^R(.))$ (i.e., at $T(C_{l_1}[t - 11], C_{l_2}[t - 9], C_{l_2}[t - 8], C_{l_3}[t - 12], C_{l_3}[t - 6])$) to decide, and so on. Even though $CS_{l_1}^R(.)$, $CS_{l_2}^R(.)$ and $CS_{l_3}^R(.)$ do not coincide, it is easy to convince oneself that the sample paths will be partitioned into three partitions as before, since we are dealing with perfect collision interference (i.e., $\gamma_l = 0, \; \forall l$).
\fi

\subsection{The Twin $DQIC$ Policies}
\input{section-The-DQIC1-DQIC2-Policies}

\subsection{Delay Performance - A Preview}
\label{section-Delay-Performance-Preview}
\input{section-Delay-Performance-Preview}

%% file: section-The-DQIC1-DQIC2-Policies.tex
With the objective of isolating the undesirable consequences of using $\tau_{max}$-delayed queue lengths (in comparison to using $\tau_{\textbf{.}, max}$-delayed queue lengths) on the delay performance of the system, we consider two scheduling policies both of which have access to instantaneous CSI but differ (only) in the delays that they use to access the queue lengths. The first policy, $DQIC1$ (for \textbf{D}elayed \textbf{Q}ueue lengths and \textbf{I}nstantaneous \textbf{C}hannel states), uses $\tau_{max}$-delayed queue lengths to stabilize the queues, whereas the second policy, $DQIC2$, uses $\tau_{\textbf{.}, max}$-delayed queue lengths. Specifically, the two policies are:


\vspace{5pt}
\noindent
$$\boldsymbol{DQIC1}: \argmax_{l}\big\{Q_{l}[t - \tau_{max}] \times C_{l}[t]\big\}_{l \in \mathcal{L}}$$\\[-0.8cm]
\noindent
$$\;\; \boldsymbol{DQIC2}: \argmax_{l}\big\{Q_{l}[t - \tau_{l, max}] \times C_{l}[t]\big\}_{l \in \mathcal{L}}$$

\noindent
\textit{Remark: Note that the only difference between the two policies is that the $DQIC1$ policy uses $\tau_{max}$-delayed QSI whereas the $DQIC2$ policy uses $\tau_{\textbf{.}, max}$-delayed QSI. Consequently, any difference in performance between the two policies is directly attributable to this difference in the delay values they use to access the QSI.}

%% file: section-Delay-Performance-Preview.tex
\input{table_Het-Delays-2-TXs-x-taulmax}
Towards developing a feel for what to expect in terms of the difference in delay performances of the different policies outlined in this section, we evaluate these policies through numerical simulation for the following specific instance of a two-transmitter network. We consider a network with two links $l_1$ and $l_2$ that carry transmission from their respective transmitter nodes $A$ and $B$ to their respective destinations (with $I_{l_1} = \{l_2\}$ and $I_{l_2} = \{l_1\}$), with the channels on these links modeled as independent DTMCs on the state space $\mathcal{C} = \{1, 2\}$ with crossover probability $0.1$. We consider heterogeneous delays as noted in Table \ref{tab:Het-Delays-2-TXs-x-taulmax}, where $x \geq 1$ is a parameter we vary. From this table, we see that $\tau_{max} = x, \tau_{l_1, max} = 1, \tau_{l_2, max} = x$.

The $DQIC1$ and $DQIC2$ policies specialize to this setting as follows:

\vspace{5pt}
\noindent
$$\boldsymbol{DQIC1}: \argmax \big\{Q_{l_1}[t - x] \times C_{l_1}[t], \; Q_{l_2}[t - x] \times C_{l_2}[t]\big\}$$\\[-1cm]
\noindent
$$\boldsymbol{DQIC2}: \argmax \big\{Q_{l_1}[t - 1] \times C_{l_1}[t], \; Q_{l_2}[t - x] \times C_{l_2}[t]\big\}$$

\noindent
i.e., $DQIC1$ schedules link $l_1$ if $Q_{l_1}[t - x] \times C_{l_1}[t] \geq Q_{l_2}[t - x] \times C_{l_2}[t]$, and link $l_2$ otherwise, whereas $DQIC2$ schedules link $l_1$ if $Q_{l_1}[t - 1] \times C_{l_1}[t] \geq Q_{l_2}[t - x] \times C_{l_2}[t]$, and link $l_2$ otherwise. 

Packets arrive into the two queues at the two transmitters as independent Poisson processes with rates $\lambda_1 = \lambda_2 = \frac{1}{4}$ (for this setting, it is easily seen that sum rates (i.e. $\lambda_1 + \lambda_2$) of up to 1.75 are supportable.\footnote{\label{fn:supportable_sum_rate} Considering the four possibilities of data rates for $C_{l_1}[t]$ and $C_{l_2}[t]$, a data rate of 1 unit is supportable when $C_{l_1}[t] = C_{l_2}[t] = 1$ which happens with steady-state probability 0.25, and a data rate of 2 units are supportable in all the other three cases since when the queues are saturated, both the $DQIC1$ and $DQIC2$ policies would choose the link with the largest data rate. These three cases happen with steady-state probability 0.25 each.}) Packets are time-stamped on arrival and on exit, and on servicing link $l$ successfully, a number of packets equal to $\min(C_l[t], Q_l[t])$ are removed from the queue of link $l$. Fig. \ref{fig:delay_performance_R_H_DQIC1_DQIC2}(a) depicts the average queueing delay per-packet (in units of time-slots) of the $DQIC1$, $DQIC2$, $R$ and $H$ policies. Fig. \ref{fig:delay_performance_R_H_DQIC1_DQIC2}(b) depicts the percentage reduction in the average queueing delay per-packet in the $DQIC2$ policy compared to that of the $DQIC1$ policy.
\input{fig_delay_performance_R_H_Policy1_Policy2}

\noindent
\textbf{Conclusion}: The average per-packet queueing delays of the $DQIC1$ and $R$ policies grow linearly with $\tau_{max}$ whereas that of the $DQIC2$ policy tends to become almost impervious to increase in $\tau_{max}$. \textit{Comparing the $DQIC1$, $DQIC2$ and $R$ policies and their delay performances, the fact that the $DQIC1$ and $R$ policies use $\tau_{max}$-delayed queue lengths isolates itself as the sole reason for their undesirable delay performances}.

%% file: table_Het-Delays-2-TXs-x-taulmax.tex
\begin{table}
\scriptsize
\centering
\caption{{\small An instance heterogenous delay values for a wireless network with two links}}
\label{tab:Het-Delays-2-TXs-x-taulmax}       
\begin{threeparttable}
\begin{tabular}{p{4cm}C{1.3cm}C{1.3cm}}
\hline \hline \noalign{\smallskip}
& At TX $A^{\dagger}$ & At TX $B^{\dagger}$  \\
\noalign{\smallskip}\Xhline{2.5\arrayrulewidth}\noalign{\smallskip}
Delay in obtaining NSI of link $l_1$ & $0$ & $1$   \\
Delay in obtaining NSI of link $l_2$ & $x$ & $0$   \\
\noalign{\smallskip}\hline \hline
\end{tabular}
\begin{tablenotes}
  \scriptsize
  \item ${}^{\dagger} A, B$ are transmitter nodes of links $l_1$ and $l_2$ respectively.
\end{tablenotes}
\end{threeparttable}
\end{table}

%% file: fig_delay_performance_R_H_Policy1_Policy2.tex
\begin{figure}
\centering
\ifdefined\ONECOLUMN
\includegraphics[width=.5\textwidth, keepaspectratio]{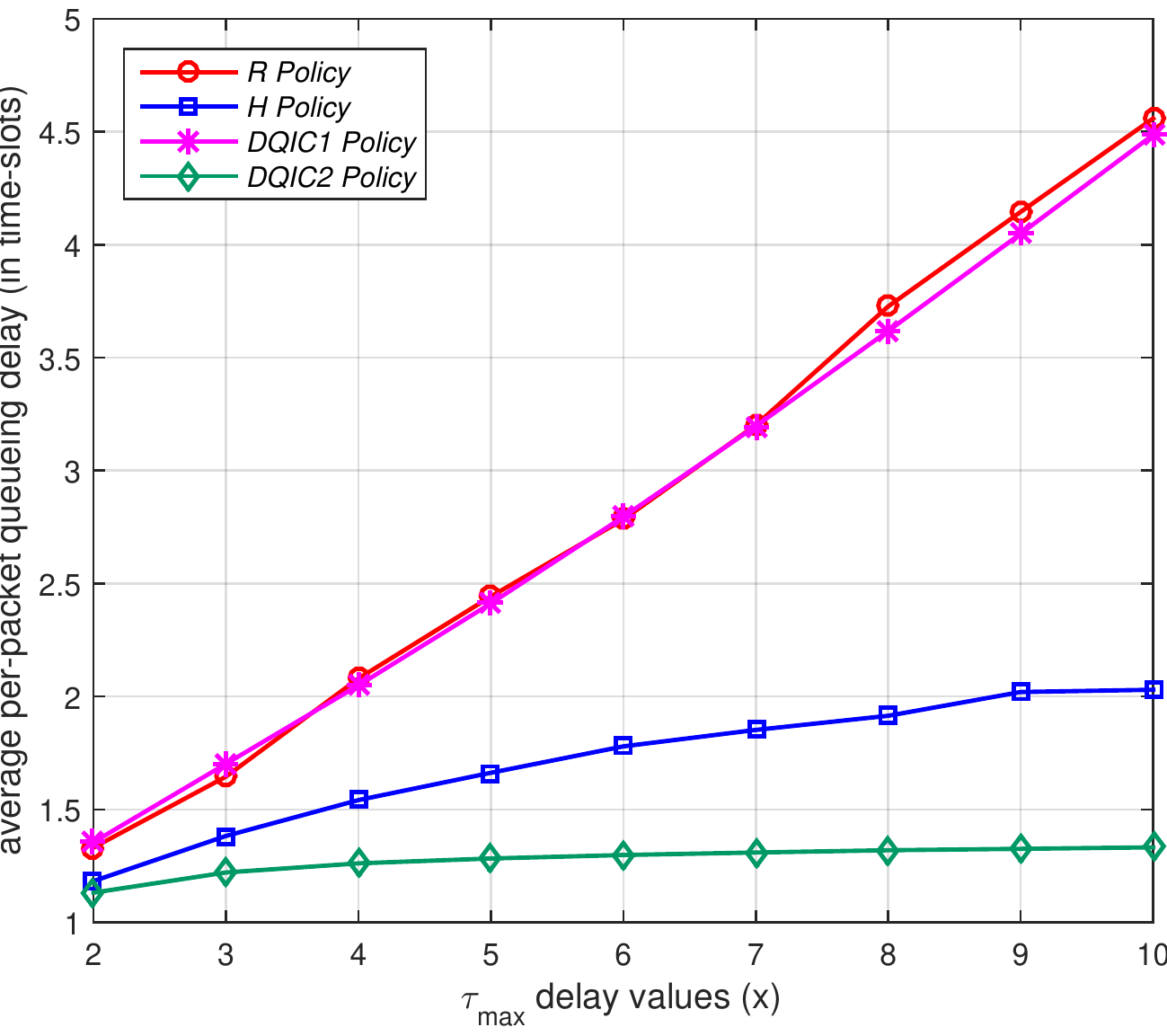}
\else
\includegraphics[width=.4\textwidth, keepaspectratio]{delay_performance_R_H_Policy1_Policy2.pdf}\\[-0.2cm]
{\centering \footnotesize (a) \normalsize}\\
\includegraphics[width=.4\textwidth, keepaspectratio]{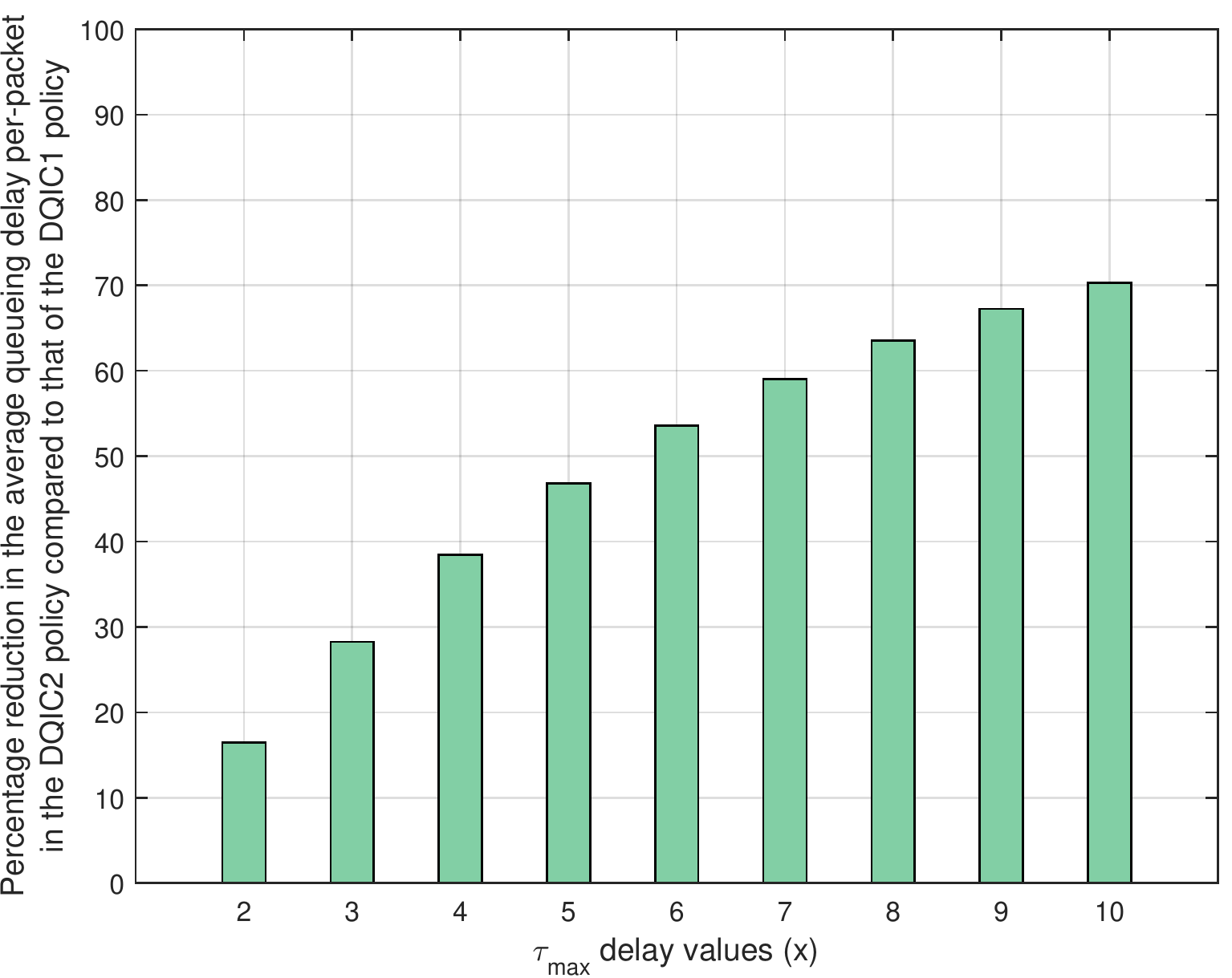}\\[-0.1cm]
{\centering \footnotesize (b) \normalsize}\\
\fi
\centering
\ifdefined\ONECOLUMN
\parbox{5in}{\caption{Comparison of average per-packet queueing delays of the $R$, $H$ (see Sec. \ref{section-The-H-Policy}), $DQIC1$, $DQIC2$, \LCVARIANTONE and \LCVARIANTTWO (see Sec. \ref{section-low-complexity-scheduling-policies})  policies. The $R$ and $DQIC1$ policies use $\tau_{max}$-delayed QSI whereas the $H$, $DQIC2$, \LCVARIANTONE and \LCVARIANTTWO policies use $\tau_{\textbf{.}, max}$-delayed QSI.\protect\footnotemark{}}\label{fig:delay_performance_R_H_DQIC1_DQIC2}
}
\else
\caption{(a) Comparison of average queueing delay per-packet in the $DQIC1$, $DQIC2$, $R$, and $H$ policies. The $R$ and $DQIC1$ policies use $\tau_{max}$-delayed QSI whereas the $H$ and $DQIC2$ policies use $\tau_{\textbf{.}, max}$-delayed QSI. The average per-packet queueing delays of the $DQIC1$ and $R$ policies grow linearly with $\tau_{max}$ whereas that of the $DQIC2$ policy tends to become almost unaffected by increase in $\tau_{max}$. Comparing the $DQIC1$, $DQIC2$ and $R$ policies and their delay performances, the fact that the $DQIC1$ and $R$ policies use $\tau_{max}$-delayed queue lengths isolates itself as the sole reason for their undesirable delay performances. The (relatively) small difference in the queueing delay performances of the $H$ and $DQIC2$ policies is due to the fact that the $DQIC2$ policy has access to instantaneous CSI whereas the $H$ policy only has access to delayed CSI. (b) Percentage reduction in the average queueing delay per-packet in the $DQIC2$ policy compared to that in the $DQIC1$ policy.}
\label{fig:delay_performance_R_H_DQIC1_DQIC2}
\fi
\end{figure}
\footnotetext{The \LCVARIANTONE and \LCVARIANTTWO policies are identical for a network with only two links (see Algorithm \ref{LC1Listing} in Sec. \ref{section-low-complexity-scheduling-policies}).}

%% file: section-The-H-Policy.tex
In Sec. \ref{section-Structure-of-Heterogeneously-Delayed-NSI}, we conjectured that $\{C_{l}[t - \tau_{l,max}]\}_{l \in \mathcal{L}}$, being less delayed compared to $\{C_{l}[t - \tau_{max}]\}_{l \in \mathcal{L}}$, and being commonly available at all transmitters in the network, it may be preferable that each transmitter base its transmit/no-transmit decision on this information rather than on $\{C_{l}[t - \tau_{max}]\}_{l \in \mathcal{L}}$. In addition to testing this hypothesis, we wish to reduce the computational complexity and the average per-packet queueing delay in comparison with that of the $R$ policy. Motivated by these, we define the $H$ policy to be the following scheduling policy. In each time slot,

\begin{itemize}[leftmargin=*]
	\item \textbf{Step 1:} using the common NSI available at all transmitters, each transmitter determines the optimal threshold function vector, by solving the following optimization problem:
	\begin{equation}
			\label{eqn:Opt_Step1_H_policy}
			{\displaystyle \argmax_{\textbf{T}} \sum_{l \in \mathcal{L}} Q_l[t - \tau_{l, max}] R_{l, \tau_{\textbf{.}, max}}(\textbf{T}) },
	\end{equation}	
\ifdefined\ONECOLUMN
	where
		\begin{equation}
		\label{eqn:R_l_tau_l_max}
		\begin{split}
			\textstyle R_{l, \tau}\left(\textbf{T}\right) := \textstyle \Ex[C_l\left[t\right] \idoperator_{\{C_l[t] \geq T_l(.)\}} (\gamma_l + \left(1 - \gamma_l\right) \textstyle \prod_{m \in I_l} \idoperator_{\{C_m[t] < T_m(.)\}} ) \; | \; \boldsymbol{\mathrm{C}}\left[t - \tau \right] \; ], 
		\end{split}
		\end{equation}
		\begin{equation*}
			\boldsymbol{\mathrm{C}}[t - \tau_{\textbf{.}, max}] := \{C_l[t - \tau_{l, max}]\}_{l \in \mathcal{L}}, \quad T_l: CS_l(\boldsymbol{\mathrm{C}}[t](0 : \tau_{\textbf{.}, max})) \rightarrow \mathbb{R}, \nonumber
		\end{equation*}
		\begin{equation}
		\begin{split}
		\label{equation-critical-set-of-link-l-H-policy}
		\mathcal{CS}_l(\boldsymbol{\mathrm{C}}[t](0 : \tau_{\textbf{.}, max})) := \mathcal{CS}(\boldsymbol{\mathrm{C}}[t](0 : \tau_{\textbf{.}, max})) \cap \mathcal{P}_l(\boldsymbol{\mathrm{C}}[t](0 : \tau_{\textbf{.}, max})),
		\end{split}
		\end{equation}
\begin{equation}
	\label{equation-critical-set-H-policy}
	\mbox{and } \mathcal{CS}(\boldsymbol{\mathrm{C}}[t](0 : \tau_{\textbf{.}, max})) := \{ \{ C_l[t - \tau_k(l)] \}_{k \in \mathcal{L}, k \neq l} \}_{l \in \mathcal{L}}.
\end{equation}
\else
	where
		\begin{equation}
		\label{eqn:R_l_tau_l_max}
		\begin{split}
			\textstyle R_{l, \tau}\left(\textbf{T}\right) := & \textstyle \Ex\Bigg[C_l\left[t\right] \; \One_{\{C_l[t] \geq T_l(.)\}} \Big(\gamma_l + \left(1 - \gamma_l\right) \\
			&\quad \prod_{m \in I_l}
			\One_{\{C_m[t] < T_m(.)\}} \Big) \; \bigg\rvert \; 
			 \boldsymbol{\mathrm{C}}\left[t - \tau \right] \; \Bigg], 
		\end{split}
		\end{equation}
	$\quad \boldsymbol{\mathrm{C}}[t - \tau_{\textbf{.}, max}] := \{C_l[t - \tau_{l, max}]\}_{l \in \mathcal{L}}$,
\begin{flalign}
&\quad T_l: CS_l(\boldsymbol{\mathrm{C}}[t](0 : \tau_{\textbf{.}, max})) \rightarrow \mathbb{R}, \nonumber &
\end{flalign}
\begin{equation}
\begin{split}
\label{equation-critical-set-of-link-l-H-policy}
\mathcal{CS}_l(\boldsymbol{\mathrm{C}}[t](0 : \tau_{\textbf{.}, max})) & :=  \mathcal{CS}(\boldsymbol{\mathrm{C}}[t](0 : \tau_{\textbf{.}, max})) \\ 
&\quad\quad \cap \mathcal{P}_l(\boldsymbol{\mathrm{C}}[t](0 : \tau_{\textbf{.}, max})),
\end{split}
\end{equation}
\begin{equation}
	\label{equation-critical-set-H-policy}
	\mbox{and } \mathcal{CS}(\boldsymbol{\mathrm{C}}[t](0 : \tau_{\textbf{.}, max})) := \{ \{ C_l[t - \tau_k(l)] \}_{k \in \mathcal{L}, k \neq l} \}_{l \in \mathcal{L}}.
\end{equation}
\fi

	\item \textbf{Step 2:} each transmitter observes its current critical NSI, evaluates its threshold function (found in Step 1) at this critical NSI, and attempts to transmit if and only if its current channel rate exceeds the threshold value, i.e., when 
	$C_l[t] \geq T_l(CS_l(\boldsymbol{\mathrm{C}}[t](0 : \tau_{\textbf{.}, max})))$. \vspace{0.005cm}
\end{itemize}

\vspace{0.2cm}
\noindent
\footnotesize 
\textcolor{darkgray}{\textit{$Example \text{ } \stepcounter{ExampleCtr}\theExampleCtr$}: Continuing Example \ref{Ex:NSI-H-policy}, for the $H$ policy, the critical NSI of the network $\mathcal{CS}(\boldsymbol{\mathrm{C}}[t](0 : \tau_{\textbf{.}, max}))$ is the set $\{C_{l_1}[t-1], C_{l_1}[t-3], C_{l_2}[t-2], C_{l_2}[t-4], C_{l_3}[t-1], C_{l_3}[t-2]\}$ and the critical NSI available at the transmitter nodes of links $l_1, l_2, l_3$, namely $\mathcal{CS}_{l_1}(\boldsymbol{\mathrm{C}}[t](0 : \tau_{\textbf{.}, max})), \mathcal{CS}_{l_2}(\boldsymbol{\mathrm{C}}[t](0 : \tau_{\textbf{.}, max})), \mathcal{CS}_{l_3}(\boldsymbol{\mathrm{C}}[t](0 : \tau_{\textbf{.}, max}))$, respectively are $\{C_{l_1}[t-1], C_{l_1}[t-3], C_{l_2}[t-2], C_{l_2}[t-4], C_{l_3}[t-1], C_{l_3}[t-2]\}$, $\{C_{l_1}[t-1], C_{l_1}[t-3], C_{l_2}[t-2], C_{l_2}[t-4], C_{l_3}[t-2]\}$, and $\{C_{l_1}[t-3], C_{l_2}[t-4], C_{l_3}[t-1], C_{l_3}[t-2]\}$. \hfill$\Box$}\\
\normalsize

\vspace{0.1cm}
\noindent
\textit{Remark: It is to be noted that the $R$ and $H$ policies differ only in the delay values  that they use to access queue lengths and channel state information, with the $R$ policy using the delay value $\tau_{max}$ and the $H$ policy using the delay value $\tau_{l, max}$ to access the NSI of link $l$}.

\subsection{Computational Complexity}
\label{section-Computational Complexity of the $H$ Policy}
In this section, similar to the way that we dealt with the $R$ policy in Sec. \ref{section-A-Note-on-the-Computation-Complexity-of-R-Policy}, we first consider the complexity of functional evaluation and obtain an expression for the number of threshold function vectors required in the domain of optimization in the general case. Next, we characterize the structure on the delay values (in the table of delay values) that produces the worst-case functional evaluation complexity in the $H$ policy. We then consider the complexity of functional evaluation in this worst-case setting, and obtain an expression for the number of threshold function vectors required in the domain of optimization in the worst case.
\begin{proposition}
	\label{lemma-characterization-of-general-case-func-eval-complexity-of-H-policy}
	For the $H$ policy, the total number of threshold functions that are needed to be considered in the domain of optimization in \Expr (\ref{eqn:Opt_Step1_H_policy}), in general, is $(\mathcal{C}+1)^{\sum_{l_i \in \mathcal{L}}\mathcal{C}^{|\mathcal{W}_i|}}$, where $\mathcal{C}$ is the number of channel states, $\mathcal{L}$ is the set of links in the network, and $\mathcal{W}_i \coloneqq CS_{l_i}(\boldsymbol{\mathrm{C}}[t](0 : \tau_{\textbf{.}, max})) \setminus \{C_{l_i}[t - \tau_{l_i, max}]\}_{l_i \in \mathcal{L}}$ is the set of input parameters to the threshold function $T_{l_i}(.)$.
\end{proposition}
We present a proof of this result in Appendix \ifelseEV{\ref{appendix-proof-characterization-of-general-case-func-eval-complexity-of-H-policy}}{A}. We illustrate this result in Example \ref{Ex:func-eval-and-sample-path-compl-H-policy}. Next, we present the following result that will be required later:
\begin{proposition}
\label{lemma-critical-set-of-link-l-R-policy-equals-critical-set-of-link-l-H-policy}
$CS_l^R(\boldsymbol{\mathrm{C}}[t](0 : \tau_{max})) = CS_l(\boldsymbol{\mathrm{C}}[t](0 : \tau_{\textbf{.}, max}))$.
\end{proposition}
We provide a proof of this proposition in Appendix \ifelseEV{\ref{appendix-proof-lemma-critical-set-of-link-l-R-policy-equals-critical-set-of-link-l-H-policy}}{D}. We next characterize the structure on the delay values that brings out the worst-case functional evaluation complexity in the $H$ policy.
\begin{proposition}
\label{lemma-characterization-of-worst-case-delay-values-for-H-policy}
The worst-case complexity of functional evaluation in the $H$ policy is realized when the delays in the table of delay values are all distinct, and the delay values at positions $(i,j)$ in row $i$ of the table of delay values, for $j = 1, 2, \dots, i-1, i+1, \ldots, L$ appear in descending order, for all $i$.
\end{proposition}
\indent
We highlight a minor difference in the proof of this proposition compared to that of Proposition \ref{lemma-characterization-of-worst-case-delay-values-for-R-policy}, in Appendix \ifelseEV{\ref{appendix-proof-lemma-characterization-of-worst-case-delay-values-for-H-policy}}{E}. We now characterize the functional evaluation complexity of the $H$ policy for the worst-case setting noted in Proposition \ref{lemma-characterization-of-worst-case-delay-values-for-H-policy}.
\begin{proposition}
\label{lemma-characterization-of-worst-case-func-eval-complexity-of-H-policy}
For the $H$ policy, the total number of threshold functions that are needed to be considered in the domain of optimization in \Expr (\ref{eqn:Opt_Step1_H_policy}), for the worst-case scenario noted in Proposition \ref{lemma-characterization-of-worst-case-delay-values-for-H-policy}, is $(\mathcal{C}+1)^{\sum_{l_i \in \mathcal{L}}\mathcal{C}^{i(L-2)}}$, where $\mathcal{C}$ is the number of channel states and $L$ is the number of links in the network.
\end{proposition}
We present a proof of this result in Appendix \ifelseEV{\ref{appendix-proof-lemma-characterization-of-worst-case-func-eval-complexity-of-H-policy}}{F}. Next, we characterize the sample-path complexity of the $H$ policy.
\begin{proposition}
\label{lemma-characterization-of-sample-path-complexity-of-H-policy}
For the $H$ policy, the number of sample paths that are needed to be considered in the evaluation of the conditional expectation in \Expr (\ref{eqn:R_l_tau_l_max}) is given by $\mathcal{C}^{\sum_{l \in \mathcal{L}} \tau_{l, max}}$, where $\mathcal{C}$ is the number of channel states.
\end{proposition}
\indent
The proof of this proposition is similar to the proof of Proposition \ref{lemma-characterization-of-sample-path-complexity-of-R-policy} and is therefore omitted. We now illustrate with an example the reduction in computational complexity that the $H$ policy achieves over that of the $R$ policy. 

\vspace{0.1cm}
\noindent
\footnotesize
\textcolor{darkgray}{\textit{Example \refstepcounter{ExampleCtr}\theExampleCtr\label{Ex:func-eval-and-sample-path-compl-H-policy}}: Consider the same network as in Example \ref{Ex:func-eval-and-sample-path-compl}. After computing $CS_{l_1}(.)$, $CS_{l_2}(.)$ and $CS_{l_3}(.)$ using Eq. (\ref{equation-critical-set-of-link-l-H-policy}), we see that given $\{C_l[t - \tau_{l, max}]\}_{l \in \mathcal{L}}$, $T_{l_1}(.), T_{l_2}(.), T_{l_3}(.)$ are functions of $1, 2$ and $3$ variables respectively. Using the same arguments as in the proofs of Propositions \ref{lemma-characterization-of-worst-case-func-eval-complexity-of-R-policy} and \ref{lemma-characterization-of-worst-case-func-eval-complexity-of-H-policy} in Appendices \ifelseEVWithoutCite{\ref{appendix-proof-characterization-of-worst-case-func-eval-complexity-of-R-policy}}{B} and \ifelseEV{\ref{appendix-proof-lemma-characterization-of-worst-case-func-eval-complexity-of-H-policy}}{F}, the number of choices of threshold functions for $T_{l_1}(.), T_{l_2}(.)$ and $T_{l_3}(.)$ are $3^{2}$, $3^{4}$ and $3^8$ respectively. Therefore, the total number of threshold function vectors \textbf{T} in the domain of optimization is $3^{14}$ (the $R$ policy required $3^{56}$ threshold functions; see Example \ref{Ex:func-eval-and-sample-path-compl}) and the number of sample paths required to be considered in evaluating the conditional expectation is $2^{32}$ (the $R$ policy required considering $2^{36}$ sample paths; see Example \ref{Ex:func-eval-and-sample-path-compl}), yielding an enormously massive reduction in computational complexity over that of the $R$ policy. We note that this vast reduction is mainly due to the fact that for the $H$ policy, the exponent in the double exponential in the expression in Proposition \ref{lemma-characterization-of-worst-case-func-eval-complexity-of-H-policy} is smaller than that of the $R$ policy.}
\normalsize

\subsection{Delay Performance}
Fig. \ref{fig:delay_performance_R_H_DQIC1_DQIC2} shows the average per-packet queueing delay (in units of time-slots) in the $H$ policy for the setting of the example in Sec. \ref{section-Delay-Performance-Preview}. The average per-packet queueing delay of the $R$ policy grows linearly with $\tau_{max}$ whereas that of the $H$ policy tends to flatten out. Comparing the $R$, $H$, $DQIC1$ and $DQIC2$ policies and their delay performances, it is clearly evident that the use of $\tau_{\textbf{.}, max}$-delayed queue lengths in the $H$ and $DQIC2$ polices is what gives these policies their better delay performances in comparison to those of the $R$ and $DQIC1$ policies which use $\tau_{max}$-delayed queue lengths.



\subsection{Throughput Optimality}
In this section we show that, like the $R$ policy, the $H$ policy too is throughput optimal. First, we prove that if an arrival process is supportable, then the expected arrival rate of this process should lie within the system throughput region $\Lambda$ defined in Sec. \ref{section-Characterization-of-Throughput-Region}. This would then mean that the system throughput region $\Lambda$ is the region that encompasses all supportable arrival rates given the NSI structure in Sec. \ref{section-Structure-of-Heterogeneously-Delayed-NSI}. Next, we show that the $H$ policy stabilizes all arrival rate vectors in the system throughput region $\Lambda$. These two together would then imply that the $H$ policy is throughput optimal.

\begin{lemma}
\label{lemma-Lambda-encompasses-all-supportable-arrival-rates}
Under the NSI structure noted in Sec. \ref{section-Structure-of-Heterogeneously-Delayed-NSI}, if the traffic arrival process $\{A[t]\}_t$ is supportable, then $E[A[t]] \in \Lambda$.
\end{lemma}

The proof of this lemma, which is similar to the proof of Lemma 4.1 in \cite{Reddy_et_al_12}, is available in Appendix \ifelseEV{\ref{appendix-proof-lemma-Lambda-encompasses-all-supportable-arrival-rates}}{G}.

\begin{corollary}
The system throughput regions defined in Equation (\ref{eqn:Throughput_Region}), and in Sec. 4.1 in \cite{Reddy_et_al_12}, are identical.
\end{corollary}

This is an immediate consequence of Lemma 4.1 in \cite{Reddy_et_al_12} and Lemma \ref{lemma-Lambda-encompasses-all-supportable-arrival-rates} stated above. See Sec. 4.1 in \cite{Reddy_et_al_12} for the definition of the system throughput region considered in \cite{Reddy_et_al_12}.

\begin{theorem}
\label{theorem-H-policy-is-throughput-optimal}
The $H$ policy is throughput optimal.
\end{theorem}

The proof of this theorem, which is similar to the proof of the corresponding theorem for the $R$ policy in \cite{Reddy_et_al_12} (but with significantly more details) is available in Appendix \ifelseEV{\ref{appendix-proof-of-theorem-H-policy-is-throughput-optimal}}{H}. \textit{An important implication of this theorem is that using $\tau_{l, max}$-delayed NSI instead of $\tau_{max}$-delayed NSI (for each link $l$) does not harm the achievable system throughput, a fact that we exploit crucially in designing our computationally efficient near-throughput-optimal policies in Sec. \ref{section-low-complexity-scheduling-policies}}.


%% file: section-Analytical-Characterization-of-Delay-Performance.tex
In this section, we characterize the average queueing delay per packet in the $DQIC1$, $DQIC2$, and $R$ policies analytically. To do this, we first fix a time-horizon, consider all possible packet arrival patterns and all possible channel states in each slot within this time horizon and compute the average delay experienced by a packet in each of the above scheduling policies in the limit as the time-horizon tends to infinity. We need the following additional notation:\\

\hspace{-0.75cm}
\begin{tabular}{rp{0.35\textwidth}}
 $T \coloneqq$ & Time horizon (i.e, the number of time slots over which we measure the average delay per packet) \\[0.25cm] 

 $N_i \coloneqq$ & Number of packets that have been serviced from time $t = 0$ to $t = T$ from the queue associated with link $l_i$ \\[0.1cm] 
 
 $ N \coloneqq$ $\displaystyle \sum_{i=1}^{\mathcal{L}} N_i$	& is the total number of packets in the system that have been serviced from time $t = 0$ to $t = T$ from the queue associated with any link in the network \\[0.2cm] 
\end{tabular}

\begin{tabular}{rp{0.35\textwidth}}
 $a_{l,j} \coloneqq$ & $\twopartdef{ \begin{minipage}{0.2\textwidth}The arrival time of the $j$th packet from the head of line in the queue associated with link $l$ \vspace{0.1in} \end{minipage}}{j \leq Q_l[t]}{\infty}{\text{otherwise}}$         \\[0.9cm]
 
 $\One_{\{a\}} \coloneqq$ & $\twopartdef{ \begin{minipage}{0.001\textwidth} 1  \end{minipage}}{a \text{ is true} }{0}{\text{otherwise}}$         \\[0.4cm]

 $\One_{\{a,b\}} \coloneqq$ & $ \One_{\{a\}} \times \One_{\{b\}}$ \\[0.2cm]

 $x^{+} \coloneqq $ & $\max\{0, x\}$ \\[0.1cm] 
	&  \\ 
\end{tabular} 







With these definitions in place, and with the assumption that the queues are purged of the packets that were transmitted in a particular time slot at the end of that time slot, we are now ready to characterize the average per-packet queueing delay in these policies. First, we characterize $\bar{D}_{arrivals, CSI}$ -- the long-term average queueing delay conditioned on the knowledge of the channel states of all links in the network for all time slots (i.e., given $\{C_l[t]\}_{l \in \mathcal{L}}$, for all $t$ such that $-\tau_{max} \leq t \leq T-1$) and conditioned on the availability of the actual arrivals at all links in the network for all time slots (i.e., $\{A_l[t]\}_{l \in \mathcal{L}}$, for all $t$ such that $-\tau_{max} \leq t \leq T-1$) as follows:

\begin{equation}
\label{Eqn:D_bar_arrivals_CSI}
\displaystyle \bar{D}_{arrivals,CSI} \coloneqq \lim\limits_{T \rightarrow \infty} \sum_{t = -\tau_{max}}^{T-1} \sum_{i = 1}^{\mathcal{L}} \One_{\{.\}} \sum_{j = 1}^{C_{l_i}[t]} ( t - a_{l_i,j})^{+} \times \frac{1}{N},
\end{equation}

\noindent
where the indicator function in Eq. (\ref{Eqn:D_bar_arrivals_CSI}) indicates the condition under which link $l_i$ is scheduled for transmission in time slot $t$, and is is defined, for the various policies, as follows:\footnote{\label{footnote:minus-tau-max} Note that the time index $t$ starts from $-\tau_{max}$ in Eq. (\ref{Eqn:D_bar_arrivals_CSI}) since at time $t = 0$, $\tau_{max}$-delayed NSI would be the NSI at time $t = -\tau_{max}$. Note that $N_i$ (and hence $N$) does not include the packets serviced in time slots $t = -\tau_{max}$ to $t = -1$.}\\

\noindent
For the $DQIC1$ Policy:
\[\One_{\{.\}} \coloneqq
\left\{
	\begin{array}{l}
		\One_{ \left\{Q_{l_i}[t] \times C_{l_i}[t] \, > \, \max\limits_{k<i} \{Q_{l_k}[t] \times C_{l_k}[t]\}, \right. } \\ 
		{}_{\left. \hspace{0.4in} Q_{l_i}[t] \times C_{l_i}[t] \, \geq \, \max\limits_{k>i} \{Q_{l_k}[t] \times C_{l_k}[t]\} \right\}  \quad \mbox{if } t < 0 } \\
		\One_{ \left\{Q_{l_i}[t - \tau_{max}] \times C_{l_i}[t] \, > \, \max\limits_{k<i} \{Q_{l_k}[t - \tau_{max}] \times C_{l_k}[t]\}, \right. } \\ 
		{}_{\left. \hspace{-0.1in} Q_{l_i}[t - \tau_{max}] \times C_{l_i}[t] \, \geq \, \max\limits_{k>i} \{Q_{l_k}[t - \tau_{max}] \times C_{l_k}[t]\} \right\}  \;\; \mbox{if } t \geq 0 }
	\end{array}
\right.
\]

In the indicator function in the expressions above, we note that the splitting of the comparison of the product of queue-length and channel state on link $l_i$ with that on the other links, into two -- namely, (i) comparison with that on links $l_k, k < i$, and (ii) comparison with that on links $l_k, k > i$, creates a lexicographic ordering among the links. This is required to consistently resolve the ``winner'' in case there is a tie in the queue-length channel-state product on multiple links -- we always resolve in favor of the smallest $i$ (as a convention) in case of a tie.\\

\noindent
For the $DQIC2$ Policy:
\[\One_{\{.\}} \coloneqq
\left\{
	\begin{array}{l}
		\One_{ \left\{Q_{l_i}[t] \times C_{l_i}[t] \, > \, \max\limits_{k<i} \{Q_{l_k}[t] \times C_{l_k}[t]\}, \right. } \\ 
		{}_{\left. \hspace{0.4in} Q_{l_i}[t] \times C_{l_i}[t] \, \geq \, \max\limits_{k>i} \{Q_{l_k}[t] \times C_{l_k}[t]\} \right\}  \quad \mbox{if } t < 0 } \\
		\One_{ \left\{Q_{l_i}[t - \tau_{l, max}] \times C_{l_i}[t] \, > \, \max\limits_{k<i} \{Q_{l_k}[t - \tau_{l, max}] \times C_{l_k}[t]\}, \right. } \\ 
		{}_{\left. \hspace{-0.1in} Q_{l_i}[t - \tau_{l, max}] \times C_{l_i}[t] \, \geq \, \max\limits_{k>i} \{Q_{l_k}[t - \tau_{l, max}] \times C_{l_k}[t]\} \right\}  \; \mbox{if } t \geq 0 }
	\end{array}
\right.
\]

\noindent
For the $R$ Policy: 
$\One_{\{.\}} \coloneqq
\One_{\big\{C_{l_i}[t] \geq T_{l_i}\big(CS_{l_i}^R\left(\boldsymbol{\mathrm{C}}[t]\left(0 : \tau_{max} \right) \right)\big) \big\} }
$

\noindent
For the $H$ Policy:
$\One_{\{.\}} \coloneqq
\One_{\big\{C_{l_i}[t] \geq T_{l_i}\big(CS_{l_i}\left(\boldsymbol{\mathrm{C}}[t]\left(0 : \tau_{\textbf{.}, max} \right) \right) \big) \big\} }
$\\

Note that the knowledge of the number of packet arrivals into the queue associated with link $l_i$ in each time slot $t$, for $-\tau_{max} \leq t \leq T-1$ is subsumed in the $a_{l_i,j}$ term and the knowledge of the channel state on link $l_i$ in each time slot $t$, for $-\tau_{max} \leq t \leq T-1$ is accounted for in the fact that the queues are purged of the packets transmitted in a time slot at the end of that time slot, which in turn affects the $a_{l_i,j}$ term. Also note that any packets remaining in the queues after time slot $T-1$ are not serviced and hence do not contribute to the average queueing delay.

Now, removing the conditioning on knowledge of CSI, we get the expression for $\bar{D}_{arrivals}$ -- the average per-packet queueing delay (conditioned only on the arrivals) as follows:

\begin{equation}
\begin{aligned}
\label{Eqn:D_bar_arrivals}
\displaystyle \bar{D}_{arrivals} \coloneqq \hspace{-1.5cm} \sum\limits_{\substack{\qquad\quad\big\{\{c_l[t]\}_{l \in \mathcal{L}}\big\} \operatorname*{}\limits_{t = -\tau_{max}}^{{}^{{}^{T-1}} \hfill}} } \hspace{-1.4cm} \bar{D}_{arrivals, CSI} & \prod_{m = 1}^{\mathcal{L}} \Biggl(\pi \left(c_m[-\tau_{max}] \right)  \\
 & \prod_{n = -\tau_{max}}^{T-2} p_{c_m[n] \, c_m[n+1]} \Biggr)
\end{aligned}
\end{equation}

\noindent
where $\pi(c)$ is the steady-state probability of being in state $c$ in the channel-state Markov chain, and $p_{ij}$ is the one-step transition probability from state $i$ to  state $j$ in the channel-state Markov chain.

Finally, removing the conditioning on the knowledge of arrivals on each link in each time slot $t, -\tau_{max} \leq t \leq T-1$, we get the expression for $\bar{D}$ -- the required average per-packet queueing delay as follows:

\begin{equation}
\label{Eqn:D_bar}
\displaystyle \bar{D} \coloneqq \hspace{-1.4cm} \sum\limits_{ \substack{\qquad\quad \big\{\{a_l[t]\}_{l \in \mathcal{L}} \big\} \operatorname*{}\limits_{t = -\tau_{max}}^{{}^{{}^{T-1}} \hfill} } } \hspace{-1.0cm} \bar{D}_{arrivals} \; \prod_{l = 1}^{\mathcal{L}} \Biggl(\prod_{t = -\tau_{max}}^{T-1} \Pr \left( A_l[t] = a_l[t] \right) \Biggr)
\end{equation}

\subsection{Computational Complexity}
\label{subsec:computationalcomplexity}
Since Expr. (\ref{Eqn:D_bar}) is not in closed form, we will have to resort to numerical evaluation of this expression for a fixed finite value of $T$. For a fixed finite $T$, the number of time slots over which Expr. (\ref{Eqn:D_bar}) needs to be evaluated is $\tau_{max} + T$. At each link in the network, the number of packet arrivals in each time slot can be any of the values in the set $\{0, 1, \ldots, A_{max}\}$, where $A_{max}$ is the maximum number of packets that can arrive in each time slot (see Sec. \ref{subsec:Traffic-Model-and-Queue-Dynamics}). Thus, the number of possible arrival streams at a link (over all the $\tau_{max} + T$ time slots), is given by $(A_{max}+1)^{(\tau_{max}+T)}$, and therefore the number of possible arrival streams taking all the links together is $(A_{max}+1)^{L(\tau_{max}+T)}$. Further, in case of Markov chains where any state can be reached from any other state in one step (as in the setting considered in Sec. \ref{subsec:Results-Discussion-Queueing-Delay}), the channel state on each link can be any of the $\mathcal{C}$ states, where $\mathcal{C}$ is the number of states in the channel Markov chain. Thus, for each possible arrival stream pattern at each link, over the fixed finite time horizon $T$, the number of possible channel state transition patterns at a link is given by $\mathcal{C}^{(\tau_{max}+T)}$, and therefore the number of possible channel state transition patterns taking all the links together is $\mathcal{C}^{L(\tau_{max}+T)}$. Thus, using Eq. (\ref{Eqn:D_bar}), the computational complexity of comprehensively evaluating the average queueing delay per-packet over a fixed finite $T$ is $(A_{max}+1)^{L(\tau_{max}+T)} \times \mathcal{C}^{L(\tau_{max}+T)}$ or $((A_{max}+1)\mathcal{C})^{L(\tau_{max}+T)}$ operations. Thus, for example, in a network with 2 transmitters, where in each time slot, the number of packet arrivals can only be $0$ or $1$, and where the channel state Markov chain has only $2$ states, for a time horizon of $T=10$, and for $\tau_{max} = 2$, the computational complexity is $2^{48}$ (or, roughly $10^{14}$) operations.

%% file: section-low-complexity-scheduling-policies.tex
The $H$ policy, despite the immense reduction in computational complexity that it achieves in comparison to the $R$ policy, is still computationally complex, and therefore impractical. In this section, we propose and evaluate two fast and near-throughput-optimal scheduling policies -- namely, \LCVARIANTONE (for \textbf{E}liminate link with \textbf{L}east \textbf{D}ata \textbf{R}ate), and \LCVARIANTTWO (for \textbf{E}liminate link that \textbf{R}educes \textbf{D}elays for \textbf{M}aximum number of \textbf{C}hannels) -- for the heterogeneously delayed NSI setting, confining our attention to the more pragmatic case of perfect collision interference (i.e., $\gamma_l = 0, \; \forall l \in \mathcal{L}$). Initially, we consider the case of complete interference (i.e., $I_l = \mathcal{L} \setminus \{l\} \; \forall l \in \mathcal{L}$), and consider extension to the case of multiple interference sets subsequently. These low-complexity scheduling policies derive their main idea from the $H$ policy; the idea being that the common delay value that all the contending transmitters (i.e., transmitters corresponding to contending links) can use to access the NSI of some link $l$ (of at least one link), could be reduced (and hence more recent [i.e., more reliable] delayed link-statistics could be made use of in computing the schedule) by carefully choosing to eliminate one link from among the contending links. These policies are iterative (we call each iteration, a ``round'') and they operate by choosing to eliminate one link in each round; the two policies differ only in the criterion they use for selecting the link that they eliminate in each round. Further, eliminating one link in each round accords these policies polynomial running times as we show in Sec. \ref{subsec:computational-complexity-of-LC1-LC2-policies}. We list the \LCVARIANTONE policy in Algorithm \ref{LC1Listing} and demonstrate its working in detail in Sec. \ref{subsec:dynamics-of-LC-ELDR-policy}. The \LCVARIANTTWO policy is identical to the \LCVARIANTONE policy listed in Algorithm \ref{LC1Listing} except for step \ref{step:S} which is listed separately in Algorithm \ref{LC2Listing}. We derive analytical expressions for the exact expected saturated system throughputs of the \LCVARIANTONE and \LCVARIANTTWO policies in Sec. \ref{subsection:throughput-near-optimality}, and plot these expressions in Sec. \ref{section-numerical-results} and show that these policies are near-optimal.

\begin{algorithm}[p]
\small
\caption{}\label{LC1Listing}
\begin{algorithmic}[1]
\Procedure{\LCVARIANTONE}{\textit{ActiveSet}} \vspace{.1cm}
\State \parbox[t]{\dimexpr\linewidth-\algorithmicindent-0.6cm}{$\textit{I} \gets$ all links in the network for which this node\footnotemark{} is the transmitter \strut} \vspace{.1cm}
\While{$|ActiveSet| > 2$} \vspace{.1cm}
\State \label{step:recompute_tau_l_max} \parbox[t]{\dimexpr\linewidth-\algorithmicindent-0.55cm}{Compute $\tau_{l, max} \; \forall l \in $ \textit{ActiveSet} from the delay table after suppressing the rows and columns corresponding to links not in \textit{ActiveSet} \strut} \vspace{.1cm}
\State \label{step-compute-queue-length-weighted-conditional-expected-data-rates} \parbox[t]{\dimexpr\linewidth-\algorithmicindent-0.55cm}{Compute the queue-length weighted expected data rate that will be realized if link $l$ is allowed to carry transmission given the $\tau_{l, max}$-delayed channel state of link $l$, for all links $l$ in \textit{ActiveSet}, as follows: $Q_l[t - \tau_{l, max}] \times E\big[C_l[t] \, | \, C_l[t - \tau_{l, max}] = c_{l, \tau_{l, max}} \big]$, where $c_{l, \tau_{l, max}}$ is the channel-state realization of link $l$ at time $t - \tau_{l, max}$\strut} \vspace{.18cm}
\State \label{step:H} \parbox[t]{\dimexpr\linewidth-\algorithmicindent-0.55cm}{Let $H$ be the link with the largest queue-length weighted expected data rate computed in Step \ref{step-compute-queue-length-weighted-conditional-expected-data-rates} (arbitrarily chosen if more than one satisfy this criterion)\strut} \vspace{.15cm}
\State \label{step:EC} \parbox[t]{\dimexpr\linewidth-\algorithmicindent-0.55cm}{Let $EC$ (for \textbf{E}limination \textbf{C}andidates) be the subset of $ActiveSet \setminus \{H\}$ such that if link $K \in ActiveSet \setminus \{H\}$ then link $K \in EC$ if, on recomputing the delays $\tau_{l, max} \; \forall l \in $ \textit{ActiveSet} after masking the rows and columns corresponding to link $K$ and the links not in \textit{ActiveSet} from the table of delay values, there is a reduction in the $\tau_{l, max}$ value of at least one link $l \in ActiveSet \setminus \{K\}$ \strut} \vspace{.18cm}
\If {$EC = \phi$ (the empty set)} 
	\If {$H \in I$} 
		\State \parbox[t]{\dimexpr\linewidth-\algorithmicindent-1.6cm}{Set transmit decision of $H$ = \textsc{Transmit}; For all $l \in I \setminus \{H\}$, set transmit decision of $l$ = \textsc{NoTransmit} \strut} 
	\Else 
		\State \parbox[t]{\dimexpr\linewidth-\algorithmicindent-1.6cm}{For all $l \in I$, set transmit decision of $l$ = \textsc{NoTransmit}} \vspace{.1cm}
	\EndIf \vspace{.1cm}
\State Exit.
\Else \vspace{.1cm}
\State \label{step:S} \parbox[t]{\dimexpr\linewidth-\algorithmicindent-1.05cm}{Let $S$ be the link in $EC$ with the lowest queue-length weighted expected data rate computed in Step \ref{step-compute-queue-length-weighted-conditional-expected-data-rates} (arbitrarily chosen if more than one satisfy this criterion)\strut} \vspace{.15cm}
	\If {$S \in I$} 
		\State \parbox[t]{\dimexpr\linewidth-\algorithmicindent-1.7cm}{Set transmit decision of $S$ = \textsc{NoTransmit}. $I \gets I \setminus \{S\}$ \strut} 
		\State \textbf{if} $I = \phi$ \textbf{then} Exit.
	\EndIf \vspace{.1cm}
\State $ActiveSet \gets ActiveSet \setminus \{S\}$ \vspace{.1cm}
\EndIf
\EndWhile \vspace{.1cm}
\State \label{step-recompute-delays-and-expected-rates-one-last-time}\parbox[t]{\dimexpr\linewidth-\algorithmicindent}{Recompute the delays $\tau_{l, max}$ as in Step \ref{step:recompute_tau_l_max}. Recompute the queue-length weighted expected data rates as in Step \ref{step-compute-queue-length-weighted-conditional-expected-data-rates}. Let $T \in ActiveSet$ be the link with the largest expected data rate (arbitrarily chosen if more than one satisfy this criterion)\strut} \vspace{.1cm}
\If {$T \in I$} 
\State \parbox[t]{\dimexpr\linewidth-\algorithmicindent-0.7cm}{Set transmit decision of $T$ = \textsc{Transmit}. For all $l \in I \setminus \{T\}$, set transmit decision of $l$ = \textsc{NoTransmit}\strut} 
\Else 
\State \parbox[t]{\dimexpr\linewidth-\algorithmicindent-0.7cm}{For all $l \in I$, set transmit decision of $l$ = \textsc{NoTransmit}\strut} 
\EndIf \vspace{.1cm}
\EndProcedure
\end{algorithmic}
\end{algorithm}
\footnotetext{the node where this algorithm is being run}

\subsection{Dynamics of the \LCVARIANTONE Policy}
\label{subsec:dynamics-of-LC-ELDR-policy}
We now demonstrate the working of the \LCVARIANTONE policy for a network with four wireless links -- $l_1$, $l_2$, $l_3$, $l_4$ (with $A$, $B$, $C$, $D$ as the transmitter nodes on these links, respectively), all contending for transmission in the current time slot. We will assume that the queues at the transmitter node of these links are saturated, and that the heterogeneous delays are as in Table \ref{tab:Het-Delays-for-Example-Network}. Let each of the wireless links be modeled as a Markov chain on the state space $\mathcal{C} = \{1, 2\}$ with crossover probability $0.1$ (the case of \textit{very slow varying channel}). The $n$-step transition probability matrices for this Markov chain are as below (shown for only those delays [$n$-values] that are required in our computation):
\input{Het-Delays-for-Example-Network}
\ifdefined\ONECOLUMN
\begin{flalign*}
\quad\quad\quad\quad P^{(1)} &= \left[ \begin{array}{cc}
0.9 & 0.1 \\
0.1 & 0.9 \end{array} \right],
P^{(2)} =  \left[ \begin{array}{cc}
0.82 & 0.18 \\
0.18 & 0.82 \end{array} \right], 
P^{(3)} =  \left[ \begin{array}{cc}
0.756 & 0.244 \\
0.244 & 0.756 \end{array} \right], 
P^{(4)} = \left[ \begin{array}{cc}
0.7048 & 0.2952 \\
0.2952 & 0.7048 \end{array} \right], \\ 
P^{(5)} &=  \left[ \begin{array}{cc}
0.6638 & 0.3362 \\
0.3362 & 0.6638 \end{array} \right] &
\end{flalign*}
\else
\begin{flalign*}
\quad \small P^{(1)} &= \small \left[ \begin{array}{cc}
0.9 & 0.1 \\
0.1 & 0.9 \end{array} \right], \quad\quad
P^{(2)} =  \left[ \begin{array}{cc}
0.82 & 0.18 \\
0.18 & 0.82 \end{array} \right], &\\
\small P^{(3)} &= \small \left[ \begin{array}{cc}
0.756 & 0.244 \\
0.244 & 0.756 \end{array} \right], 
P^{(4)} = \left[ \begin{array}{cc}
0.7048 & 0.2952 \\
0.2952 & 0.7048 \end{array} \right], &\\
\small P^{(5)} &= \small \left[ \begin{array}{cc}
0.6638 & 0.3362 \\
0.3362 & 0.6638 \end{array} \right] &
\end{flalign*}
\fi


\begin{figure*}[!htb]
    \centering
    \begin{subfigure}[t]{0.23\textwidth}
        \centering
        \includegraphics[width=1\linewidth]{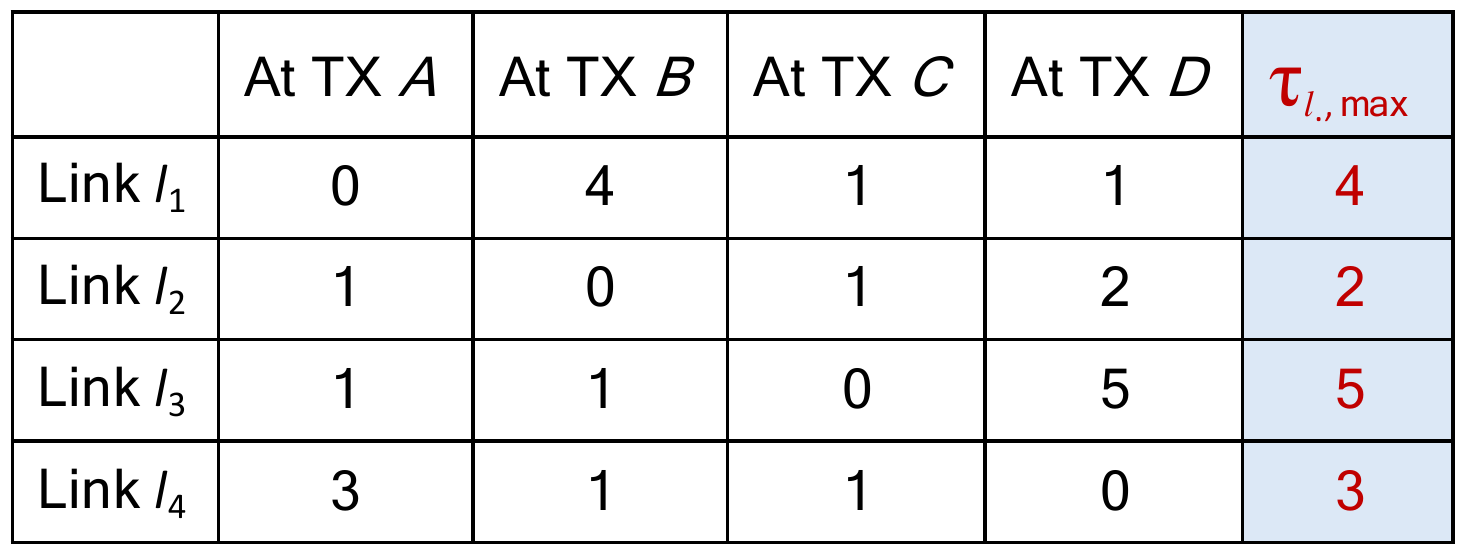}
        \caption{}
    \end{subfigure}%
    \quad 
    \begin{subfigure}[t]{0.23\textwidth}
        \centering
        \includegraphics[width=1\linewidth]{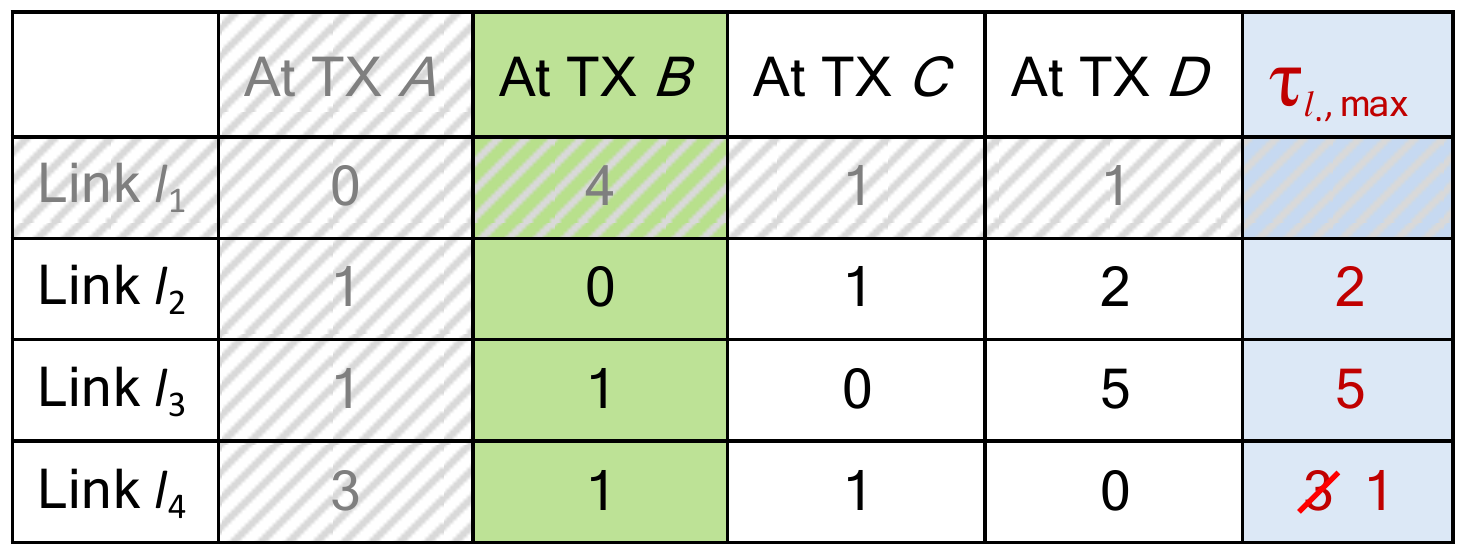}
        \caption{}
    \end{subfigure}
    \begin{subfigure}[t]{0.23\textwidth}
        \centering
        \includegraphics[width=1\linewidth]{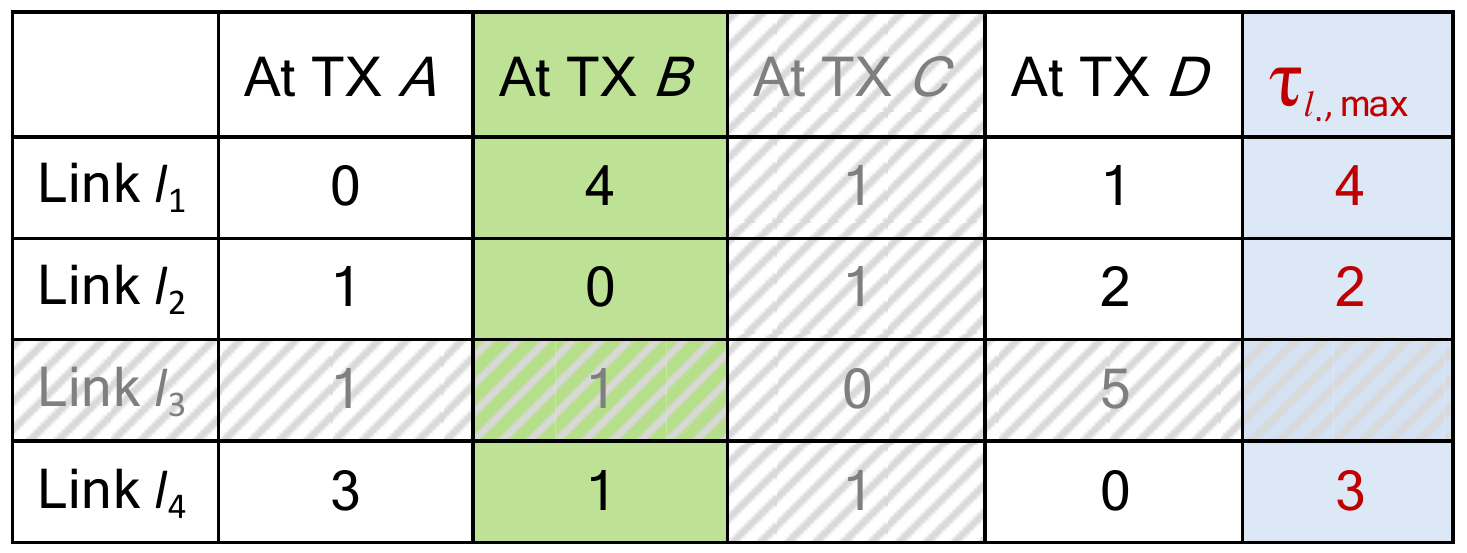}
        \caption{}
    \end{subfigure}%
    \quad 
    \begin{subfigure}[t]{0.23\textwidth}
        \centering
        \includegraphics[width=1\linewidth]{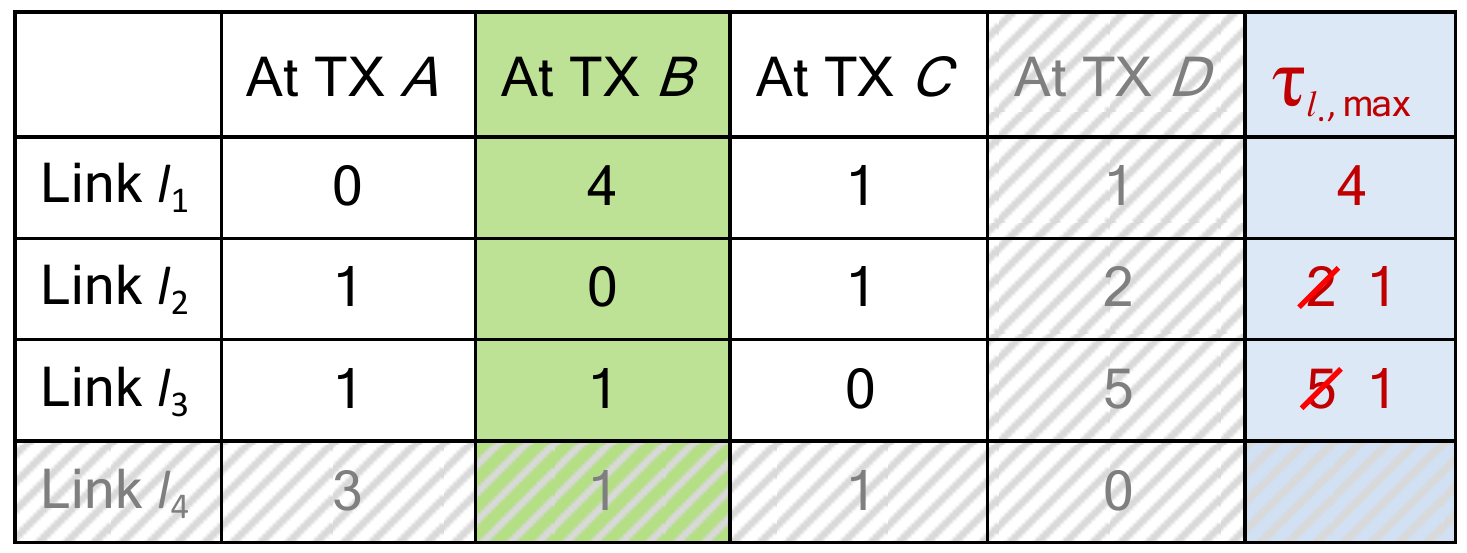}
        \caption{}
    \end{subfigure}
    \begin{subfigure}[t]{0.23\textwidth}
        \centering
        \includegraphics[width=1\linewidth]{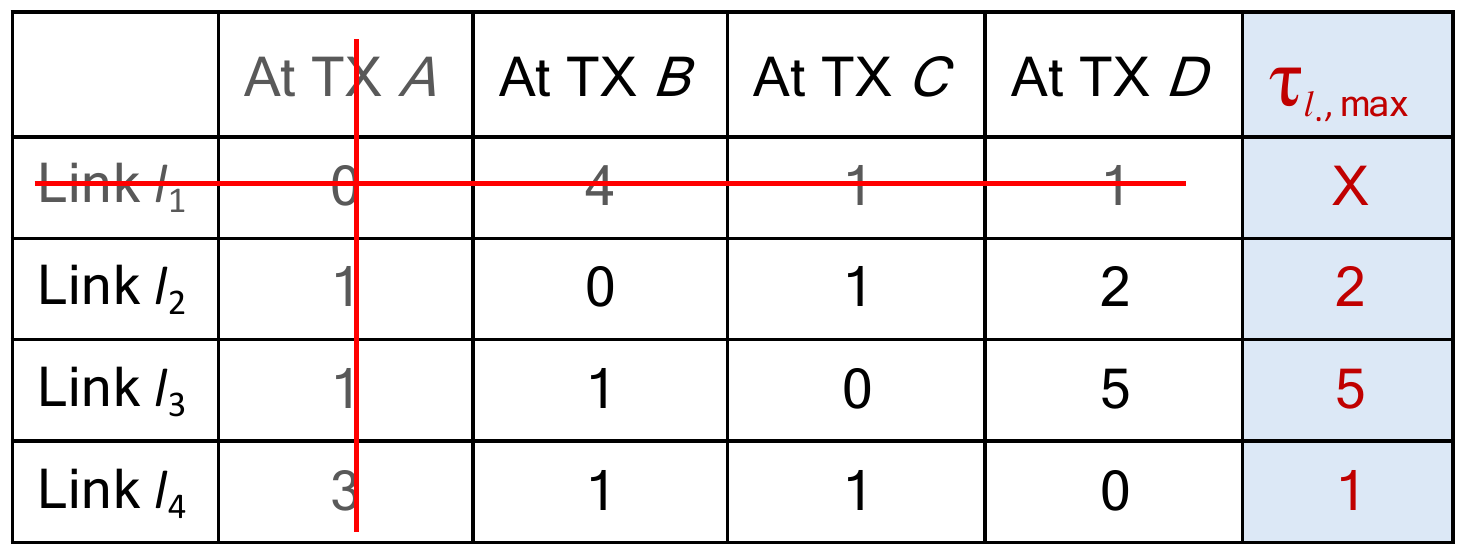}
        \caption{}
    \end{subfigure}%
    \quad 
    \begin{subfigure}[t]{0.23\textwidth}
        \centering
        \includegraphics[width=1\linewidth]{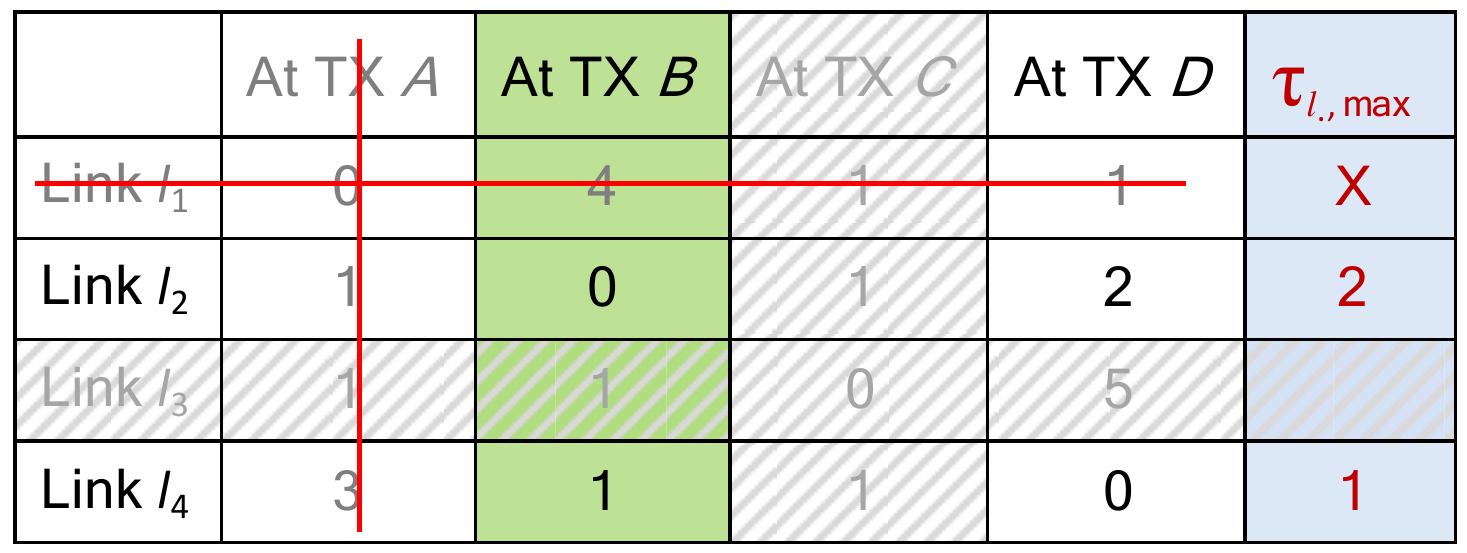}
        \caption{}
    \end{subfigure}
    \begin{subfigure}[t]{0.23\textwidth}
        \centering
        \includegraphics[width=1\linewidth]{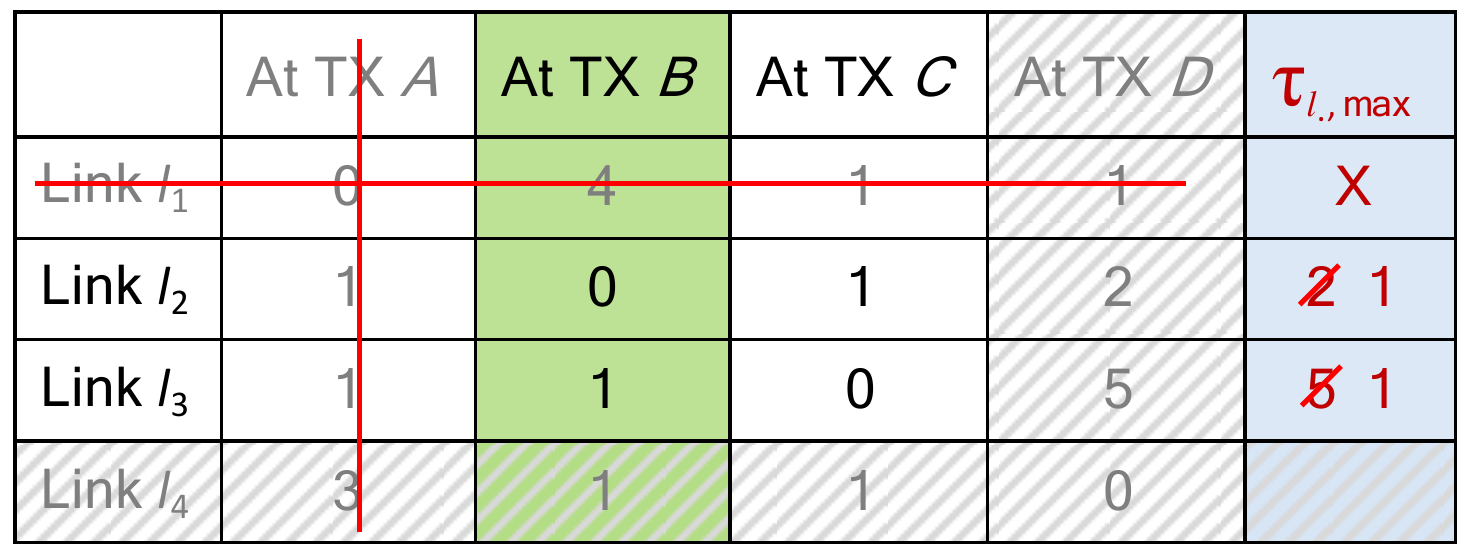}
        \caption{}
    \end{subfigure}%
    \quad 
    \begin{subfigure}[t]{0.23\textwidth}
        \centering
        \includegraphics[width=1\linewidth]{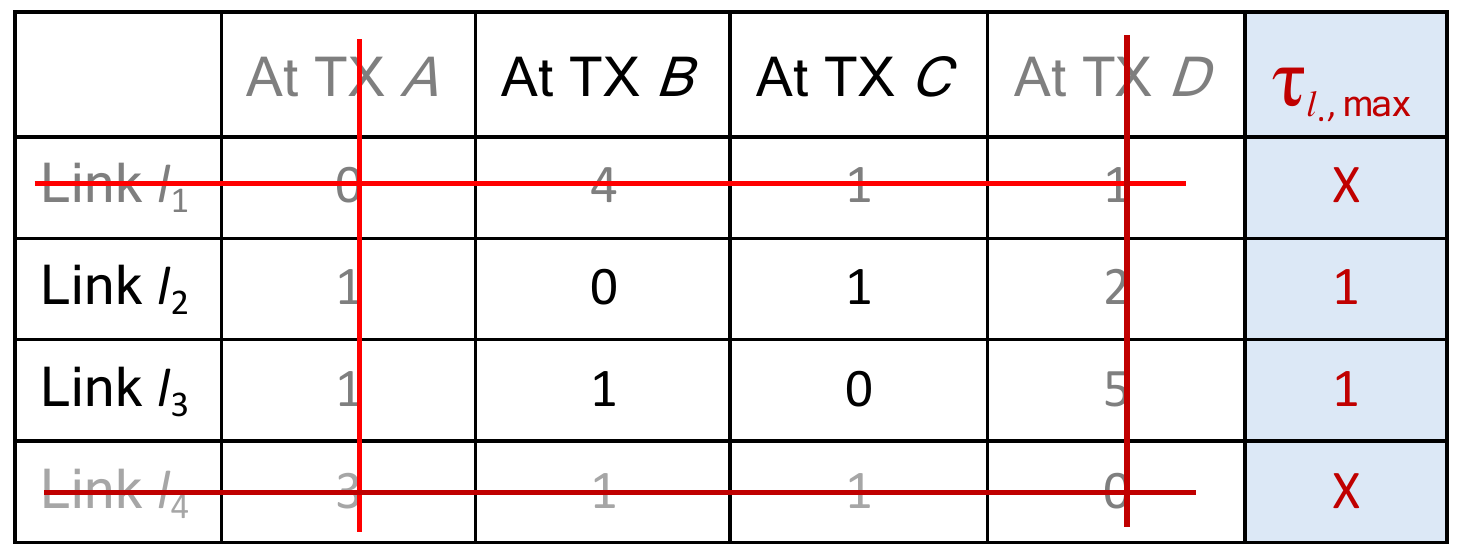}
        \caption{}
    \end{subfigure}
    \caption{Illustration of the dynamics of the \LCVARIANTONE policy. In the beginning, \textit{ActiveSet} = $\{l_1, l_2, l_3, l_4\}$. Subfigure (a) shows the original table of delay values and $\tau_{l, max}$ values computed in step 4 of Algorithm \ref{LC1Listing} in round 1. Subfigures (b), (c), (d) show the computation of the set $EC$ in round 1 (step 7). In subfigure (b), after setting aside the link with the largest expected data rate (link $l_2$; shown by highlighting in green the column of transmitter B [the transmitter node of link $l_2$]), we see whether eliminating link $l_1$ reduces the $\tau_{l, max}$ delays for any other link in \textit{ActiveSet}. This is done by temporarily masking the row and column pertaining to link $l_1$ (shown hatched), recomputing the $\tau_{l, max}$ delays and comparing with previously computed values. The value of $\tau_{l_4, max}$ reduces (from 3 to 1), and hence link $l_1$ is a candidate for elimination, and therefore belongs to $EC$. Similarly, in subfigure (c), we determine that link $l_3$ does not belong to $EC$, and in subfigure (d), we determine that link $l_4$ belongs to $EC$. In subfigure (e), we eliminate link $l_1$ (shown by striking-through in red the row and column pertaining to link $l_1$) since only $l_1$  and $l_4$ belong to $EC$ and $l_1$ has smaller expected data rate than $l_4$ (steps 16, 21). This reduces \textit{ActiveSet} to $\{l_2, l_3, l_4\}. $Subfigures (f), (g), (h) show similar calculations for round 2, at the end of which $l_4$ is eliminated, leaving behind only $l_2$ and $l_3$ in \textit{ActiveSet}.}
	\label{fig:Illustration-LC-ELDR-policy} \vspace{-1em}
\end{figure*}
\vspace{0.1in}
\noindent
We assume the following delayed channel-realizations for our illustration: $C_{l_1}[t - 4] = 2$, $C_{l_1}[t - 1] = 1$, $C_{l_2}[t - 2] = 2$, $C_{l_2}[t - 1] = 1$, $C_{l_3}[t-5] = 2$, $C_{l_3}[t-1] = 2$, $C_{l_4}[t - 3] = 2$, $C_{l_4}[t - 1] = 1$. We first illustrate the working of the \LCVARIANTONE policy when it is executed at transmitter $C$, and towards the end we mention the minor differences in the dynamics when it is executed at the other transmitters. Since we are initially concerned with the case of complete interference, we set $ActiveSet = \{l_1, l_2, l_3, l_4\}$ before invoking Algorithm \ref{LC1Listing}, where \textit{ActiveSet} is the set of links that are (still) contending for transmission in the current time slot.\\

\noindent
(\textbf{step 2}) transmitter $C$ sets $I = \{l_3\}$\\

\noindent
\textbf{ROUND 1}:\begin{verbbox}\footnotesize ``\verb|while|''  \end{verbbox}\footnote{We will call the \verbBox $\,$ loop body from line 3 to line 23, and also the computation in lines 24 -- 29 in Algorithm 1, a ``round''.} \vspace{0.1cm}

\noindent
(\textbf{step 4}) transmitter $C$ computes $\tau_{l_{i}, max}$, for $i = 1, 2, 3, 4$, which results in $\tau_{l_1, max} = 4$, $\tau_{l_2, max} = 2$, $\tau_{l_3, max} = 5$, $\tau_{l_4, max} = 3$. This is illustrated in the last column of the table in Fig. \ref{fig:Illustration-LC-ELDR-policy}(a) \vspace{0.1cm}

\noindent
(\textbf{step 5}) transmitter $C$ computes the expected data rate on each link $l_i$ in \textit{ActiveSet} given the $\tau_{l_{i}, max}$-delayed channel-state realization of that link, as follows:\footnote{we ignore the multiplication of the delayed queue-lengths with the expected data rates since we are considering saturated queues for this illustration.} $E\big[C_{l_1}[t] \, | \, C_{l_1}[t - 4] = 2\big] = 1.7048$, $E\big[C_{l_2}[t] \, | \, C_{l_2}[t - 2] = 2\big] = 1.82$, $E\big[C_{l_3}[t] \, | \, C_{l_3}[t - 5] = 2\big] = 1.6638$, $E\big[C_{l_4}[t] \, | \, C_{l_4}[t - 3] = 2\big] = 1.756$ \vspace{0.1cm}

\noindent
(\textbf{step 6}) transmitter $C$ sets aside link $l_2$ by setting $H = l_2$, since link $l_2$ has the largest expected data rate in this round, as computed in the previous step (this is done so that the link with the largest expected data rate in this round is not eliminated). This is shown by highlighting transmitter $B$'s ($B$ is the transmitter node on link $l_2$) column in green in Fig. \ref{fig:Illustration-LC-ELDR-policy}(b). \vspace{0.1cm}

\noindent
(\textbf{step 7}) transmitter $C$, to decide on the link it wants to eliminate in this round, computes $EC$ as follows:
\begin{enumerate}[label=(\roman*)]
  \item first, transmitter $C$ temporarily masks the row and column pertaining to link $l_1$ (shown by hatching the corresponding row and column, in Fig. \ref{fig:Illustration-LC-ELDR-policy}(b)), computes the $\tau_{l_{i}, max}$ values for links $l_i$, $i=2,3,4$ (i.e., for all links other than link $l_1$), compares them to the previously computed $\tau_{l_{i}, max}$ values for these links, finds that upon masking the row and column pertaining to link $l_1$, the $\tau_{l_{i}, max}$ value for at least one link (namely, that of link $l_4$) reduces (from 3 to 1; see Fig.  \ref{fig:Illustration-LC-ELDR-policy}(b)), and hence decides that link $l_1$ is a candidate for elimination, and adds it to the set of elimination candidates, $EC$. What this means is that if transmitter $C$ does choose to eliminate link $l_1$, each of the transmitters corresponding to the remaining links would then be able to access the delayed channel state of link $l_4$ at a reduced common delay value of 1 unit (instead of 3 units, as was required previously) since the other transmitters assuredly possess the delayed channel-state of link $l_4$ at this new reduced delay of 1 unit
  
  \item transmitter $C$ skips this procedure for link $l_2$ since $l_2$, being the link with the largest expected data rate, was set aside in the previous step
  
  \item next, transmitter $C$ temporarily masks the row and column pertaining to link $l_3$ (i.e., its own link; shown again by hatching the row and column of link $l_3$, in Fig. \ref{fig:Illustration-LC-ELDR-policy}(c)), computes the $\tau_{l_{i}, max}$ values for links $l_i$ for $i=1,2,4$ (i.e., for all links other than link $l_3$), compares them to the previously computed $\tau_{l_{i}, max}$ values for these links, finds that upon masking the row and column pertaining to link $l_3$, the $\tau_{l_{i}, max}$ value for none of the links $l_i$ for $i=1,2,4$ reduce, and hence decides that link $l_3$ (i.e., its own link) is not a  candidate for elimination
  
  \item lastly, transmitter $C$ temporarily masks the row and column pertaining to link $l_4$ (shown by hatching the row and column of link $l_4$, in Fig. \ref{fig:Illustration-LC-ELDR-policy}(d)), computes the $\tau_{l_{i}, max}$ values for links $l_i$ for $i=1,2,3$ (i.e., for all links other than link $l_4$), compares them to the previously computed $\tau_{l_{i}, max}$ values for these links, finds that upon masking the row and column pertaining to link $l_4$, the $\tau_{l_{i}, max}$ value for at least one link (namely, that of links $l_2$ and $l_3$) reduces (from 2 to 1 for link $l_2$, and from 5 to 1 for link $l_3$; see Fig.  \ref{fig:Illustration-LC-ELDR-policy}(d)), and hence decides that link $l_4$ is a candidate for elimination, and adds it to the set of elimination candidates, $EC$. This means that if transmitter $C$ does choose to eliminate link $l_4$, it is assured that each of the links remaining in \textit{ActiveSet} would then be able to access the channel state of links $l_2$ and $l_3$ at reduced delays of 1 unit and 1 unit respectively (instead of 2 and 5 units respectively, as was required previously)
\end{enumerate}
Thus, transmitter $C$ has computed the set of elimination candidates to be $EC = \{l_1, l_4\}$
\vspace{0.1cm}

\noindent
(\textbf{step 16}) Among the elimination candidates in set $EC$, transmitter $C$ chooses link $l_1$ to be the one that will be eliminated in this round by setting $S = l_1$, since link $l_1$ has the lowest expected data rate among $l_1$ and $l_4$ (which have expected data rates of 1.7048 and 1.756, respectively; see step 5 above). This is shown by striking out the row and column pertaining to link $l_1$ in Fig. \ref{fig:Illustration-LC-ELDR-policy}(e)
\vspace{0.1cm}

\noindent
\noindent
(\textbf{step 21}) Transmitter $C$ removes link $l_1$ from \textit{ActiveSet} (since link $l_1$, having been eliminated in this round, is no longer in contention with the other links in \textit{ActiveSet} for carrying transmission in the current time slot). Thus, $ActiveSet = \{l_2, l_3, l_4\}$ is the revised set of links still contending\\

\noindent
\textbf{ROUND 2}: \vspace{0.1cm}

\noindent
Note: For all purposes, the table of delay values for this round is the original table minus the row and column corresponding to link $l_1$ since it was eliminated in the previous round. \vspace{0.1cm}

\noindent
(\textbf{step 4}) transmitter $C$ sets $\tau_{l_1, max} = $ X (don't care) since link $l_1$ does not belong to \textit{ActiveSet} anymore. Transmitter $C$ then calculates the new values of $\tau_{l_{i}, max}$, for $i = 2, 3, 4$. All these values could potentially be smaller than the corresponding values in round 1, since when calculating these values in round 2, transmitter $C$ ignores the delay values in the row and column pertaining to link $l_1$ which was eliminated in the previous round. The new values are: $\tau_{l_2, max} = 2, \tau_{l_3, max} = 5, \tau_{l_4, max} = 1$. Note that $\tau_{l_4, max}$ has reduced from 3 (in the previous round) to 1 (see Fig. \ref{fig:Illustration-LC-ELDR-policy}(e)).
\vspace{0.1cm}

\noindent
(\textbf{step 5}) as in round 1, transmitter $C$ computes the expected data rate on each link $l_i$ in \textit{ActiveSet} given the $\tau_{l_{i}, max}$-delayed channel-state realization of that link (as computed in step 4 above), as follows: $E\big[C_{l_2}[t] \, | \, C_{l_2}[t - 2] = 2\big] = 1.82$, $E\big[C_{l_3}[t] \, | \, C_{l_3}[t - 5] = 2\big] = 1.6638$, $E\big[C_{l_4}[t] \, | \, C_{l_4}[t - 1] = 1\big] = 1.1$ \vspace{0.1cm}

\noindent
(\textbf{step 6}) as in round 1, transmitter $C$ sets aside the link that has the largest expected data rate in this round, as computed in the previous step. Coincidentally, link $l_2$ turns out to be that link in this round also. Therefore, transmitter $C$ sets $H = l_2$. This is shown by highlighting transmitter $B$'s column in green in Fig. \ref{fig:Illustration-LC-ELDR-policy}(f). \vspace{0.1cm}

\noindent
(\textbf{step 7}) once again, transmitter $C$ goes about computing $EC$ (to decide on the link it wants to eliminate in this round) as follows:
\begin{enumerate}[label=(\roman*)]
  \item first, transmitter $C$ decides to skip this procedure for link $l_1$ (since it is not in \textit{ActiveSet} anymore), and for link $l_2$ (since $l_2$, being the link with the largest expected data rate, was set aside in step 6 above)
  
  \item next, following similar arguments as in round 1, transmitter $C$ decides that link $l_3$ (i.e., its own link) is not a candidate for elimination (see Fig. \ref{fig:Illustration-LC-ELDR-policy}(f))
  
  \item lastly, again following similar arguments as in round 1, transmitter $C$ decides that link $l_4$ is a candidate for elimination, and adds it to the set of elimination candidates, $EC$ (see Fig.  \ref{fig:Illustration-LC-ELDR-policy}(g)).
\end{enumerate}
Thus, transmitter $C$ has computed the set of elimination candidates to be $EC = \{l_4\}$
\vspace{0.1cm}

\noindent
(\textbf{step 16}) since there is only one elimination candidate (namely, link $l_4$) in set $EC$, transmitter $C$ chooses to eliminate link $l_4$ by setting $S = l_4$. This is shown by striking out the row and column pertaining to link $l_4$ in Fig. \ref{fig:Illustration-LC-ELDR-policy}(h) \vspace{0.1cm}

\noindent
(\textbf{step 21}) Transmitter $C$ removes link $l_4$ from \textit{ActiveSet} (since link $l_4$, having been eliminated in this round, is no longer in contention with the other links in \textit{ActiveSet} for carrying transmission in the current time slot). Thus, $ActiveSet = \{l_2, l_3\}$ is the revised set of links still contending \\

\noindent
\textbf{ROUND 3}: \vspace{0.1cm}

\noindent
(\textbf{step 24}) transmitter $C$ sets $\tau_{l_1, max}$ and $\tau_{l_4, max}$ to X (don't care) since links $l_1$ and $l_4$ do not belong to \textit{ActiveSet} anymore. Transmitter $C$ then calculates the new values of $\tau_{l_{i}, max}$, for $i = 2, 3$ to be $\tau_{l_2, max} = 1$, $\tau_{l_3, max} = 1$. Transmitter $C$ computes the expected data rate on each link $l_i$ in \textit{ActiveSet} given the $\tau_{l_{i}, max}$-delayed channel-state realization of that link (where the new $\tau_{l_{i}, max}$ value is as computed only just), as follows: $E\big[C_{l_2}[t] \, | \, C_{l_2}[t - 1] = 1\big] = 1.1$, $E\big[C_{l_3}[t] \, | \, C_{l_3}[t - 1] = 2\big] = 1.9$. Note that $\tau_{l_2, max}$ and $\tau_{l_3, max}$ have changed from 2 and 5 respectively (in the previous round) to 1 each (see Fig. \ref{fig:Illustration-LC-ELDR-policy}(h)). Transmitter $C$ sets $T = l_3$ since link $l_3$ (i.e., its own link) has the largest expected data rate as computed above
\vspace{0.1cm}

\noindent
(\textbf{step 26}) Transmitter $C$ sets the transmit decision for link $l_3$ (i.e, for its own link) to \textsc{Transmit} \vspace{0.1cm}

\noindent
(\textbf{step 30}) Transmitter $C$ terminates this run of the \LCVARIANTONE policy. \\

\indent 
It is easy to convince oneself that all the transmitters running the \LCVARIANTONE policy in the current time slot would arrive at the same decision -- links other than $l_3$ would set their transmit decision to \textsc{NoTransmit} while link $l_3$ sets its transmit decision to \textsc{Transmit}. To illustrate this in short, transmitter $D$ for example, when running the \LCVARIANTONE policy in the current time slot, would make the same calculations as shown above till step 16 in round 2, whereupon transmitter $D$ would execute step 18 (note that $I = \{l_4\}$ in this case) and thereby set the transmit decision for link $l_4$ (i.e., its own link) to \textsc{NoTransmit}, and terminate this run of the algorithm at transmitter $D$ at step 19.

So far, we have considered the case of networks with complete interference. Extension to the case of networks with multiple interference sets is as follows: (i) Let \textit{ActiveSet} contain all the links in the network. Set $M \gets \phi$. Set $\delta_{ij} \gets 0 \; \forall l_i \notin I_{l_j} \; \forall i,j$ where $\delta_{ij}$ is the delay value in row $i$ and column $j$ of the delay table (ii) Run the \LCVARIANTONE policy. Let $l$ be the link chosen for transmission by the \LCVARIANTONE policy. Set $M \gets M \cup \{l\}$. Set $ActiveSet \gets ActiveSet \setminus (\{l\} \cup I_l)$. (iii) Repeat step (ii) while \textit{ActiveSet} is non-empty. When $ActiveSet$ becomes empty, the links in $M$ are the set of links that will be allowed to carry transmission in the current time slot.

The policy \LCVARIANTTWO is a minor variation of \LCVARIANTONE, where only step $16$ of \LCVARIANTONE is modified as noted in Algorithm \ref{LC2Listing}. In the example above, in step 16 of round 1, the \LCVARIANTONE policy chose to eliminate link $l_1$ since it had a lower expected data rate compared to that of link $l_4$, whereas the \LCVARIANTTWO policy chooses to eliminate link $l_4$ since eliminating link $l_4$ reduces the delays with which the channel state of two channels (namely, $l_2$ and $l_3$) can be accessed, whereas eliminating link $l_1$ would reduce the delay of only one channel (namely, $l_4$).

\begin{algorithm}[t]
\small
\caption{}\label{LC2Listing}
\begin{algorithmic}[16]
\Procedure{\LCVARIANTTWO}{\textit{ActiveSet}} 
\Statex \hspace{-0.8cm} \parbox{\algorithmicindent}{{\scriptsize 16:}\strut} \hspace{0.5cm} 
\parbox[t]{\dimexpr\linewidth-\algorithmicindent}{Let $S$ be the link in $EC$ with the lowest expected data rate among those links that, upon their elimination, reduce the maximum of the delay values in a row, for the largest number of rows (i.e., channels) [corresponding to links in \textit{ActiveSet}] in the table of delay values [after suppressing the rows and columns corresponding to links that are not in \textit{ActiveSet}]\strut} \vspace{-.1cm}
\EndProcedure
\end{algorithmic}
\end{algorithm}

\subsection{Throughput Near-Optimality}
\label{subsection:throughput-near-optimality}
\noindent
In this section, we derive analytical expressions for the expected saturated system throughputs of the \LCVARIANTONE and \LCVARIANTTWO policies. We evaluate these expressions in Sec. \ref{section-numerical-results} for various network settings, and demonstrate that these expressions approximate the optimal throughput values very closely. We need a few definitions, which we introduce here rather informally, and make these definitions precise mathematically later in Appendices \ifelseEVWithoutCite{\ref{appendix-proof-saturated-system-throughput-of-LC1-policy}}{J} and \ifelseEV{\ref{appendix-proof-saturated-system-throughput-of-LC2-policy}}{K}. Let $N$ be the number of links and $T \coloneqq \{1, 2, \dots, N\}$ be the set of links in \textit{ActiveSet} before calling the \LCVARIANTONE policy. Further, let
\begin{align*}
\One \big\{ a \big\} & \coloneqq \left\{ \bfrac{1 \quad \mbox{if $a$ is true}}{\hspace{-0.14cm}0 \quad \mbox{otherwise}} \right. \\
\One_{+} {\big\{ {a_i}\big\} }_{i \in I} & \coloneqq \max{ \big\{ \One \{a_i\} \big\}}_{i \in I}
\end{align*}

We will call the ``\verb|while|'' loop body from line 3 to line 23, and also the computation in lines 24--29 in the algorithm listing of the \LCVARIANTONE policy in Sec. \ref{section-low-complexity-scheduling-policies}, a ``round''. Thus, if the \LCVARIANTONE policy terminates at line 26 or at line 28, then it would have executed $N-1$ rounds (specifically, $N-2$ rounds in the body of the ``\verb|while|'' loop, and the last round [round $N-1$] in lines 24--29), and it would have executed $r \, (r < N-1)$ rounds if it terminates at line 14 or 19. Thus, the \LCVARIANTONE policy can terminate after $r$ rounds, $1 \leq r \leq N-1$.

Let $\tau_j^{(r)}$ (not to be confused with $\tau_l(h)$ defined in Sec. \ref{section-Structure-of-Heterogeneously-Delayed-NSI}) be equal to $\tau_{l_{j}, max}$ (see Sec. \ref{section-Structure-of-Heterogeneously-Delayed-NSI}) at the beginning of round $r$, where $\tau_{l_{j}, max}$ is calculated after masking the rows and columns pertaining to the links that have been eliminated in rounds 1 to $r-1$ (as illustrated in Figs. \ref{fig:Illustration-LC-ELDR-policy}(a), \ref{fig:Illustration-LC-ELDR-policy}(e), and \ref{fig:Illustration-LC-ELDR-policy}(h)). Let $c_{k, \tau}$ be the realization of channel-state on link $l_k$ at time $t - \tau$.  

\noindent
$\bm{e^{(r)}}$: Let $e^{(r)}$ be the link that was (or that will be) eliminated (by the \LCVARIANTONE policy) in round $r$ (thus, ``$e^{(r)} = k$'' [for $k \in T$] implies that link $k$ was (or will be) eliminated in round $r$, and ``$e^{(r)} = 0$'' implies that there is no candidate link to eliminate, and hence that the \LCVARIANTONE policy terminates).

\indent
Let $i \in T$ be the link chosen by the \LCVARIANTONE policy in a particular time slot, say $t$, when the \LCVARIANTONE policy terminates after executing $r$ rounds, $1 \leq r \leq N-1$. We now consider the working of the \LCVARIANTONE policy when it is executed at the  transmitter node of link $i$ in time slot $t$. Consider a particular round $\tilde{r}$ ($1 \leq \tilde{r} \leq r$) of the \LCVARIANTONE policy. If link $i$ has survived (i.e., was not eliminated in) the first $\tilde{r}-1$ rounds, then it will survive round $\tilde{r}$ if one of the following happens:
\begin{enumerate}[(i)]
  \item $\bm{s(i, \tilde{r}, 1)}$: link $i$ has the largest expected data rate in round $\tilde{r}$ (hence $i = H$ and therefore $i \notin EC$; see steps 6 and 7 of Algorithm \ref{LC1Listing}). We denote this condition as $s(i, \tilde{r}, 1)$.
  
  \item $\bm{s(i, \tilde{r}, 2)}$: link $i$ does not have the largest expected data rate in round $\tilde{r}$ (hence $i \neq H$), and link $i$ does not reduce the maximum of the delay values in a row, for any row (corresponding to a link in \textit{ActiveSet}) in the table of delay values (after suppressing the rows and columns corresponding to links that have been eliminated up to round $\tilde{r}$) if it is eliminated in round $\tilde{r}$ (hence $i \notin EC$; see step 7 of Algorithm \ref{LC1Listing}), and some link (other than the one with the largest expected data rate in round $\tilde{r}$) reduces the maximum of the delay values in a row, for at least one row if it is eliminated in round $\tilde{r}$ (i.e., $EC \neq \phi$). We denote this condition as $s(i, \tilde{r}, 2)$.
  
  \item $\bm{s(i, \tilde{r}, 3)}$: link $i$ does not have the largest expected data rate in round $\tilde{r}$ (hence $i \neq H$), and link $i$ reduces the maximum of the delay values in a row, for at least one row (corresponding to a link in \textit{ActiveSet}) in the table of delay values (after suppressing the rows and columns corresponding to links that have been eliminated up to round $\tilde{r}$) if it is eliminated in round $\tilde{r}$ (hence $i \in EC$), and link $i$ does not have the smallest expected data rate among the links that reduce the maximum of the delay values in a row, for at least one row if that link is eliminated in round $\tilde{r}$ (hence $i \neq S$; see step 16 of Algorithm \ref{LC1Listing}). We denote this condition as  $s(i, \tilde{r}, 3)$.
\end{enumerate}
With the required definitions in place, we are now ready to state the following crucial result:

\begin{proposition}
\label{lemma-saturated-system-throughput-of-LC1-policy}
The expected saturated system throughput of the \LCVARIANTONE policy is exactly
\begin{equation*}
\begin{split}
\sum_{i=1}^{N} \sum_{r=1}^{N-1} \sum_{\substack{c_{j, \tau_{j}^{(q)}}\\ j \in T \\ q \in \{1, 2, \dots, r\}}} \kern-1em \Ex \Bigg[ &  C_{l_i}[t] \; \prod_{\tilde{r}=1}^{r-1} \Big(\OnePlus \big\{s(i, \tilde{r}, n)\big\}_{n \in \{ 1, 2, 3 \}} 
\end{split}
\end{equation*}
\begin{equation*}
\begin{split}
& \qquad \qquad \;\; \times \One\big\{e^{(\tilde{r})} \neq 0 \big\} \Big) \One \big\{s(i,r,1) \big\} \One\big\{e^{(r)} = 0\big\} 
\end{split}
\end{equation*}
\begin{equation*}
\begin{split}
& \kern-0.5em \bigg\rvert \; C_{l_i} \big[ t - \tau_i^{(r)} \big] = c_{i, \tau_i^{(r)}},  
\end{split}
\end{equation*}
\begin{equation*}
\begin{split}
\hspace{3cm} & \kern-1.6em \Big\{C_{l_i} \big[ t - \tau_i^{(q)} \big] = c_{i, \tau_i^{(q)}} \Big\}_{q \in \{ 1, 2, \dots, r-1 \}}, 
\end{split}
\end{equation*}
\begin{equation*}
\begin{split}
\hspace{3cm} & \kern-2em \Big\{C_{l_p} \big[ t - \tau_p^{(q)} \big] = c_{p, \tau_p^{(q)}} \Big\}_{\substack{ p \in T \setminus \{i\} \\ q \in \{ 1, 2, \dots, r \}}} \Bigg] 
\end{split}
\end{equation*}
\begin{equation*}
\begin{split}
\hspace{3cm} & \kern-5em \times \prod_{m = 1}^{N} \left( \pi_{m} \big( c_{m, \tau_{m}^{(1)}} \big) \prod_{n=1}^{r-1} p_{{}_{{}_{m,}} \;c_{m, \tau_{m}^{(n)}} \; c_{m, \tau_{m}^{(n+1)}}}^{(\tau_m^{(n)} - \tau_m^{(n+1)})} \right)
\end{split}
\end{equation*}
where $\pi_{m}(c)$ is the steady-state probability of being in state $c$ in the channel-state Markov chain of link $m$, and $p_{m, \, ij}^{(k)}$ is the transition probability of reaching state $j$ from state $i$ in $k$ steps in the channel-state Markov chain of link $m$, with $p_{m, \, ij}^{(0)} = 1 \; \forall i, j, m.$\footnote{$\pi_{m}(c) = \pi(c) \text{ and } p_{m, \, ij}^{(k)} = p_{ij}^{(k)}  \; \forall m \in \mathcal{L}$ in our model; see Sec. \ref{section-Network-Model}}
\end{proposition}
We present mathematically precise definitions of the symbols that were defined informally in the paragraphs preceding Proposition \ref{lemma-saturated-system-throughput-of-LC1-policy}, and the proof itself in Appendix \ifelseEV{\ref{appendix-proof-saturated-system-throughput-of-LC1-policy}}{J}.

\begin{proposition}
\label{lemma-saturated-system-throughput-of-LC2-policy}
The expression for the expected saturated system throughput of the \LCVARIANTTWO policy is the same as the expression in Proposition \ref{lemma-saturated-system-throughput-of-LC1-policy}, except that the expression for $e^{(r)}$ is redefined to be as noted in Appendix \ifelseEV{\ref{appendix-proof-saturated-system-throughput-of-LC2-policy}}{K}.
\end{proposition}

The proof of this proposition is essentially the same as that for Proposition \ref{lemma-saturated-system-throughput-of-LC1-policy}; the minor departures are noted in Appendix \ifelseEV{\ref{appendix-proof-saturated-system-throughput-of-LC2-policy}}{K}. We evaluate the expressions in Propositions \ref{lemma-saturated-system-throughput-of-LC1-policy} and \ref{lemma-saturated-system-throughput-of-LC2-policy} numerically and plot them for various network settings in Sec. \ref{section-numerical-results}, and demonstrate that these expressions approximate the optimal throughput values very closely and hence that the \LCVARIANTONE and \LCVARIANTTWO policies are near-throughput-optimal.

\subsection{Computational Complexity}
\label{subsec:computational-complexity-of-LC1-LC2-policies}
\noindent
We have the following result: 
\begin{proposition}
\label{lemma-computational-complexity-LC1-LC2-policies}
The running times of both of the \LCVARIANTONE and \LCVARIANTTWO policies are $\BigO(\mathcal{C}L^2 + L^4)$.
\end{proposition}
We provide a proof of this result in Appendix \ifelseEV{\ref{appendix-proof-run-time-complexity-of-LC1-and-LC2-policies}}{I}. Thus from this proposition, for the case of a network with complete interference, the computational complexities of the \LCVARIANTONE and \LCVARIANTTWO policies are polynomial in the number links in the network and linear in the number of channel states. For a network with multiple interference sets, in the worst case (which happens when $I_l = \phi, \;\; \forall l \in \mathcal{L}$), there are $L$ calls to Algorithm \ref{LC1Listing}, thus there is an additional multiplicative factor of $L$, which does not alter the (orders of) polynomial and linear dependencies on the number of links and the number of channel states respectively. Thus, the \LCVARIANTONE and \LCVARIANTTWO policies offer an enormous improvement over the computational complexities of the $R$ and $H$ policies. We compare the running times of these policies in Table \ref{tab:run_time_comparision}.

%% file: Het-Delays-for-Example-Network.tex
\begin{table}
\centering
\scriptsize
\caption{{\small Heterogeneous delays for the illustration in Sec. \ref{subsec:dynamics-of-LC-ELDR-policy}}}
\label{tab:Het-Delays-for-Example-Network}
\ifdefined\ONECOLUMN
\begin{tabular}{p{3.3cm}C{1cm}C{1cm}C{1cm}C{1cm}}
\else
\begin{tabular}{p{3.3cm}C{0.85cm}C{0.85cm}C{0.84cm}C{0.85cm}}
\fi
\hline \hline \noalign{\smallskip}
& At TX $A$ & At TX $B$ & At TX $C$ & At TX $D$ \\
\noalign{\smallskip}\Xhline{2.5\arrayrulewidth}\noalign{\smallskip}
Delay in obtaining NSI of link $l_1$ & $0$ & $4$ & $1$ & $1$  \\
Delay in obtaining NSI of link $l_2$ & $1$ & $0$ & $1$ & $2$ \\
Delay in obtaining NSI of link $l_3$ & $1$ & $1$ & $0$ & $5$ \\
Delay in obtaining NSI of link $l_4$ & $3$ & $1$ & $1$ & $0$ \\
\noalign{\smallskip}\hline \hline
\end{tabular}
\end{table}

%% file: section-numerical-results.tex
In this section, we present numerical results comparing the saturated system throughputs of the $R$, $H$, \LCVARIANTONE and \LCVARIANTTWO policies. In addition, for purposes of comparison, we define two new policies that we call $O$ and $IC$:

\noindent
\ifdefined\ONECOLUMN
\begin{flalign*}
&\mbox{Policy } O \;\;: \argmax\{Q_l[t - \tau_{l, max}] \times \Ex[C_l[t] \, | \, C_l[t - \tau_{l, max}] = .]\}_{l \in \mathcal{L}}.\\
\end{flalign*}
\else
\\
Policy $O$: $\argmax\{Q_l[t - \tau_{l, max}] \times \Ex[C_l[t] \, | \, C_l[t - \tau_{l, max}] = c_{l, \tau_{l, max}}]\}_{l \in \mathcal{L}}$.\\
\fi

\noindent
\ifdefined\ONECOLUMN
\begin{flalign*}
&\mbox{Policy } IC: \argmax\{Q_l[t] \times C_l[t]\}_{l \in \mathcal{L}}.
\end{flalign*}
\else
Policy $IC$: $\argmax\{Q_l[t] \times C_l[t]\}_{l \in \mathcal{L}}$.\\
\fi \\
\indent 
The $O$ policy chooses the link $l$ that has the largest value of queue-length weighted expected data rate on the link, computed using the $\tau_{l, max}$-delayed NSI of that link. Note that the $O$ policy corresponds to stopping after round 1 in the \LCVARIANTONE and \LCVARIANTTWO policies and declaring link $H$ the ``winner'' (see step \ref{step:H} of Algorithm \ref{LC1Listing}). The $IC$ policy chooses the link $l$ that has the largest value of queue-length weighted data rate on the link, computed using instantaneous NSI of that link, with the assumption that each link has access to the instantaneous NSI for all channels. \emph{We emphasize that the $IC$ policy does not conform to the structure of delayed NSI noted in Sec. \ref{section-Structure-of-Heterogeneously-Delayed-NSI}, but it serves as an upper bound on the system throughput that can be achieved in the delayed NSI regime}.

\subsection{Methodology}
As in \cite{Reddy_et_al_12}, we consider networks with complete interference (i.e, $I_l = \mathcal{L} \setminus \{l\}, \; \forall l \in \mathcal{L}$) and perfect collision (i.e., $\gamma_l = 0, \; \forall l \in \mathcal{L}$). The channel state on each link is modeled as a DTMC on the state space $\mathcal{C} = \{1, 2\}$. We only consider single-hop transmissions in the network. We implemented all the policies in a custom C++ simulator. All the numbers from simulation that we quote in this section were averaged over $10^7$ trials.
\input{table_VSD_delay_profile}
\input{fig_plot1_sat_thpt_comp_VSD_3TXs_vary_channels}

\subsection{Results and Discussion - Throughput}
Fig. \ref{figure-plot1-sat-thpt-comp-VSD_3TXs_vary_channels} shows the expected saturated system throughputs of the various policies for a network with three links, for the \textit{very small delay} (VSD) delay profile (see Table \ref{tab:VSD_Delay_Profile}), for different channel profiles. For this purpose, we define five channel profiles -- namely, \textit{very slow varying channel} (VSVC), \textit{slow varying channel} (SVC), \textit{medium varying channel} (MVC), \textit{fast varying channel} (FVC) and \textit{very fast varying channel} (VFVC), with channel-state crossover probabilities of $0.1, 0.3, 0.5, 0.7$ and $0.9$ respectively. We observe that all the policies perform equally well when the delay values are very small (the plots for all but the $IC$ policy overlap in Fig.  \ref{figure-plot1-sat-thpt-comp-VSD_3TXs_vary_channels}).

\input{fig_plot2_sat_thpt_comp_3TXs_VSVC_vary_delays}
Fig. \ref{figure-plot2-sat-thpt-comp-3TXs_VSVC_vary_delays} shows the expected saturated system throughputs of the \LCVARIANTONE, \LCVARIANTTWO, $O$ and $IC$ polices for a network with three links and VSVC channel profile, for different delay profiles. For this purpose, in addition to the VSD delay profile noted in Table \ref{tab:VSD_Delay_Profile}, we define four other delay profiles -- namely, \textit{small delays} (SD), \textit{medium delays} (MD), \textit{large delays} (LD), and \textit{very large delays} (VLD) as shown below:
\ifdefined\ONECOLUMN
\[SD : \left[ \begin{array}{ccc}
0 & 1 & 3 \\
2 & 0 & 4 \\
1 & 2 & 0 
\end{array} \right], \; \;
MD : \left[ \begin{array}{ccc}
0 & 7 & 11 \\
8 & 0 & 9 \\
12 & 6 & 0 
\end{array} \right], \; \;
LD : \left[ \begin{array}{ccc}
0 & 20 & 15 \\
24 & 0 & 17 \\
12 & 28 & 0 
\end{array} \right], \; \;
VLD : \left[
\begin{array}{ccc}
0 & 78 & 36 \\
59 & 0 & 88 \\
45 & 92 & 0 
\end{array} \right]
\] 
\else
\[ \;\; \small SD : \left[ \begin{array}{ccc}
0 & 1 & 3 \\
2 & 0 & 4 \\
1 & 2 & 0 
\end{array} \right], \; \;
\quad\quad MD : \left[ \begin{array}{ccc}
0 & 7 & 11 \\
8 & 0 & 9 \\
12 & 6 & 0 
\end{array} \right], \; \;
\]
\[ \small LD : \left[ \begin{array}{ccc}
0 & 20 & 15 \\
24 & 0 & 17 \\
12 & 28 & 0 
\end{array} \right], \; \;
VLD : \left[
\begin{array}{ccc}
0 & 78 & 36 \\
59 & 0 & 88 \\
45 & 92 & 0 
\end{array} \right]
\] 
\fi
Similar to the entries in Table \ref{tab:VSD_Delay_Profile}, the entry at position $ij$ in each of the delay profiles is to be interpreted as the least delay with which the NSI of link $l_i$ is available at transmitter $j$ (i.e., at the transmitter node of link $l_j$). We do not show the system throughputs of the $R$ and $H$ polices in Fig. \ref{figure-plot2-sat-thpt-comp-3TXs_VSVC_vary_delays} since the delay values for all but the VSD profile are such that it is computationally hard to evaluate the saturated system throughputs of these policies. We note that the \LCVARIANTONE and \LCVARIANTTWO policies perform almost equally well, and both of them outperform the $O$ policy. Fig. \ref{figure-plot2-sat-thpt-comp-3TXs_VSVC_vary_delays} also depicts the degradation of system throughput with increasing delay values.

\ifExtendedVersion
\input{table_very_small_delays_for_10_TXs}
\input{fig_plot4_sat_thpt_comp_VSVC_VSD_vary_TXs}
Fig. \ref{figure-plot4-sat-thpt-comp-VSVC_VSD_vary_TXs} shows the expected saturated system throughputs of the $R$, $H$, \LCVARIANTONE, \LCVARIANTTWO, $O$ and $IC$ polices for the case of VSVC channel profile and VSD delay profile, for different number of links in the network. The VSD delay profile, noted in Table \ref{table-VSD-Delay-Profile}, is for a network with ten links. The VSD delay profile for a network with $n$ links, $2 \leq n \leq 10$, is derived from this table by taking the $n \times n$ sub-matrix hinged at the top-left corner of this table. We show analysis plots of the \LCVARIANTONE and \LCVARIANTTWO policies only up to \#links = 5 since computing these values for networks with larger number of links is computationally costly. We note that the throughputs of both the \LCVARIANTONE and \LCVARIANTTWO policies overlap with the optimal throughput values of the $R$ and $H$ policies.
\fi

\input{table_very_medium_delays_for_10_TXs}
\input{fig_plot3_sat_thpt_comp_VSVC_MD_vary_TXs}
Fig. \ref{figure-plot3-sat-thpt-comp-VSVC_MD_vary_TXs} shows the expected saturated system throughputs of the \LCVARIANTONE, \LCVARIANTTWO, $O$ and $IC$ polices for the case of VSVC channel profile and MD delay profile, for different number of links in the network. The MD delay profile for a network with ten links is shown in Table \ref{table-MD-Delay-Profile}.
\ifExtendedVersion
As before, the MD delay profile for a network with $n$ links, $2 \leq n \leq 10$, is derived from this table by taking the $n \times n$ sub-matrix hinged at the top-left corner of this table.
\fi
We do not show the system throughputs of the $R$ and $H$ polices in Fig. \ref{figure-plot3-sat-thpt-comp-VSVC_MD_vary_TXs} since the MD delay profile is such that it is computationally hard to evaluate the saturated system throughputs of these policies. 
\ifExtendedVersion
Again, as before, we show analysis plots of the \LCVARIANTONE and \LCVARIANTTWO policies only up to \#links = 5 since computing these values for networks with more links is computationally costly. 
\fi
We note that both of the \LCVARIANTONE and \LCVARIANTTWO policies outperform the $O$ policy. We also note, more importantly, that there are situations (\#links = 3, 4) where \LCVARIANTONE outperforms \LCVARIANTTWO and other situations (\#links = 6 to 10) where \LCVARIANTTWO outperforms \LCVARIANTONE, implying that none of these policies could be throughput optimal.\footnote{With reference to the low-complexity scheduling policies in Sec. \ref{section-low-complexity-scheduling-policies}, in the process of obtaining a link that would maximize the expected system throughput, in each iteration, the \LCVARIANTONE policy discards the link with the lowest expected data rate, and the \LCVARIANTTWO policy discards the link that, upon its elimination, reduces the delay values for the largest number of channels (see step \ref{step:S} of the algorithms). Even though taking recourse to eliminating one link in each iteration gives the \LCVARIANTONE and \LCVARIANTTWO policies their short running times, the approaches that these policies take to achieve this, as alluded to in Fig. \ref{figure-plot3-sat-thpt-comp-VSVC_MD_vary_TXs}, are not optimal. It remains unresolved as to whether there is an optimal strategy to isolate a link that would be eliminated in each iteration, and if one exists, what its structure should be. This needs further exploration.}\\
\indent
\input{fig_near_optimality}
Fig. \ref{figure-near-optimality} shows the expected saturated system throughputs of the $R$, $H$, \LCVARIANTONE and \LCVARIANTTWO polices for networks with four links, VSVC channel profile and different delay profiles (DP\ifExtendedVersion $\kern-0.1em$ for short; see below\fi). We first comment that the high computational cost of the $R$ and $H$ policies hinder us from measuring their performances for very many delay profiles, and for all the delay profiles that we could measure, the expected saturated system throughputs of the  \LCVARIANTONE and \LCVARIANTTWO policies were less than 1.1\% (and in many cases less than 0.5\%) away from the optimal throughput values of the $R$ and $H$ policies, thus demonstrating the near-optimality of throughputs of the \LCVARIANTONE and \LCVARIANTTWO policies.
\ifExtendedVersion
The following are the definitions of the six delay profiles referred to in Fig. \ref{figure-near-optimality} (the entry at position $ij$ has the same interpretation as before): 
\[ \;\; \footnotesize DP1 : \left[ \begin{array}{cccc}
0 & 1 & 1 & 1 \\
1 & 0 & 1 & 1 \\
1 & 1 & 0 & 1 \\
1 & 1 & 1 & 0
\end{array} \right], \; \;
\quad\quad DP2 : \left[ \begin{array}{cccc}
0 & 1 & 1 & 2 \\
2 & 0 & 2 & 2 \\
2 & 2 & 0 & 1 \\
2 & 2 & 2 & 0
\end{array} \right], \; \; 
\]
\[ \;\; \footnotesize DP3 : \left[ \begin{array}{cccc}
0 & 1 & 1 & 2 \\
2 & 0 & 2 & 2 \\
2 & 2 & 0 & 1 \\
2 & 1 & 1 & 0
\end{array} \right], \; \;
\quad\quad DP4 : \left[ \begin{array}{cccc}
0 & 2 & 2 & 3 \\
1 & 0 & 2 & 2 \\
2 & 2 & 0 & 2 \\
1 & 2 & 1 & 0
\end{array} \right], \; \;
\]
\[ \;\; \footnotesize DP5 : \left[ \begin{array}{cccc}
0 & 2 & 2 & 2 \\
1 & 0 & 2 & 2 \\
2 & 3 & 0 & 3 \\
1 & 2 & 1 & 0
\end{array} \right], \; \;
\quad\quad DP6 : \left[ \begin{array}{cccc}
0 & 4 & 4 & 4 \\
4 & 0 & 4 & 4 \\
4 & 4 & 0 & 4 \\
4 & 4 & 4 & 0
\end{array} \right] \; \; \;\,
\]
\else
\fi
\normalsize

\vspace{0.2cm}
\indent
In Table \ref{tab:run_time_comparision}, we compare the running times of the $R$, \LCVARIANTONE and \LCVARIANTTWO policies for a network with varying number of links, VSVC channel profile, and MD delay profile. \textit{Evidently, in stark contrast to the $R$ policy, the \LCVARIANTONE and \LCVARIANTTWO policies are computationally very efficient, validating the computational complexity analyses in Propositions \ref{lemma-characterization-of-worst-case-func-eval-complexity-of-R-policy}, \ref{lemma-characterization-of-sample-path-complexity-of-R-policy} and \ref{lemma-computational-complexity-LC1-LC2-policies}}.
\input{table_Runtime_comparision}

\subsection{Results and Discussion -- Queueing Delay}
\label{subsec:Results-Discussion-Queueing-Delay}
In this section, we evaluate the queueing delay  performances of the $DQIC1$, $DQIC2$, $R$ and $H$ policies numerically and compare these with the queueing delay values obtained by evaluating the analytical expression in Eq. (\ref{Eqn:D_bar}) for these policies. In Sec. \ref{subsec:computationalcomplexity}, we saw that the computational cost of comprehensively evaluating Expr. (\ref{Eqn:D_bar}) is prohibitively expensive even for a small network that has only 2 transmitters, when there are only two channel states for any link, and when the number of packet arrivals in any slot is only 0 or 1, for a small number of time slots $T = 10$. While reducing $T$ to smaller values (say $T = 5$) will allow for a comprehensive evaluation of Expr. (\ref{Eqn:D_bar}), small values of $T$ do not allow us to capture the dynamics of packet arrivals (and hence of queue lengths) and of channel state transitions precisely enough. In fact, the number of time slots ($T$) over which we evaluate Expr. (\ref{Eqn:D_bar}) should be at least of the order of a few hundred to be able to capture the queue and channel transition dynamics with reasonable accuracy. Therefore, we resort to evaluating Expr. (\ref{Eqn:D_bar}) not comprehensively but for a few arrival streams and for a few channel state sample paths. We do this in two ways -- (i) by selecting these sample arrival streams and channel state sample paths randomly, and (ii) by selecting typical (i.e., high probability) sample arrival streams and typical channel state sample paths. This gives us leeway in evaluating Expr. (\ref{Eqn:D_bar}) for larger values of $T$. 

\input{fig_delay_performance_2TXs_DQIC1_DQIC2_Analysis_Sim_Comparison}
First, as in Sec. \ref{section-Delay-Performance-Preview}, we consider a network with two transmitters (links), with the channels on all links modeled as independent Markov chains on the state space $\{1, 2\}$ with channel state transition probability $0.1$. We consider heterogeneous delays as noted in Table \ref{tab:Het-Delays-2-TXs-x-taulmax}, where $x \geq 1$ is a parameter we vary. Packets arrive into the two queues at the two transmitters as independent Poisson processes with rates $\lambda_1 = \lambda_2 = \frac{1}{4}$ (for this setting, it is easily seen that sum rates (i.e. $\lambda_1 + \lambda_2$) of up to 1.75 are supportable; see footnote \ref{fn:supportable_sum_rate}). In Fig. \ref{fig:delay_perf_2TXs_DQIC12_T2k_100Arrls_100CSI}, we compare the average queueing delay performances of the $DQIC1$ and $DQIC2$ policies. For the analytical plots we take the time horizon $T = 2000$ and for the simulation plots we average the delay values over $10^7$ trials. For reasons mentioned in the previous paragraph, for the analytical plot in Fig. \ref{fig:delay_perf_2TXs_DQIC12_T2k_100Arrls_100CSI}(a), at each of the two links in the network, we consider a 100 random arrival streams and a 100 random channel state sample paths, and for the analytical plot in Fig. \ref{fig:delay_perf_2TXs_DQIC12_T2k_100Arrls_100CSI}(b), we consider a 100 typical arrival streams and 100 typical channel state sample paths (this gives a total of $100^4$ or $10^8$ different arrival stream - channel sample path pairs). We observe that the match between analytical and simulation results is very precise when using typical sequences for arrival streams and channel transitions.

\input{fig_delay_performance_R_H_analytical_sim}
Next, in Fig. \ref{fig:delay_performance_R_H_analytical_sim}(a), we compare the average queueing delay performances of the $R$ and $H$ policies, and in Fig. \ref{fig:delay_performance_R_H_analytical_sim}(b), that of $R$, \LCVARIANTONE and \LCVARIANTTWO policies.\footnote{The \LCVARIANTONE and \LCVARIANTONE are identical for the case of networks with only two links.} Because of the prohibitively large computational complexity involved in evaluating the $R$ and $H$ policies, we consider only 10 typical arrival streams and 10 typical channel state sample paths (for a total of $10^4$ different arrival stream - channel sample path pairs). Again, the match between analytical and simulation results is precise validating the gains in delay performance that can be obtained by using $\tau_{\textbf{.}, max}$-delayed QSI.

Finally, in Fig. \ref{fig:delay_performance_R_H_analytical_sim}(c), concentrating on the delayed queue-lengths at link $l_1$, we compare the correlation coefficients between the queue-lengths at $Q_{l_1}[t]$ (i.e., the instantaneous queue-length) and $Q_{l_1}[t-x]$ for $x = 1, \dots, 10$ in the $DQIC1$ and $DQIC2$ policies. Since in our example, for the case $\tau_{max} = 2$, the $DQIC2$ policy uses the queue-length at time $t-1$ and the $DQIC1$ policy uses the queue-length at time $t-2$, we have highlighted these values with small filled circles. As we would expect, we see that the queue-length at time $t - 1$ (i.e. at time $t-\tau_{l_1, max}$) used by the $DQIC2$ policy is more positively correlated with the instantaneous queue-length than is the queue-length at time $t-x$ for $x = 2, \ldots, 10$ used by the $DQIC1$ policy, thus explaining the delay performance gains of the $DQIC2$ policy over that of the $DQIC1$ policy.

%% file: table_VSD_delay_profile.tex
\begin{table}
\ifdefined\ONECOLUMN
\else
\scriptsize
\fi
\centering
\caption{{\small The \textit{Very Small Delays} (VSD) delay profile for a wireless network with three links}}
\label{tab:VSD_Delay_Profile}       
\ifdefined\ONECOLUMN
\begin{tabular}{p{4cm}C{1.5cm}C{1.5cm}C{1.5cm}}
\else
\begin{tabular}{p{3.3cm}C{0.95cm}C{0.95cm}C{0.95cm}}
\fi
\hline \hline \noalign{\smallskip}
& At TX $A$ & At TX $B$ & At TX $C$  \\
\noalign{\smallskip}\Xhline{2.5\arrayrulewidth}\noalign{\smallskip}
Delay in obtaining NSI of link $l_1$ & $0$ & $1$ & $1$  \\
Delay in obtaining NSI of link $l_2$ & $1$ & $0$ & $1$  \\
Delay in obtaining NSI of link $l_3$ & $1$ & $2$ & $0$  \\
\noalign{\smallskip}\hline \hline
\end{tabular}
\end{table}

%% file: fig_plot1_sat_thpt_comp_VSD_3TXs_vary_channels.tex
\begin{figure}
\centering
\ifdefined\ONECOLUMN
\includegraphics[width=0.5\textwidth]{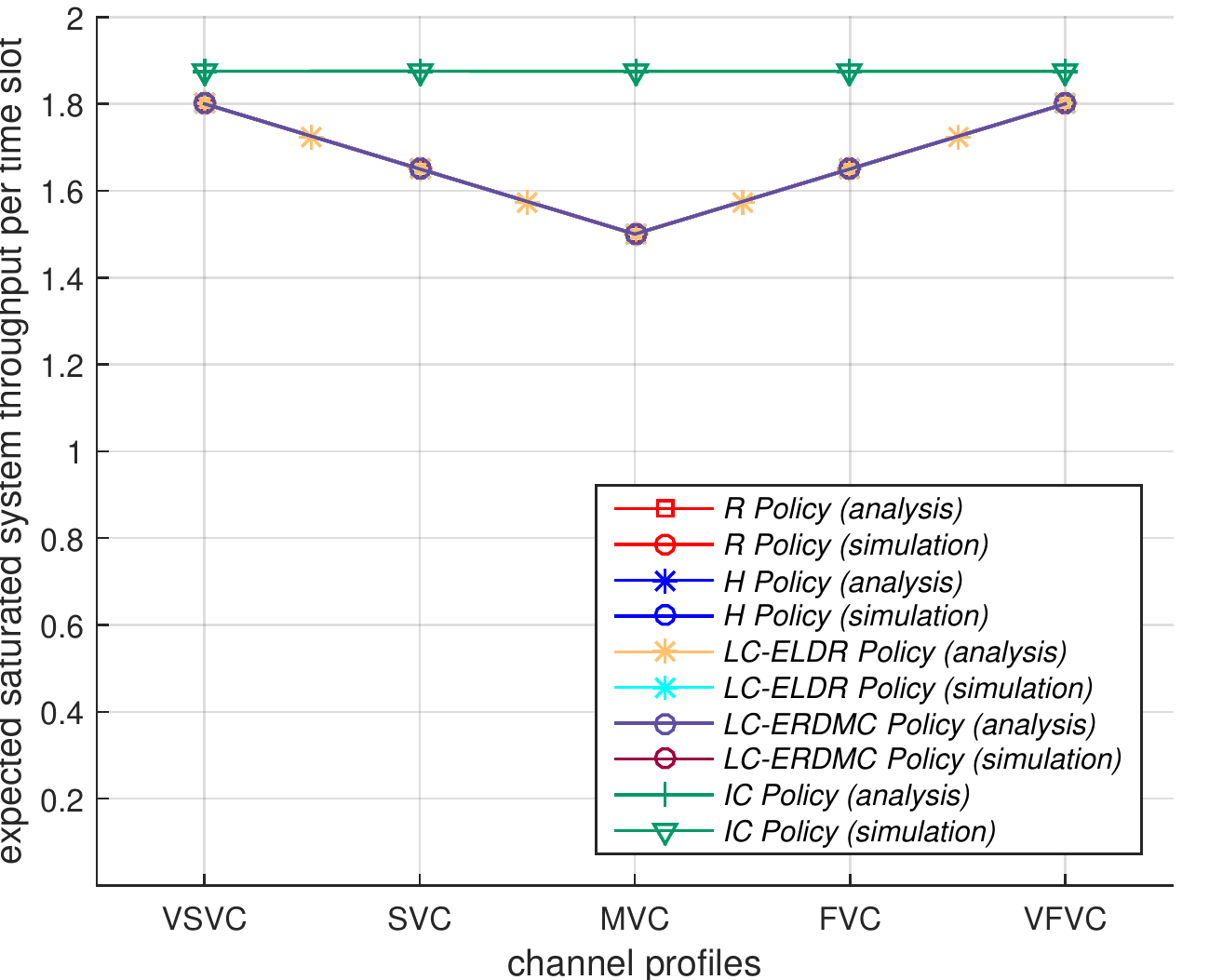}
\parbox{5in}{\caption{Expected saturated system throughputs (in data units transmitted) per time slot of the various policies for a network with three links, VSD delay profile and different channel profiles. The curves of all but the IC policy overlap.} \label{figure-plot1-sat-thpt-comp-VSD_3TXs_vary_channels}}
\else
\includegraphics[width=0.4\textwidth]{plot1_VSD_3TXs_vary_channels.pdf}
\caption{Expected saturated system throughputs (in data units transmitted) per time slot of the various policies for a network with three links, VSD delay profile and different channel profiles. The curves of all but the IC policy overlap.}
\label{figure-plot1-sat-thpt-comp-VSD_3TXs_vary_channels}       
\fi
\end{figure}

%% file: fig_plot2_sat_thpt_comp_3TXs_VSVC_vary_delays.tex
\begin{figure}
\centering
\ifdefined\ONECOLUMN
\includegraphics[width=0.5\textwidth]{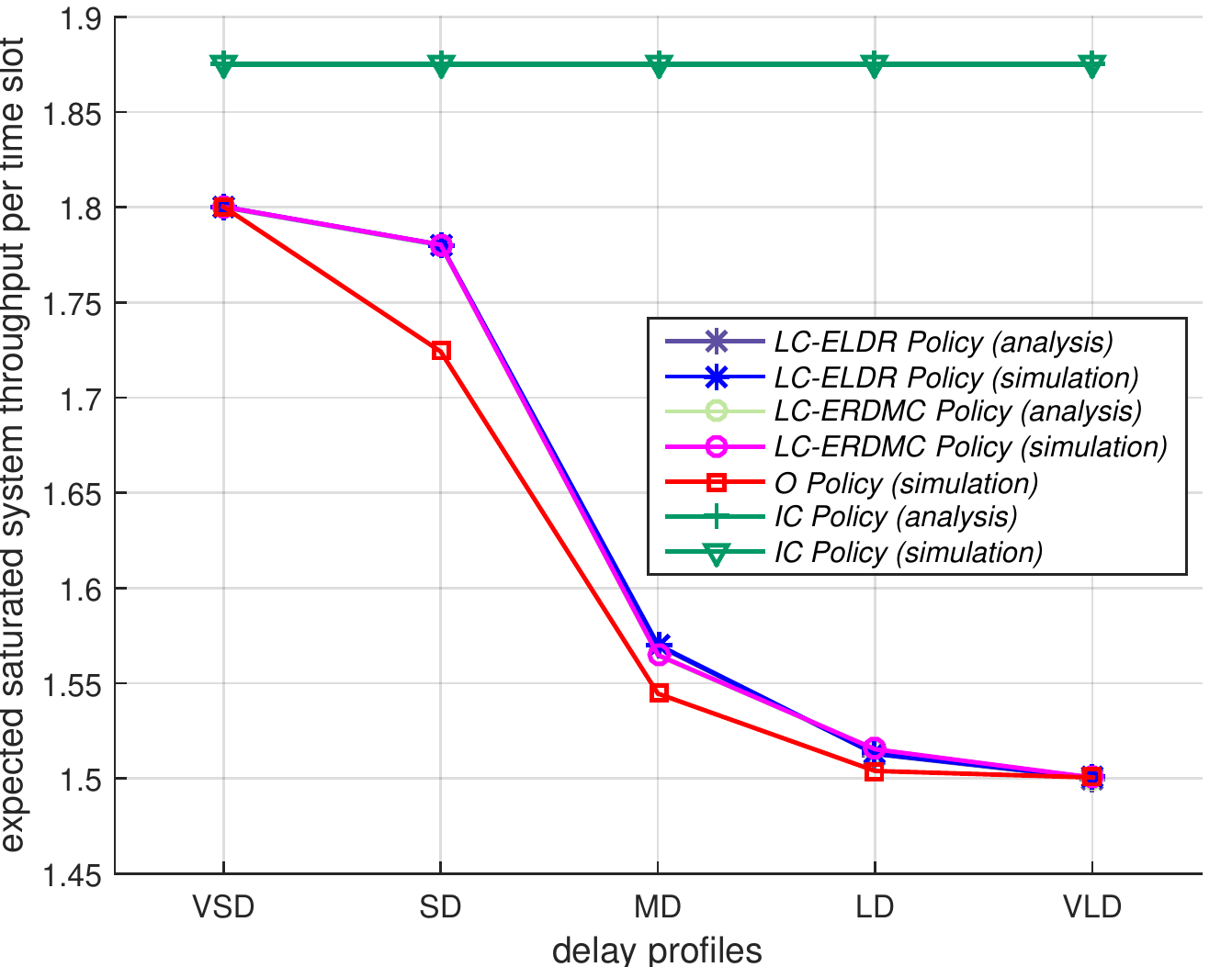}
\parbox{5in}{\caption{Expected saturated system throughputs (in data units transmitted) per time slot of \LCVARIANTONE, \LCVARIANTTWO, $O$ and $IC$ policies for a network with three links, VSVC channel profile and different delay profiles. The analysis and simulation curves of each policy overlap. Throughput of \LCVARIANTTWO is marginally below and marginally above that of \LCVARIANTONE for the MD and LD delay profiles respectively.} \label{figure-plot2-sat-thpt-comp-3TXs_VSVC_vary_delays}}
\else
\includegraphics[width=0.4\textwidth]{plot2_3TXs_VSVC_vary_delays.pdf}
\caption{Expected saturated system throughputs (in data units transmitted) per time slot of \LCVARIANTONE, \LCVARIANTTWO, $O$ and $IC$ policies for a network with three links, VSVC channel profile and different delay profiles. The analysis and simulation curves of each policy overlap. Throughput of \LCVARIANTTWO is marginally below and marginally above that of \LCVARIANTONE for the MD and LD delay profiles respectively.}
\label{figure-plot2-sat-thpt-comp-3TXs_VSVC_vary_delays}       
\fi
\end{figure}

%% file: table_very_small_delays_for_10_TXs.tex
\begin{table}
\footnotesize
\centering
\caption{{\small Table of heterogeneous delay values for the \textit{Very Small Delays} (VSD) delay profile for a network with ten links}} \label{table-VSD-Delay-Profile}       
\begin{tabular}{l >{\centering\arraybackslash}m{0.3cm} >{\centering\arraybackslash}m{0.3cm}>{\centering\arraybackslash}m{0.3cm}>{\centering\arraybackslash}m{0.3cm}>{\centering\arraybackslash}m{0.3cm}>{\centering\arraybackslash}m{0.3cm}>{\centering\arraybackslash}m{0.3cm}>{\centering\arraybackslash}m{0.3cm}>{\centering\arraybackslash}m{0.3cm}>{\centering\arraybackslash}m{0.3cm}}
\hline\hline\noalign{\smallskip}
& At TX $T_1$ & At TX $T_2$ & At TX $T_3$ & At TX $T_4$ & At TX $T_5$ & At TX $T_6$ & At TX $T_7$ & At TX $T_8$ & At TX $T_9$ & At TX $T_{10}$ \\
\noalign{\smallskip}\Xhline{2.5\arrayrulewidth}\noalign{\smallskip}
Link $l_1$  & 0 & 1 & 1 & 1 & 1 & 1 & 1 & 1 & 1 & 1  \\
Link $l_2$  & 1 & 0 & 1 & 1 & 1 & 1 & 1 & 1 & 1 & 1  \\ 
Link $l_3$  & 1 & 2 & 0 & 1 & 1 & 1 & 1 & 1 & 2 & 1  \\
Link $l_4$  & 2 & 1 & 1 & 0 & 1 & 1 & 1 & 2 & 1 & 1  \\
Link $l_5$  & 1 & 1 & 1 & 1 & 0 & 1 & 2 & 1 & 1 & 2  \\
Link $l_6$  & 1 & 2 & 1 & 1 & 1 & 0 & 1 & 1 & 1 & 1  \\
Link $l_7$  & 1 & 1 & 1 & 1 & 1 & 1 & 0 & 2 & 1 & 1  \\
Link $l_8$  & 1 & 2 & 1 & 1 & 2 & 1 & 1 & 0 & 2 & 1  \\
Link $l_9$  & 2 & 1 & 1 & 1 & 1 & 1 & 1 & 2 & 0 & 1  \\
Link $l_{10}$ & 1 & 1 & 2 & 1 & 1 & 1 & 2 & 1 & 1 & 0  \\
\noalign{\smallskip}\hline \hline
\end{tabular}
\end{table}

%% file: fig_plot4_sat_thpt_comp_VSVC_VSD_vary_TXs.tex
\begin{figure}
\centering
\ifdefined\ONECOLUMN
\includegraphics[width=0.5\textwidth]{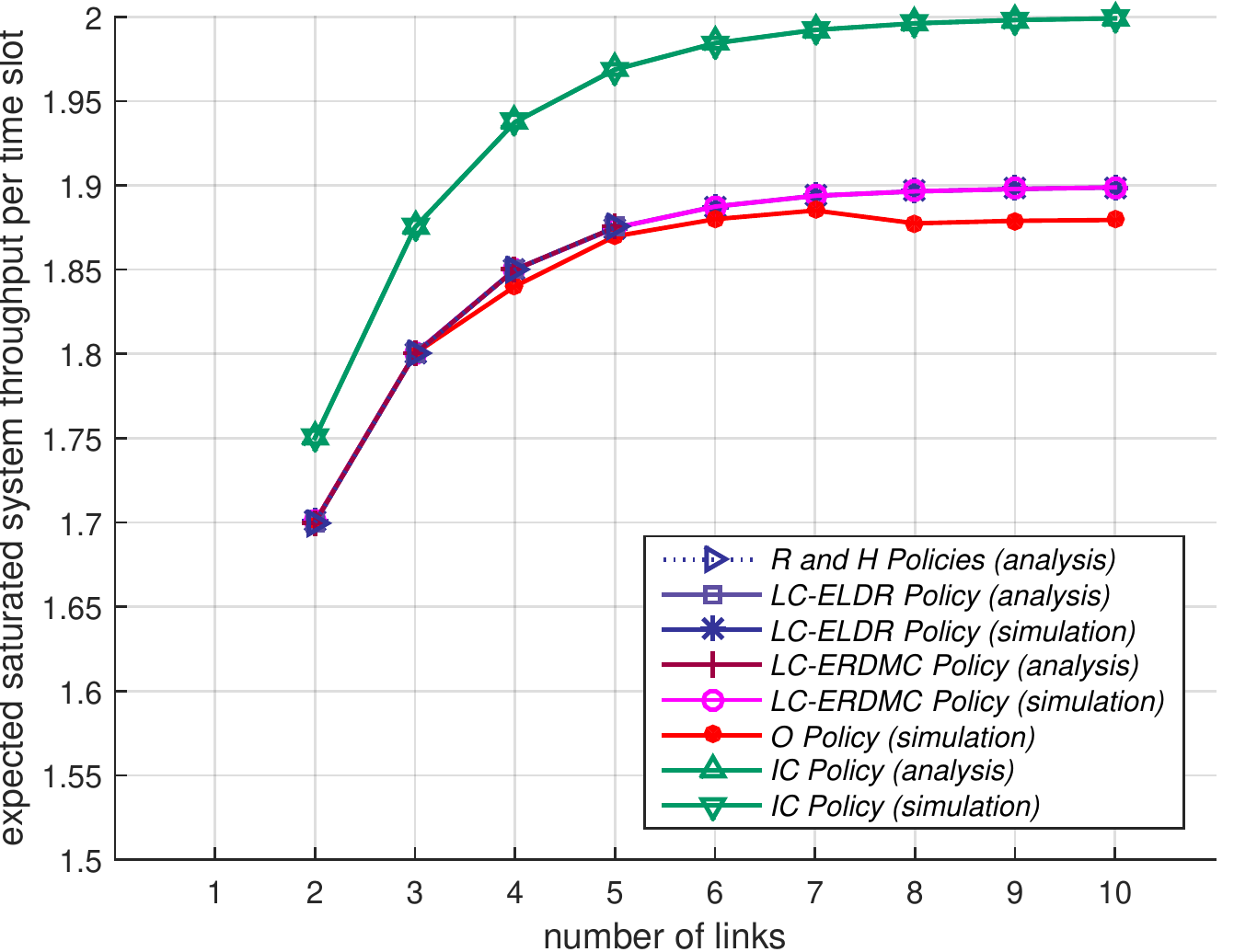}
\parbox{5in}{\caption{Expected saturated system throughputs (in data units transmitted) per time slot of the $R$, $H$, \LCVARIANTONE, \LCVARIANTTWO, $O$ and $IC$ policies for a network with VSVC channel profile and VSD delay profile, for different number of links in the network (analysis curves of \LCVARIANTONE and \LCVARIANTTWO are shown only up to \#links = 5). The analysis and simulation curves of each policy overlap.} \label{figure-plot4-sat-thpt-comp-VSVC_VSD_vary_TXs}}
\else
\includegraphics[width=0.4\textwidth]{plot4_VSVC_VSD_vary_TXs.pdf}
\caption{Expected saturated system throughputs (in data units transmitted) per time slot of the $R$, $H$, \LCVARIANTONE, \LCVARIANTTWO, $O$ and $IC$ policies for a network with VSVC channel profile and VSD delay profile, for different number of links in the network (analysis curves of \LCVARIANTONE and \LCVARIANTTWO are shown only up to \#links = 5). The analysis and simulation curves of each policy overlap.}
\label{figure-plot4-sat-thpt-comp-VSVC_VSD_vary_TXs}       
\fi
\end{figure}

%% file: table_very_medium_delays_for_10_TXs.tex
\begin{table}
\footnotesize
\centering
\caption{{\small Table of heterogeneous delay values for the \textit{Medium Delays} (MD) delay profile for a network with ten links}} \label{table-MD-Delay-Profile}       
\begin{tabular}{l >{\centering\arraybackslash}m{0.3cm} >{\centering\arraybackslash}m{0.3cm}>{\centering\arraybackslash}m{0.3cm}>{\centering\arraybackslash}m{0.3cm}>{\centering\arraybackslash}m{0.3cm}>{\centering\arraybackslash}m{0.3cm}>{\centering\arraybackslash}m{0.3cm}>{\centering\arraybackslash}m{0.3cm}>{\centering\arraybackslash}m{0.3cm}>{\centering\arraybackslash}m{0.3cm}}
\hline\hline\noalign{\smallskip}
& At TX $T_1$ & At TX $T_2$ & At TX $T_3$ & At TX $T_4$ & At TX $T_5$ & At TX $T_6$ & At TX $T_7$ & At TX $T_8$ & At TX $T_9$ & At TX $T_{10}$ \\
\noalign{\smallskip}\Xhline{2.5\arrayrulewidth}\noalign{\smallskip}
Link $l_1$  & 0 & 7 & 11 & 6 & 5 & 3 & 9 & 7 & 11 & 4  \\
Link $l_2$  & 8 & 0 & 9 & 7 & 3 & 2 & 5 & 9 & 2 & 8  \\ 
Link $l_3$  & 12 & 6 & 0 & 11 & 3 & 4 & 11 & 7 & 2 & 9  \\
Link $l_4$  & 9 & 11 & 2 & 0 & 5 & 7 & 2 & 1 & 10 & 5  \\
Link $l_5$  & 2 & 5 & 11 & 3 & 0 & 7 & 8 & 9 & 10 & 4  \\
Link $l_6$  & 1 & 9 & 2 & 4 & 8 & 0 & 11 & 6 & 5 & 2  \\
Link $l_7$  & 12 & 1 & 3 & 5 & 9 & 11 & 0 & 7 & 9 & 1  \\
Link $l_8$  & 7 & 7 & 1 & 2 & 11 & 8 & 4 & 0 & 11 & 8  \\
Link $l_9$  & 4 & 1 & 4 & 4 & 9 & 12 & 11 & 7 & 0 & 1  \\
Link $l_{10}$ & 1 & 1 & 12 & 4 & 7 & 1 & 1 & 9 & 12 & 0  \\
\noalign{\smallskip}\hline \hline
\end{tabular}
\end{table}

%% file: fig_plot3_sat_thpt_comp_VSVC_MD_vary_TXs.tex
\begin{figure}
\centering
\ifdefined\ONECOLUMN
\includegraphics[width=0.5\textwidth]{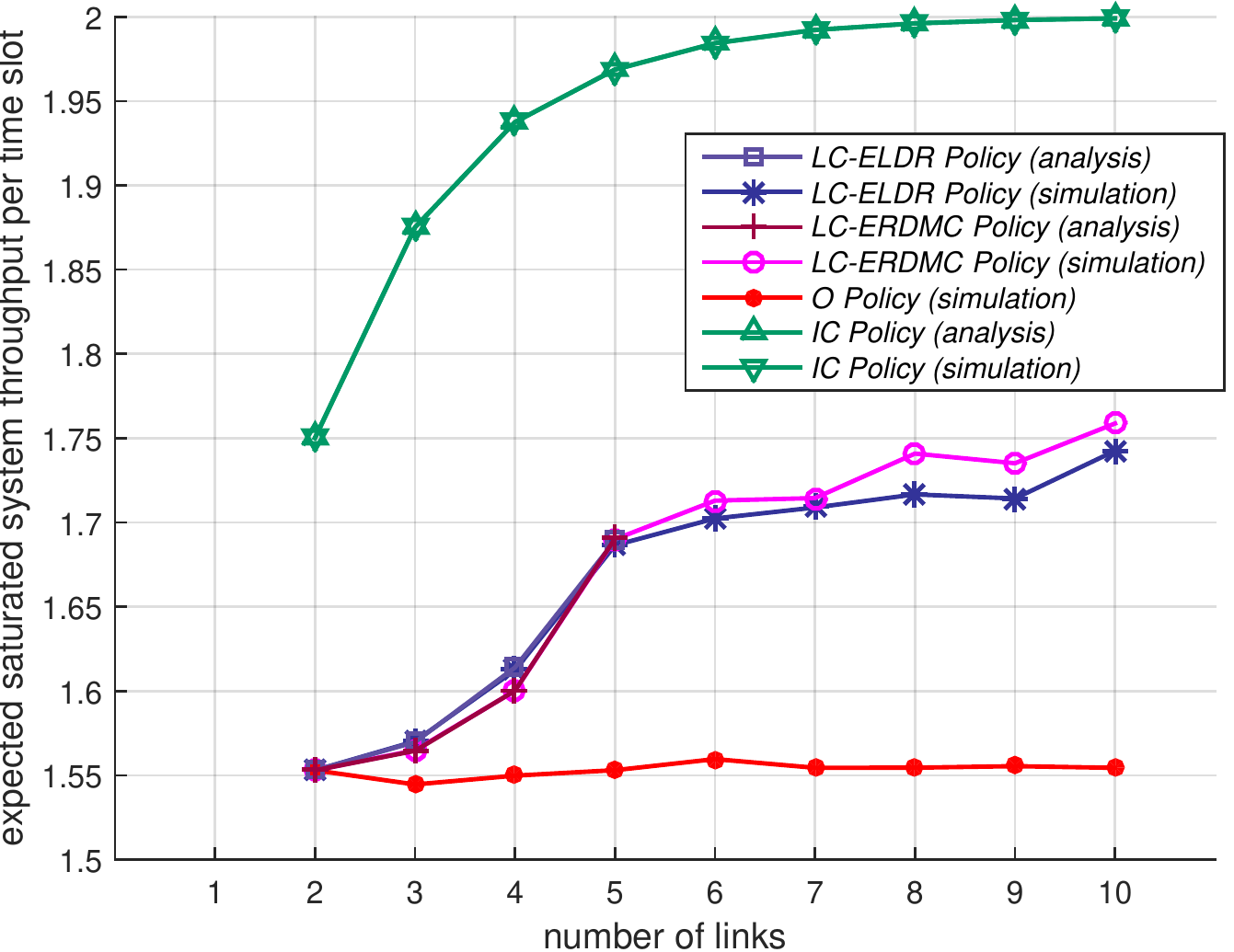}
\parbox{5in}{\caption{Expected saturated system throughputs (in data units transmitted) per time slot of the  \LCVARIANTONE, \LCVARIANTTWO, $O$ and $IC$ policies for a network with VSVC channel profile and MD delay profile, for different number of links in the network (analysis curves of \LCVARIANTONE and \LCVARIANTTWO are shown only up to \#links = 5). The analysis and simulation curves of each policy overlap.} \label{figure-plot3-sat-thpt-comp-VSVC_MD_vary_TXs}}
\else
\includegraphics[width=0.4\textwidth]{plot3_VSVC_MD_vary_TXs.pdf}
\caption{Expected saturated system throughputs (in data units transmitted) per time slot of the \LCVARIANTONE, \LCVARIANTTWO, $O$ and $IC$ policies for a network with VSVC channel profile and MD delay profile, for different number of links in the network (analysis curves of \LCVARIANTONE and \LCVARIANTTWO are shown only up to \#links = 5). The analysis and simulation curves of each policy overlap.}
\label{figure-plot3-sat-thpt-comp-VSVC_MD_vary_TXs}       
\fi
\end{figure}

%% file: fig_near_optimality.tex
\begin{figure}
\centering
\ifdefined\ONECOLUMN
\includegraphics[width=0.5\textwidth]{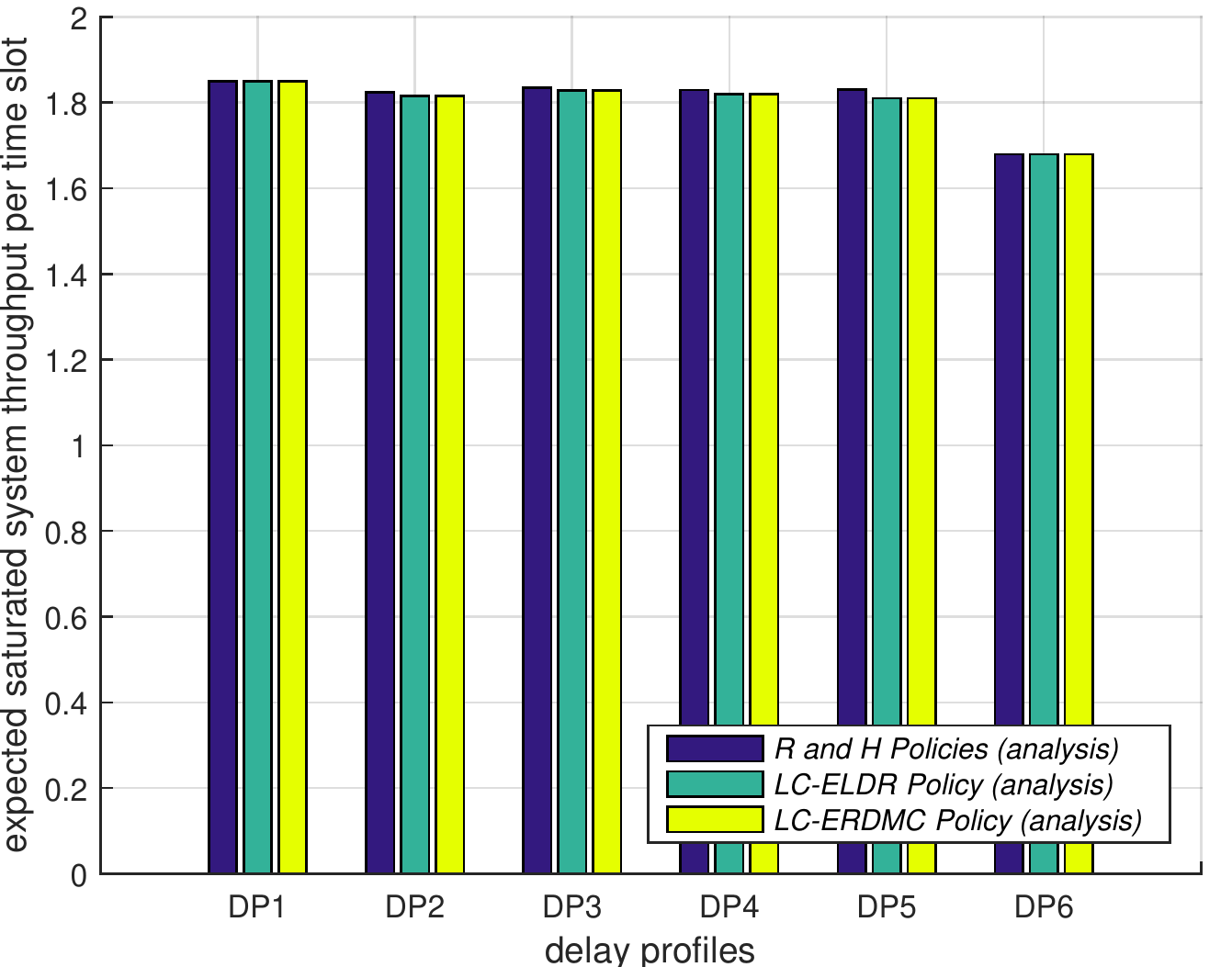}
\ifExtendedVersion
\parbox{5in}{\caption{Expected saturated system throughputs (in data units transmitted) per time slot of the $R$, $H$, \LCVARIANTONE, \LCVARIANTTWO policies for networks with four links, VSVC channel profile and different delay profiles (DP).} \label{figure-near-optimality}}
\fi
\else
\includegraphics[width=0.4\textwidth]{near-optimality.pdf}
\ifExtendedVersion
\caption{Expected saturated system throughputs (in data units transmitted) per time slot of the $R$, $H$, \LCVARIANTONE, \LCVARIANTTWO policies for networks with four links, VSVC channel profile and different delay profiles (DP).}
\fi
\label{figure-near-optimality}       
\fi
\end{figure}

%% file: table_Runtime_comparision.tex
\begin{table}
\centering
\caption{{\small Comparison of the running times of the $R$, \LCVARIANTONE and \LCVARIANTTWO policies. All times are in microseconds.}}
\label{tab:run_time_comparision}       
\begin{threeparttable}
\ifdefined\ONECOLUMN
\else
\scriptsize
\fi
\ifdefined\ONECOLUMN
\begin{tabular}{C{2cm}C{2cm}C{3cm}C{3cm}}
\else
\begin{tabular}{C{1.5cm}C{1.5cm}C{1.5cm}C{1.7cm}}
\fi
\hline \hline \noalign{\smallskip}
\# Links & $R$ Policy${}^{\dagger \S}$ & \LCVARIANTONE Policy${}^{\ddagger}$ & \LCVARIANTTWO Policy${}^{\ddagger}$  \\
\noalign{\smallskip}\Xhline{2.5\arrayrulewidth}\noalign{\smallskip}
$5$  & $4 \times 10^3$ & $11$ & $10$ \\
$10$ & $1.1 \times 10^5$ & $102$ & $96$ \\
$15$ & $3.9 \times 10^6$ & $165$ & $151$ \\
$20$ & $8.14 \times 10^7$ & $247$ & $237$ \\
\noalign{\smallskip}\hline \hline
\end{tabular}
\begin{tablenotes}
  \scriptsize
  \item $\dagger$ As reported in \cite{Reddy_et_al_12}.
  \item $\S$ Information about the delay profile used in \cite{Reddy_et_al_12} is not available. We are reasonably sure that the delay profile used in \cite{Reddy_et_al_12} uses smaller delay values than those in the MD delay profile that we have used here for the \LCVARIANTONE and \LCVARIANTTWO policies, since evaluating the $R$ policy for the MD delay profile is prohibitively time consuming. Thus the running time values that we mention here for the \LCVARIANTONE and \LCVARIANTTWO policies are relatively pessimistic, and provide a more-than-fair comparison.
  \item $\ddagger$ As measured on a computer with Intel Core 2 Duo processor clocked at 2 GHz, with 2 GB RAM.
  \item $\ddagger$ For the MD delay profile whose delay values for \#links $ = 5, 10$ are shown in Table \ref{table-MD-Delay-Profile}. The delay values in the MD delay profile for \#links $ = 15, 20$ are similar to those shown in Table \ref{table-MD-Delay-Profile}.
\end{tablenotes}
\end{threeparttable}
\end{table}

%% file: fig_delay_performance_2TXs_DQIC1_DQIC2_Analysis_Sim_Comparison.tex

\begin{figure}[!htb]
	\tiny
	\centering
	\begin{subfigure}[t]{0.9\columnwidth}
		\centering
		\includegraphics[width=1\linewidth]{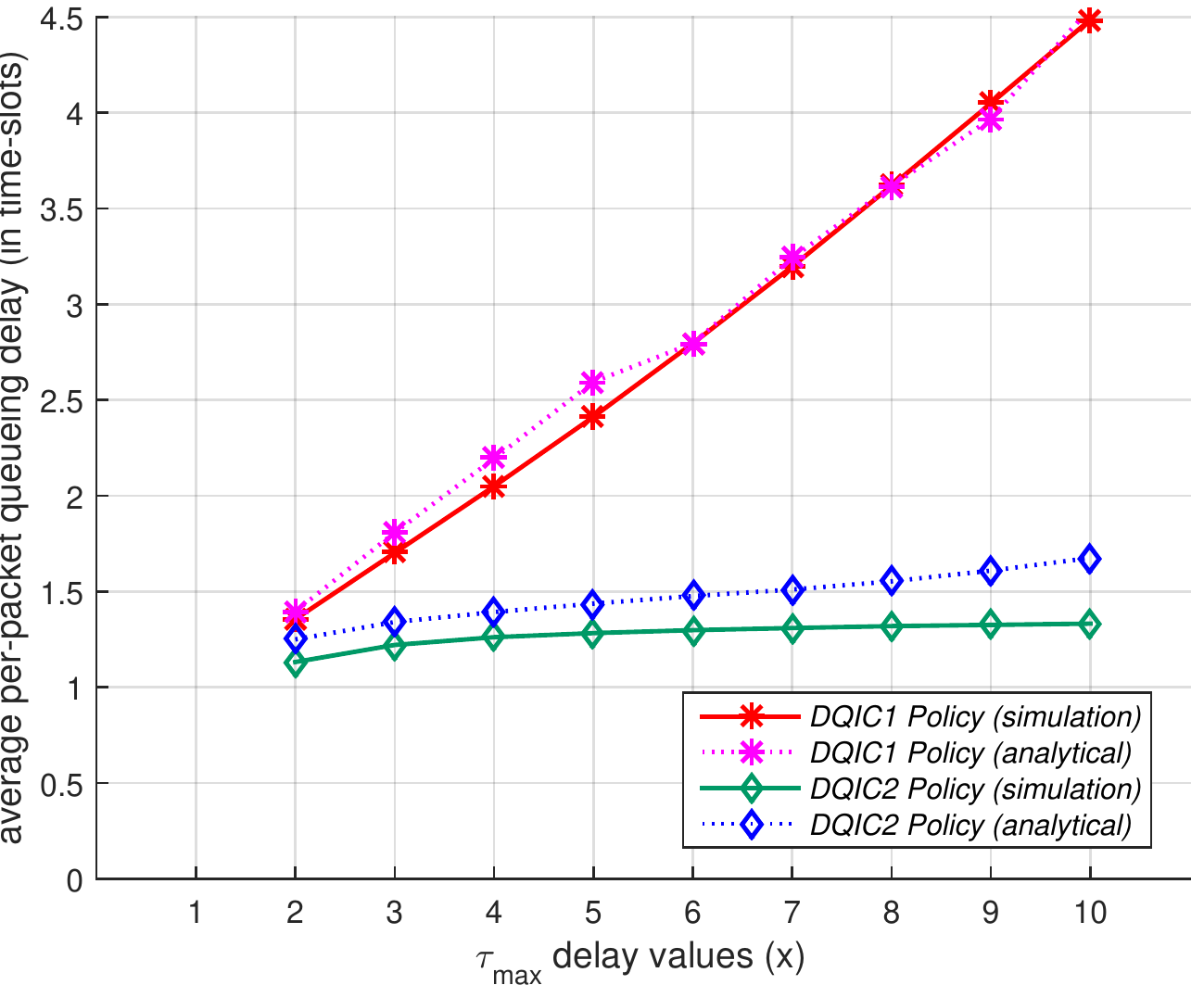}
		\label{subfig:delay_perf_2TXs_DQIC12_T2k_100Arrls_100CSI_Rand}\\[-0.4cm]
		\caption*{\scriptsize(a)}
	\end{subfigure}%
	\quad 
	\begin{subfigure}[t]{0.9\columnwidth}
		\centering
		\includegraphics[width=1\linewidth]{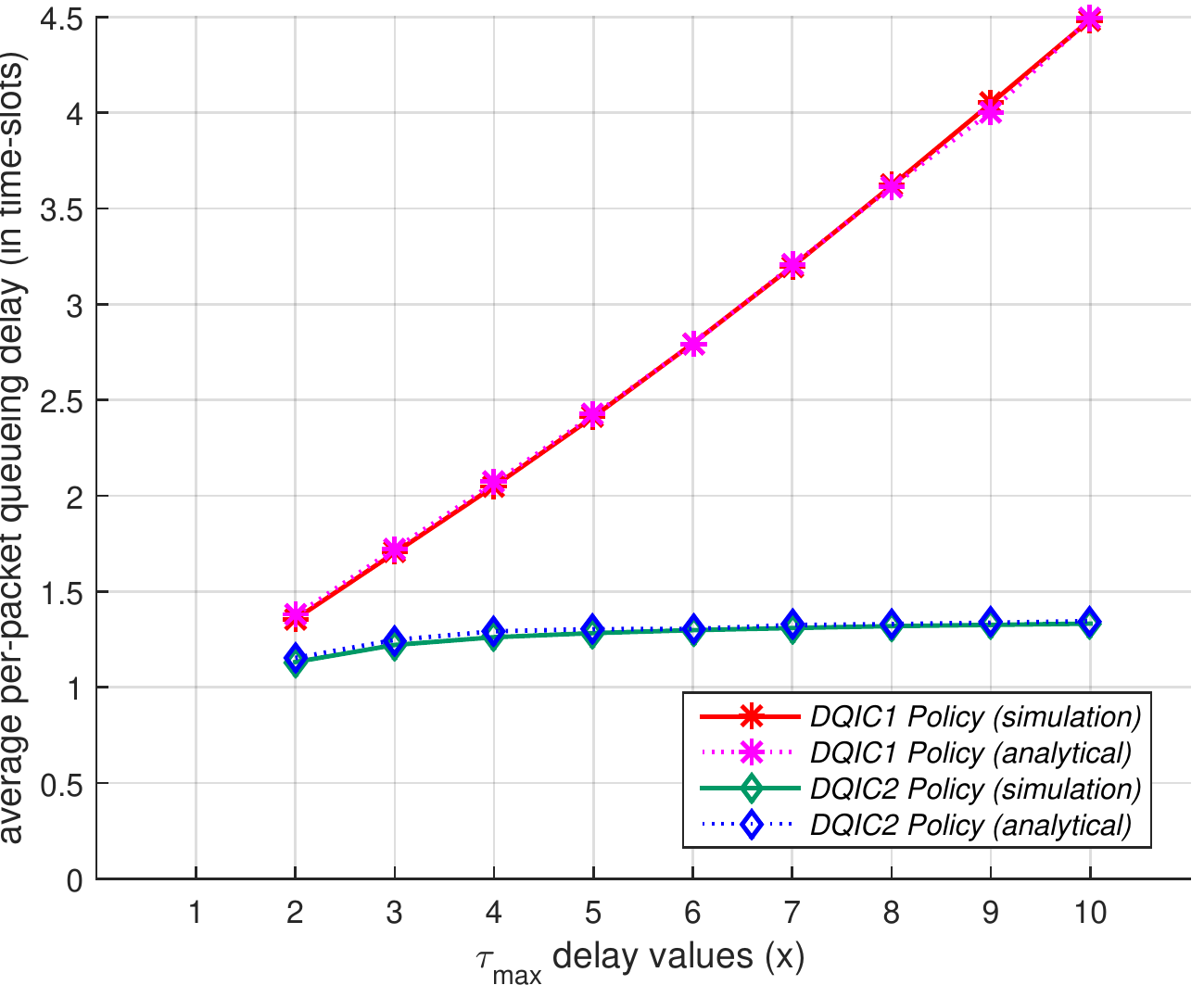}
		\label{subfig:delay_perf_2TXs_DQIC12_T2k_100Arrls_100CSI_TypSeq}\\[-0.4cm]
		\caption*{\scriptsize(b)}
	\end{subfigure}
	\caption{Comparison of analytical and simulation results for the average queueing delay per-packet in the $DQIC1$ and $DQIC2$ policies. The $DQIC1$ policy uses $\tau_{max}$-delayed QSI whereas the $DQIC2$ policy uses $\tau_{\textbf{.}, max}$-delayed QSI. The analytical queuing delay values are obtained by using 100 randomly chosen arrival streams and channel state sample paths in (a) and using 100 typical (i.e., high probability) arrival streams and channel state sample paths in (b). The match between analytical and simulation results is substantially more precise when using typical sequences for arrival streams and channel transitions.}
	\label{fig:delay_perf_2TXs_DQIC12_T2k_100Arrls_100CSI} \vspace{-1em}
\end{figure}
\vspace{0.1in}

%% file: fig_delay_performance_R_H_analytical_sim.tex
\begin{figure*}[!htb]
	\tiny
	\centering
	\begin{subfigure}[t]{0.30\textwidth}
		\centering
		\includegraphics[width=1\linewidth]{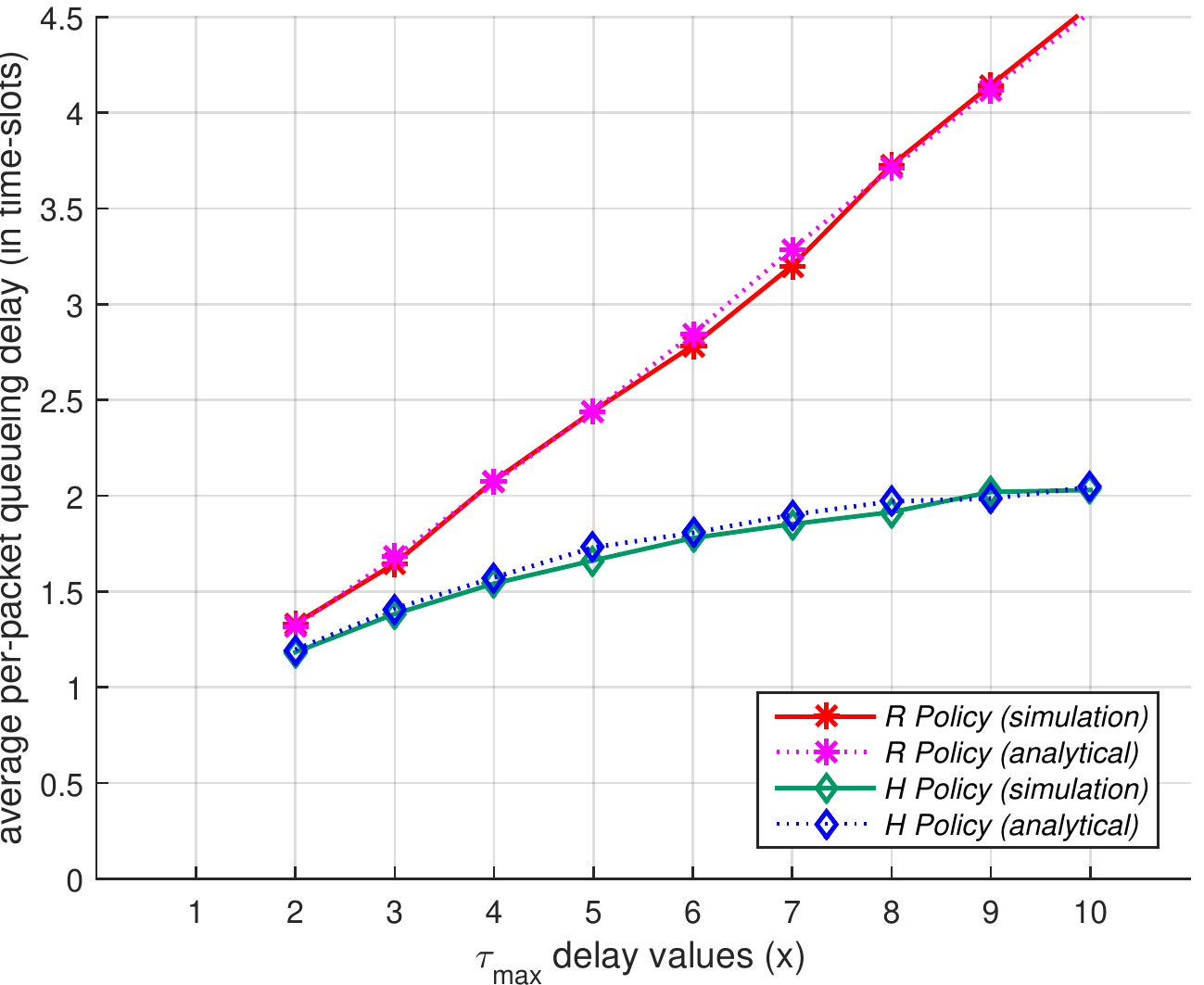}
		\caption*{\scriptsize(a)}
	\end{subfigure}%
	\quad 
	\begin{subfigure}[t]{0.30\textwidth}
		\centering
		\includegraphics[width=1\linewidth]{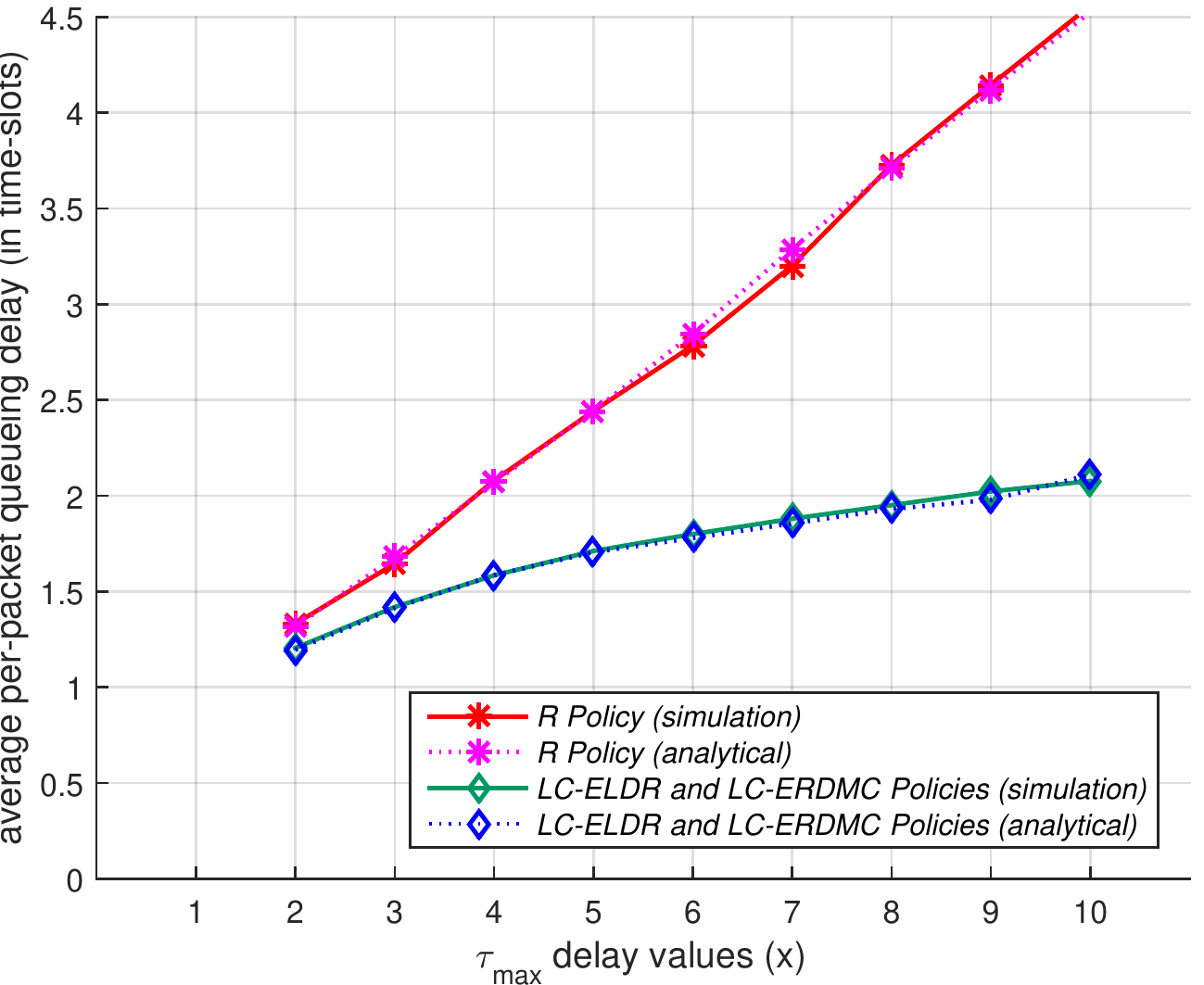}
		\caption*{\scriptsize(b)}
	\end{subfigure}
	\quad 
	\begin{subfigure}[t]{0.35\textwidth}
		\centering
		\includegraphics[width=1\linewidth]{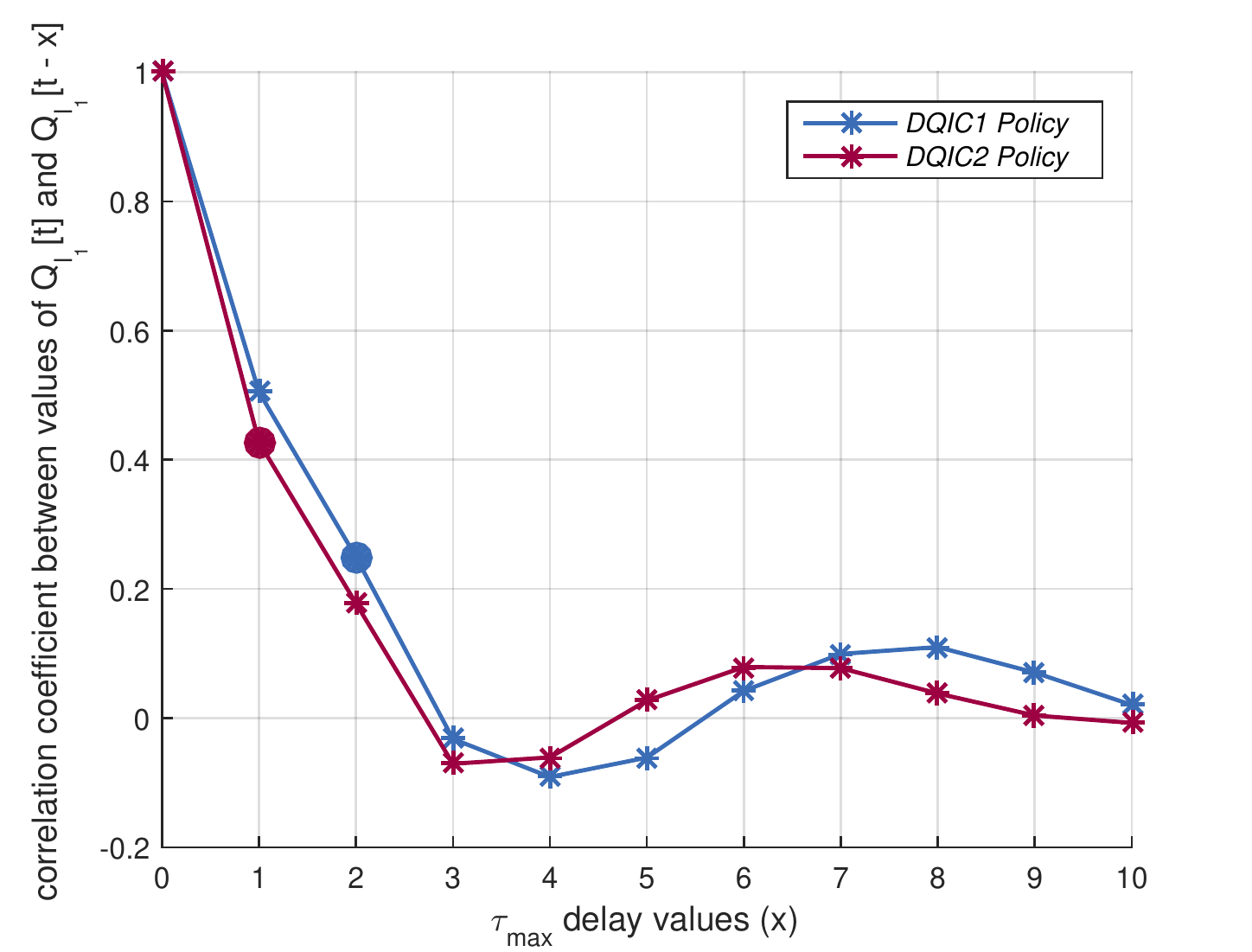}
		\caption*{\scriptsize(c)}
	\end{subfigure}
    \caption{(a) Comparison of average queueing delay per-packet in the $R$ and $H$ policies. The $R$ policy uses $\tau_{max}$-delayed QSI whereas the $H$ policy uses $\tau_{\textbf{.}, max}$-delayed QSI. The average per-packet queueing delays of the $R$ policy grows linearly with $\tau_{max}$ whereas that of the $H$ policy is sub-linear tending to become almost unaffected by increase in $\tau_{max}$. Comparing the $R$ and $H$ policies and their delay performances, the fact that the $R$ policy uses $\tau_{max}$-delayed queue lengths isolates itself as the sole reason for its undesirable delay performance. (b) Comparison of average queueing delay per-packet in the $R$ and \LCVARIANTONE/\LCVARIANTTWO policies. The \LCVARIANTONE and \LCVARIANTTWO policies, like the $H$ policy, use $\tau_{\textbf{.}, max}$-delayed QSI. (c) Comparison of correlation coefficients of queue length values in $Q_{l_1}[t]$ and $Q_{l_1}[t-x]$ in the $DQIC1$ and $DQIC2$ policies.}
    \label{fig:delay_performance_R_H_analytical_sim} \vspace{1em}
\end{figure*}
\vspace{0.1in}

%% file: section-conclusion.tex
In this work, addressing the problem of distributed scheduling in wireless networks with heterogeneously delayed NSI, we proposed, analyzed and evaluated the performances of two fast and near-throughput-optimal scheduling policies. En-route, we identified and dealt with two deficiencies (namely, non-optimal delay performance and high computational complexity) in an earlier work in \cite{Reddy_et_al_12}. We showed that the information afforded by the system model could be exploited more aggressively, and that by doing so, immense reduction in computational complexity and substantial reduction in the expected per-packet queueing delay could be obtained. We proposed a provably throughput-optimal scheduling policy that embodies these ideas. The proposed fast and near-throughput-optimal scheduling policies champion these ideas further, possess desirable queueing delay characteristics, and have running times that are of the order of microseconds, sufficiently small for use in practical deployments.



%% file: appendix-proof-characterization-of-general-case-func-eval-complexity-of-R-policy.tex
Since $T_{l_i}(.)$ is a function of $CS_{l_i}^R(\boldsymbol{\mathrm{C}}[t](0 : \tau_{max}))$ and since $\{C_{l_i}[t - \tau_{max}]\}_{l_i \in \mathcal{L}}$ is made available as part of the computation of the conditional expectation (the \textit{given}) in Eq. (\ref{eqn:R_l_tau_max}), the number of parameters to the threshold function $T_{l_i}(.)$ is the number of elements in the set $\mathcal{V}_i \coloneqq CS_{l_i}^R(\boldsymbol{\mathrm{C}}[t](0 : \tau_{max})) \setminus \{C_{l_i}[t - \tau_{max}]\}_{l_i \in \mathcal{L}}$.\\
\indent
Each of the random variables in $CS_{l_i}^R(.)$ can take any of the $\mathcal{C}$ values where $\mathcal{C}$ is the number of channel states. Thus, for $T_{l_i}(.)$, there are $\mathcal{C}^{|\mathcal{V}_i|}$ different values in the domain of $T_{l_i}(.)$. Each of these $\mathcal{C}^{|\mathcal{V}_i|}$ in the domain of $T_{l_i}(.)$ can be independently mapped to a real number. In Proposition \ref{lemma-sufficient-to-consider-c+1-real-numbers-in-range-of-Tl}, we show that it suffices to consider $\mathcal{C}+1$ carefully chosen real numbers in the range of $T_{l_i}(.)$. For example, from Proposition \ref{lemma-sufficient-to-consider-c+1-real-numbers-in-range-of-Tl}, for $\mathcal{C} = \{1, 2\}$, it suffices to consider the real numbers $0.5, 1.5, 2.5$ as the set of possible values in the range of $T_{l_i}(.)$ that each element in the domain of $T_{l_i}(.)$ can be mapped to. Thus, there are $(\mathcal{C}+1)^{\mathcal{C}^{|\mathcal{V}_i|}}$ possible choices of threshold functions for $T_{l_i}(.)$. Therefore, considering all the links in the network, the total number of threshold functions that need to be considered in the domain of optimization in Expression (\ref{eqn:Opt_Step1}) is $(\mathcal{C}+1)^{\sum_{l_i \in \mathcal{L}}\mathcal{C}^{|\mathcal{V}_i|}}$. \strut\hfill $\Box$%
\begin{proposition}
	\label{lemma-sufficient-to-consider-c+1-real-numbers-in-range-of-Tl}
	It is sufficient to consider $(\mathcal{C}+1)$ real numbers in the range of each threshold function $T_l(.)$.
\end{proposition}
\noindent
Proof:
Consider the interval $S = (c_i, c_{i+1})$ on the real number line for any integer $i \in \{0, 1, \ldots, \mathcal{C}\}$, $c_1, c_2, \ldots, c_M \in \mathcal{C}$ (where $c_M = c_{\mathcal{C}}$) are the channel states in the channel state DTMC (see Sec. \ref{section-Network-Model}) and $c_0 = c_1 - 1$.\footnote{We use $\mathcal{C}$ to denote both the set and its cardinality. Thus, as set, $\mathcal{C} = \{c_1, c_2, \ldots, c_M\}$, and as cardinality, $\mathcal{C} = M$.} Consider a fixed real number $r_1 \in S$, and a set $T \subseteq \mathcal{C}$. Evidently, $r_1$ shares the same relationship (i.e., $<, \leq, >, \geq$) with each element of $T$ as all other real numbers in $S$. Hence, it is sufficient to pick $r_1$ as a representative of all real numbers in $S$. The result follows noting that there are $\mathcal{C}+1$ intervals $S$ for each of which a representative real number $r_1$ has to be chosen. \hfill $\Box$%

%% file: appendix-proof-characterization-of-worst-case-delay-values-for-R-policy.tex
$T_l(.)$ is a function of $CS_l^R(.)$, and hence the number of choices of threshold functions for $T_l(.)$ is a function of the number of random variables in $CS_l^R(.)$, which in turn is dependent on $CS^R(.)$ through \Expr (\ref{equation-critical-set-of-link-l}). Evidently, the number of random variables in $CS^R(.)$ is maximized when the delays in the table of delay values are all distinct.\\
\indent
The number of random variables in the set $CS_{l_1}^R(.)$ pertaining to each link $l_k$ (i.e., entries of the form $C_{l_k}[t - \tau]$ in the set $CS_{l_1}^R(.)$) is maximized when $\tau_{l_1}(l_k)$ is the smallest of all the delay values in row $k$. This is because, the transmitter node of link $l_1$, having the CSI of link $l_k$ with the delay $\tau_{l_1}(l_k)$ (which is the smallest delay in row $k$), also has the CSI of link $l_k$ with delays larger than $\tau_{l_1}(l_k)$, and hence the random variables $\{C_{l_k}[t - \tau]\}_{\tau = \tau_{l_{j}}(l_k), j=1, \ldots, L, j \neq k}$ are all present in the set $CS_{l_1}^R(.)$. Similarly, after fixing the delay value at position $(k, 1)$ to be the smallest value in row $k$ as noted above, the number of random variables in the set $CS_{l_2}^R(.)$ pertaining to link $l_k$ (i.e., entries of the form $C_{l_k}[t - \tau]$ in the set $CS_{l_2}^R(.)$) is maximized when $\tau_{l_2}(l_k)$ is the second-smallest of all the delay values in row $k$ (second-smallest since, $\tau_{l_1}(l_k)$ being the smallest in row $k$, it can't be that $\tau_{l_2}(l_k) < \tau_{l_1}(l_k)$).\\
\indent
Following the structure noted above, we rearrange all the delay values in each row $i$ in ascending order (except the delay value of $0$ at position $(i,i)$ which is left as is). We note that this structure on the delays in the table of delay values characterizes a worst-case scenario for the $R$ policy. Finally, the structure on the table of delay values as needed in the statement of the proposition is obtained by swapping columns $j$ and $L-j+1$ for $j=1, \ldots, \lfloor L/2 \rfloor$, which is legitimate since the transmitters are symmetric for purposes of calculating functional evaluation complexity.\\
\ifdefined\ONECOLUMN
\begin{tabular}{p{17.2cm}p{2cm}}
& $\Box$
\end{tabular}
\else
\begin{tabular}{p{7.9cm}p{2cm}}
& $\blacksquare$
\end{tabular}
\fi

%% file: appendix-proof-characterization-of-worst-case-func-eval-complexity-of-R-policy.tex
$T_l(.)$ is a function of $CS_l^R(.)$. Given the structure of worst-case delay values noted in Proposition \ref{lemma-characterization-of-worst-case-delay-values-for-R-policy}, we see that for transmitter $1$, the set $CS_{l_1}^R(.)$ has a number of random variables that is the sum of the values in the $L$-tuple $(L-1, 1, 1, \ldots, 1)$, since transmitter $1$ has NSI of link $l_1$ with zero delay and hence possesses NSI of link $l_1$ with delays equal to all other $L-1$ values in row $1$ of the delay table, and since transmitter $1$ has NSI of link $l_i$ ($2 \leq i \leq L$) with a delay that is the maximum for link $l_i$ across all transmitters. Similarly, for transmitter $2$, the set $CS_{l_2}^R(.)$ has a number of random variables that is the sum of the values in the $L$-tuple $(1, L-1, 2, 2, \ldots, 2)$, since transmitter $2$ has NSI of link $l_1$ with a delay that is the maximum for link $l_1$ across all transmitters, and since transmitter $2$ has NSI of link $l_2$ with zero delay and hence possesses NSI of link $l_2$ with delays equal to all other $L-1$ values in row $2$ of the delay table, and since transmitter $2$ has NSI of link $l_i$ ($3 \leq i \leq L$) with a delay that is second-maximum for link $l_i$ across all transmitters. Generalizing, we see that for transmitter $i$, the set $CS_{l_i}^R(.)$ has a number of random variables that is the sum of the values in the $L$-tuple  $(i-1, i-1, \ldots, i-1, L-1, i, i, \ldots, i)$ (where the term $L-1$ is at position $i$ in the $L$-tuple). Also, given $\{C_l[t - \tau_{max}]\}_{l \in \mathcal{L}}$, the number of random variables in each of the sets $CS_{l_i}^R(.)$, $1 \leq i \leq L$, reduce by $1$. Therefore, the number of parameters for $T_{l_1}(.)$, $T_{l_2}(.)$, and $T_{l_i}(.)$ are respectively $((L-1) + (L-1) - 1)$, $(1 + (L-1) + 2(L-2) - 1)$, and $((i-1)^2 + (L-1) + i(L-i) - 1)$ which simplifies to $(i(L-2)+(L-1))$. In the language of Proposition \ref{lemma-characterization-of-general-case-func-eval-complexity-of-R-policy}, these are the values of $|\mathcal{V}_1|, |\mathcal{V}_2|$ and $|\mathcal{V}_i|$ respectively. The required result follows immediately on appealing to Proposition \ref{lemma-characterization-of-general-case-func-eval-complexity-of-R-policy}. \hfill $\Box$%

%% file: appendix-proof-characterization-of-sample-path-complexity-of-R-policy.tex
The conditional expectation in \Expr (\ref{eqn:R_l_tau_max}) is evaluated given $\boldsymbol{\mathrm{C}}[t - \tau_{max}]$. Therefore, there are $\tau_{max}$ random variables in the sample path for each link $l$ (namely, the random variables $C_l[t - (\tau_{max}-1)]$, $\ldots,$ $C_l[t]$). Hence, the total number of random variables in the sample path is $L\tau_{max}$. Each of these random variables can take $\mathcal{C}$ values. Consequently, the number of sample paths required in evaluating the conditional expectation in \Expr (\ref{eqn:R_l_tau_max}) is $\mathcal{C}^{L \tau_{max}}$.\\
\ifdefined\ONECOLUMN
\begin{tabular}{p{17.2cm}p{2cm}}
& $\Box$
\end{tabular}
\else
\begin{tabular}{p{7.9cm}p{2cm}}
& $\blacksquare$
\end{tabular}
\fi

%% file: appendix-proof-characterization-of-general-case-func-eval-complexity-of-H-policy.tex
The arguments are similar to those in the proof of Proposition \ref{lemma-characterization-of-general-case-func-eval-complexity-of-R-policy}. We present it nevertheless for purposes of completeness.

Since $T_{l_i}(.)$ is a function of $CS_{l_i}(\boldsymbol{\mathrm{C}}[t](0 : \tau_{\textbf{.}, max}))$ and since $\{C_{l_i}[t - \tau_{l_i, max}]\}_{l_i \in \mathcal{L}}$ is made available as part of the computation of the conditional expectation (the \textit{given}) in Eq. (\ref{eqn:R_l_tau_l_max}), the number of parameters to the threshold function $T_{l_i}(.)$ is the number of elements in the set $\mathcal{W}_i \coloneqq CS_{l_i}(\boldsymbol{\mathrm{C}}[t](0 : \tau_{\textbf{.}, max})) \setminus \{C_{l_i}[t - \tau_{l_i, max}]\}_{l_i \in \mathcal{L}}$.\\
\indent
Each of the random variables in $CS_{l_i}(.)$ can take any of the $\mathcal{C}$ values where $\mathcal{C}$ is the number of channel states. Thus, for $T_{l_i}(.)$, there are $\mathcal{C}^{|\mathcal{W}_i|}$ different values in the domain of $T_{l_i}(.)$. Each of these $\mathcal{C}^{|\mathcal{W}_i|}$ in the domain of $T_{l_i}(.)$ can be independently mapped to a real number. From Proposition \ref{lemma-sufficient-to-consider-c+1-real-numbers-in-range-of-Tl}, it suffices to consider $\mathcal{C}+1$ carefully chosen real numbers in the range of $T_{l_i}(.)$. For example, from Proposition \ref{lemma-sufficient-to-consider-c+1-real-numbers-in-range-of-Tl}, for $\mathcal{C} = \{1, 2\}$, it suffices to consider the real numbers $0.5, 1.5, 2.5$ as the set of possible values in the range of $T_{l_i}(.)$ that each element in the domain of $T_{l_i}(.)$ can be mapped to. Thus, there are $(\mathcal{C}+1)^{\mathcal{C}^{|\mathcal{W}_i|}}$ possible choices of threshold functions for $T_{l_i}(.)$. Therefore, considering all the links in the network, the total number of threshold functions that need to be considered in the domain of optimization in Expression (\ref{eqn:Opt_Step1_H_policy}) is $(\mathcal{C}+1)^{\sum_{l_i \in \mathcal{L}}\mathcal{C}^{|\mathcal{W}_i|}}$. \strut\hfill $\Box$%

%% file: appendix-proof-lemma-critical-set-of-link-l-R-policy-equals-critical-set-of-link-l-H-policy.tex
By definition (see \Expr (\ref{equation-critical-set}) and (\ref{equation-critical-set-H-policy})), $CS^R(\boldsymbol{\mathrm{C}}[t](0 : \tau_{max})) = CS(\boldsymbol{\mathrm{C}}[t](0 : \tau_{\textbf{.}, max}))$. The proof is complete by considering the \Expr (\ref{equation-critical-set-of-link-l}) and (\ref{equation-critical-set-of-link-l-H-policy}), and  noting that $CS^R(\boldsymbol{\mathrm{C}}[t](0 : \tau_{max}))$ and $CS(\boldsymbol{\mathrm{C}}[t](0 : \tau_{\textbf{.}, max}))$ do not have CSI of link $m$ with delay values greater than $\tau_{m, max}$, for any $m \in \mathcal{L}$, leaving the CSI of link $m$ with delays greater than $\tau_{m, max}$ in $\mathcal{P}_{lm}^{R}(\boldsymbol{\mathrm{C}}[t](0:\tau_{max}))$ (and hence in $\mathcal{P}_{l}^{R}(\boldsymbol{\mathrm{C}}[t](0:\tau_{max}))$) redundant.\\
\ifdefined\ONECOLUMN
\begin{tabular}{p{17.2cm}p{2cm}}
& $\Box$
\end{tabular}
\else
\begin{tabular}{p{7.9cm}p{2cm}}
& $\blacksquare$
\end{tabular}
\fi

%% file: appendix-proof-lemma-characterization-of-worst-case-delay-values-for-H-policy.tex
We first note from Proposition \ref{lemma-critical-set-of-link-l-R-policy-equals-critical-set-of-link-l-H-policy} that, the number of random variables in the sets $CS_l^R(.)$ and $CS_l(.)$ are the same. The rest of the arguments are similar to those in proof of Proposition \ref{lemma-characterization-of-worst-case-delay-values-for-R-policy}, and therefore omitted.\\
\ifdefined\ONECOLUMN
\begin{tabular}{p{17.2cm}p{2cm}}
& $\Box$
\end{tabular}
\else
\begin{tabular}{p{7.9cm}p{2cm}}
& $\blacksquare$
\end{tabular}
\fi

%% file: appendix-proof-lemma-characterization-of-worst-case-func-eval-complexity-of-H-policy.tex
$T_l(.)$ is a function of $CS_l(.)$. Following similar arguments as in the proof of Proposition \ref{lemma-characterization-of-worst-case-func-eval-complexity-of-R-policy} in Appendix \ref{appendix-proof-characterization-of-worst-case-func-eval-complexity-of-R-policy} we note that, for transmitter $i$, the number of random variables in the set $CS_{l_i}(.)$ is the sum of the values in the $L$-tuple  $(i-1, i-1, \ldots, i-1, L-1, i, i, \ldots, i)$ (where the term $L-1$ is at position $i$ in the $L$-tuple). Also, given $\{C_l[t - \tau_{l, max}]\}_{l \in \mathcal{L}}$, the number of random variables in each of the sets $CS_{l_i}(.)$, $1 \leq i \leq L$, reduce by $L$. Therefore, the number of parameters for the $i$th threshold function $T_{l_i}(.)$ is $((i-1)^2 + (L-1) + i(L-i) - L)$ which simplifies to $i(L-2)$. Thus, in the language of Proposition \ref{lemma-characterization-of-general-case-func-eval-complexity-of-H-policy}, $|\mathcal{W}_i| = i(L-2)$. Following similar arguments as in the proof of Proposition \ref{lemma-characterization-of-worst-case-func-eval-complexity-of-R-policy} in Appendix \ref{appendix-proof-characterization-of-worst-case-func-eval-complexity-of-R-policy}, the total number of threshold functions that are needed to be considered in the domain of optimization in \Expr (\ref{eqn:Opt_Step1_H_policy}) is $(\mathcal{C}+1)^{\sum_{l_i \in \mathcal{L}}\mathcal{C}^{i(L-2)}}$.\\
\ifdefined\ONECOLUMN
\begin{tabular}{p{17.2cm}p{2cm}}
& $\Box$
\end{tabular}
\else
\begin{tabular}{p{7.9cm}p{2cm}}
& $\blacksquare$
\end{tabular}
\fi

%% file: proof-lemma-Lambda-encompasses-all-supportable-arrival-rates.tex
The proof of this lemma is similar to the proof of Lemma 4.1 in \cite{Reddy_et_al_12}. We are given that $\{A[t]\}_t$ is supportable. This implies that there exists a policy $\mathcal{F}$ that has the following properties: (i) $\mathcal{F}$ uses $k_i$ time units delayed NSI of link $l_i$ in making its scheduling decisions at time $t$ (where $\tau_{l, max} \leq k_i \leq t$), (ii) $\mathcal{F}$ stabilizes the Markov chain $Z^{\mathcal{F}}[t] = \{\textbf{Q}[t](0 : \tau_{k, max}), \textbf{C}[t](0 : \tau_{k, max})\}$, where\footnote{we write $\boldsymbol{\mathrm{Z}}^{\mathcal{F}}[t]$ as $Z^{\mathcal{F}}[t]$ to improve legibility.} $\textbf{Q}[t](0 : \tau_{k, max}) := \{Q_{l_i}[t](0 : \tau_{k_i, max})\}_{l_i \in \mathcal{L}}$, and $\textbf{C}[t](0 : \tau_{k, max}) := \{C_{l_i}[t](0 : \tau_{k_i, max})\}_{l_i \in \mathcal{L}}$. We note that $\mathcal{F}$ could possibly use queue-state information in making its scheduling decisions.\\
\indent
First, from $\mathcal{F}$, we construct a new policy $\mathcal{F}'$, on the system state Markov chain $Y[t] = \{Q_l[t](0 : \tau_{l, max}), C[t](0 : \tau_{l, max})\}_{l \in \mathcal{L}}$ as noted next.\footnote{we write $\boldsymbol{\mathrm{Y}}[t]$ and $\boldsymbol{\mathrm{Y}}^{\mathcal{F}'}[t]$ as $Y[t]$ and $Y^{\mathcal{F}'}[t]$ respectively, to improve legibility.} Consider $z \in Z^{\mathcal{F}}[t]$ where $z = \{\textbf{q}(\tau_{\textbf{.}, max}+1 : k) \textbf{q}(0 : \tau_{\textbf{.}, max}), \textbf{c}(\tau_{\textbf{.}, max}+1 : k) \textbf{c}(0 : \tau_{\textbf{.}, max})\}$, where $\textbf{q}(\tau_{\textbf{.}, max}+1 : k) := \{q_{l_i}(\tau_{l_i, max}+1 : k_i)\}_{l_i\in \mathcal{L}}$, $\textbf{q}(0 : \tau_{\textbf{.}, max}) := \{q_l(0:\tau_{l, max})\}_{l \in \mathcal{L}}$, and $\textbf{c}(\tau_{\textbf{.}, max}+1 : k)$ and $\textbf{c}(0 : \tau_{\textbf{.}, max})$ are similarly defined. We ``collapse'' $Z^{\mathcal{F}}[t]$ onto $Y^{\mathcal{F}'}[t]$ as follows. Let $l$ be the link that policy $\mathcal{F}$ schedules in state $z$. Then $z$ is collapsed onto state $y \in Y^{\mathcal{F}'}[t]$ where $y = \{\textbf{q}(0 : \tau_{\textbf{.}, max}), \textbf{c}(0 : \tau_{\textbf{.}, max})\}$, with $\mathcal{F}'$ scheduling link $l$ in state $y$ with probability $\sum_{z'} \tilde{\pi}(z')$ ($=: \hat{\pi}(y, l)$) where $z'$ are all the states in $Z^{\mathcal{F}}[t]$ that ``collapse'' onto state $y$ (i.e. states in $Z^{\mathcal{F}}[t]$ that match the NSI $\{\textbf{q}(0 : \tau_{\textbf{.}, max}), \textbf{c}(0 : \tau_{\textbf{.}, max})\}$) such that $\mathcal{F}$ schedules link $l$ in state $z'$, and where $\tilde{\pi}(z')$ is the stationary probability of being in state $z'$ under the policy $\mathcal{F}$. Let $\hat{\pi}(y) = \sum_{l} \hat{\pi}(y,l)$. In Lemma \ref{lemma-policies-F-and-Fdash-schedule-link-l-with-same-probability}, we show that the probability with which link $l$ is scheduled in the policies $\mathcal{F}$ and $\mathcal{F}'$ is the same.\\
\indent
For $y = (\textbf{q},\textbf{c})$, we let $r(y) = \Pr(\textbf{q}|\textbf{c})$ where $\Pr(\textbf{q},\textbf{c}) = \hat{\pi}(y)$. From $\mathcal{F}'$, we construct a new policy $\mathcal{F}_s$, whose scheduling decisions are independent of all queue-state information, on the system-state Markov chain $Y[t] = \{\textbf{Q}[t](0 : \tau_{\textbf{.}, max}), \textbf{C}[t](0 : \tau_{\textbf{.}, max})\}$ as follows -- at each time, when the delayed CSI $\textbf{C}[t](0 : \tau_{\textbf{.}, max}) = \textbf{c}$, the policy  $\mathcal{F}_s$ probabilistically schedules link $l$ with probability $\sum_{y=(\textbf{q},\textbf{c})}\hat{\pi}(y,l)/\hat{\pi}(y) \times r(y)$. In Lemma \ref{lemma-policies-Fdash-and-Fs-schedule-link-l-with-same-probability}, we show that the probability with which link $l$ is scheduled in the policies $\mathcal{F}'$ and $\mathcal{F}_s$ is the same. This would then imply that $\mathcal{F}_s$ supports $A[t]$. Since $\mathcal{F}_s$ makes its scheduling decisions based on the delayed CSI $\textbf{C}[t](0 : \tau_{\textbf{.}, max}) = \textbf{c}$ (oblivious of queue state information), and since the service rates of all policies using the delayed CSI $\textbf{C}[t](0 : \tau_{\textbf{.}, max}) = \textbf{c}$ are considered in defining $\Lambda$, we have $E[A[t]] \in \Lambda$.\\
\ifdefined\ONECOLUMN
\begin{tabular}{p{17.2cm}p{2cm}}
& $\Box$
\end{tabular}
\else
\begin{tabular}{p{7.9cm}p{2cm}}
& $\blacksquare$
\end{tabular}
\fi

\begin{lemma}
\label{lemma-policies-F-and-Fdash-schedule-link-l-with-same-probability}
The policies $\mathcal{F}$ and $\mathcal{F}'$ schedule a link $l$ with the same probability.
\end{lemma}
\noindent
Proof:
Policy $\mathcal{F}$ schedules link $l$ with probability $\sum_z \tilde{\pi}(z)$, where $z$ are all the states in $Z^{\mathcal{F}}[t]$ such that $\mathcal{F}$ schedules link $l$ in state $z$. Policy $\mathcal{F}'$ schedules link $l$ with probability $\sum_y \sum_{z'} \tilde{\pi}(z')$, where $z'$ are all the states in $Z^{\mathcal{F}}[t]$ that ``collapse'' into state $y \in Y^{\mathcal{F}'}[t]$ such that $\mathcal{F}$ schedules link $l$ in state $z'$. This quantity is equal to $\sum_y \hat{\pi}(y, l)$ (alternatively, in state $y \in Y[t]$, the policy $\mathcal{F}'$ schedules link $l$ with probability $\hat{\pi}(y, l)/\hat{\pi}(y)$, and hence, overall, policy $\mathcal{F}'$ schedules link $l$ with probability $\sum_y \hat{\pi}(y, l)/ \hat{\pi}(y) \times \hat{\pi}(y) = \sum_y \hat{\pi}(y, l)$). The proof is complete by noting that $\sum_z \tilde{\pi}(z) = \sum_y \hat{\pi}(y, l)$ since states $z$ include all states in $Z^{\mathcal{F}}[t]$ such that $\mathcal{F}$ schedules link $l$ in state $z$, not just those that ``collapse'' into a particular state $y \in Y^{\mathcal{F}'}[t]$.\\
\ifdefined\ONECOLUMN
\begin{tabular}{p{17.2cm}p{2cm}}
& $\Box$
\end{tabular}
\else
\begin{tabular}{p{7.9cm}p{2cm}}
& $\blacksquare$
\end{tabular}
\fi
\begin{lemma}
\label{lemma-policies-Fdash-and-Fs-schedule-link-l-with-same-probability}
The policies $\mathcal{F}'$ and $\mathcal{F}_s$ schedule a link $l$ with the same probability.
\end{lemma}
\noindent
Proof:
From the proof of Lemma \ref{lemma-policies-F-and-Fdash-schedule-link-l-with-same-probability} we see that the policy $\mathcal{F}'$ schedules link $l$ with probability $\sum_y \hat{\pi}(y, l)$. Given that the delayed CSI $\textbf{C}[t](0 : \tau_{\textbf{.}, max}) = \textbf{c}$, the policy  $\mathcal{F}_s$ schedules link $l$ with probability:
\ifdefined\ONECOLUMN
\begin{equation*}
  \begin{split}
    \sum_{y=(\textbf{q},\textbf{c})} \frac{\hat{\pi}(y,l)}{\hat{\pi}(y)} \times r(y) &= \sum_{\textbf{q}} \frac{\hat{\pi}(y,l)}{\hat{\pi}(y)} \times r(y), \mbox{ where } y=(\textbf{q},\textbf{c}) = \sum_{\textbf{q}} \frac{\hat{\pi}(y,l)}{\hat{\pi}(y)} \times \Pr(\textbf{q}|\textbf{c}) \\
      & = \sum_{\textbf{q}} \frac{\hat{\pi}(y,l)}{\hat{\pi}(y)} \times \frac{\Pr(\textbf{q},\textbf{c})}{\Pr(\textbf{c})} = \sum_{\textbf{q}} \frac{\hat{\pi}(y,l)}{\hat{\pi}(y)} \times \frac{\hat{\pi}(y)}{\Pr(\textbf{c})} = \sum_{\textbf{q}} \frac{\hat{\pi}(y,l)}{\Pr(\textbf{c})} 
  \end{split}
\end{equation*}
\else
\begin{equation*}
  \begin{split}
    \sum_{y=(\textbf{q},\textbf{c})} & \frac{\hat{\pi}(y,l)}{\hat{\pi}(y)} \times r(y)  \\
      & = \sum_{\textbf{q}} \frac{\hat{\pi}(y,l)}{\hat{\pi}(y)} \times r(y), \mbox{ where } y=(\textbf{q},\textbf{c}) \\
      & = \sum_{\textbf{q}} \frac{\hat{\pi}(y,l)}{\hat{\pi}(y)} \times \Pr(\textbf{q}|\textbf{c}) = \sum_{\textbf{q}} \frac{\hat{\pi}(y,l)}{\hat{\pi}(y)} \times \frac{\Pr(\textbf{q},\textbf{c})}{\Pr(\textbf{c})} \\
      & = \sum_{\textbf{q}} \frac{\hat{\pi}(y,l)}{\hat{\pi}(y)} \times \frac{\hat{\pi}(y)}{\Pr(\textbf{c})} = \sum_{\textbf{q}} \frac{\hat{\pi}(y,l)}{\Pr(\textbf{c})}
  \end{split}
\end{equation*}\fi
Now, averaging over all $\textbf{c}$, we see that the policy $\mathcal{F}_s$ schedules link $l$ with probability 
\begin{equation*}
  \begin{split}
      & \sum_{\textbf{c}} \sum_{\textbf{q}} \frac{\hat{\pi}(y,l)}{\Pr(\textbf{c})} \times \Pr(\textbf{c}) = \sum_{(\textbf{q}, \textbf{c})} \hat{\pi}(y,l) = \sum_{y} \hat{\pi}(y,l).
  \end{split}
\end{equation*}
\ifdefined\ONECOLUMN
\begin{tabular}{p{17.2cm}p{2cm}}
& $\Box$
\end{tabular}
\else
\begin{tabular}{p{7.9cm}p{2cm}}
& $\blacksquare$
\end{tabular}
\fi

%% file: appendix-proof-of-theorem-H-policy-stabilizes-all-arrival-rate-vectors-in-its-thpt-region.tex
Consider the optimization formulation
	\begin{equation}
		{\displaystyle \argmax_{\textbf{F(.)}} \sum_{l \in \mathcal{L}} Q_l[t - \tau_{l, max}] R_{l, \tau_{\textbf{.}, max}}(\textbf{F(.)}) },
	\end{equation}	
\ifdefined\ONECOLUMN
	where
		\begin{equation*}
		\begin{split}
			\textstyle \hspace{0.15cm} R_{l, \tau}\left(\textbf{F(.)}\right) := \textstyle \Ex[C_l\left[t\right] F_l(.) (\gamma_l + \left(1 - \gamma_l\right) \textstyle \prod_{m \in I_l}
			(1 - F_m(.)) ) \; | \; \boldsymbol{\mathrm{C}}\left[t - \tau \right] \; ], 
		\end{split}
		\end{equation*}
		\begin{equation*}
			\boldsymbol{\mathrm{C}}[t - \tau_{\textbf{.}, max}] := \{C_l[t - \tau_{l, max}]\}_{l \in \mathcal{L}}, \mbox{ and } F_l: \mathcal{P}_l(\boldsymbol{\mathrm{C}}[t](0 : \tau_{\textbf{.}, max})) \rightarrow \{0,1\} \; \forall l \in \mathcal{{L}},
		\end{equation*}
\else
	where
		\begin{equation*}
		\begin{split}
			\textstyle \hspace{0.15cm} R_{l, \tau}\left(\textbf{F(.)}\right) := \textstyle \Ex[C_l\left[t\right] & F_l(.) (\gamma_l + \left(1 - \gamma_l\right) \\
			&\quad \textstyle  \prod_{m \in I_l}
			(1 - F_m(.)) ) \; | \; 
			 \boldsymbol{\mathrm{C}}\left[t - \tau \right] \; ], 
		\end{split}
		\end{equation*}
	$\quad \boldsymbol{\mathrm{C}}[t - \tau_{\textbf{.}, max}] := \{C_l[t - \tau_{l, max}]\}_{l \in \mathcal{L}}$, and 
\begin{flalign}
&\quad F_l: \mathcal{P}_l(\boldsymbol{\mathrm{C}}[t](0 : \tau_{\textbf{.}, max})) \rightarrow \{0,1\} \; \forall l \in \mathcal{{L}}, \nonumber &
\end{flalign}
\fi
\noindent
with the semantics that, at time $t$, the transmitter node of each link $l$ is allowed to transmit if and only if $F_{l}^{*}(\mathcal{P}_l(\boldsymbol{\mathrm{C}}[t](0 : \tau_{\textbf{.}, max}))) = 1$. From Lemma \ref{lemma-optimizing-solution-satisfies-threshold-property}, the optimizing solution (namely, $\boldsymbol{\mathrm{F^*(.)}} = \{F_{l}^*(.)\}_{l \in \mathcal{L}}$) satisfies the threshold property $F_{l}^{*}(\mathcal{P}_l(\boldsymbol{\mathrm{C}}[t](0 : \tau_{\textbf{.}, max}))) = \idoperator_{\{C_l[t] \geq T_{l}^*(\mathcal{P}_l(\boldsymbol{\mathrm{C}}[t](0 : \tau_{\textbf{.}, max})))\}}$, from Lemma \ref{lemma-optimizing-solution-depends-only-on-critical-set-NSI}, the optimizing solution depends only on the critical set NSI for each link $l$, i.e. $T_{l}^{*}(\mathcal{P}_l(\boldsymbol{\mathrm{C}}[t](0 : \tau_{\textbf{.}, max}))) = T_{l}^{*}(\mathcal{CS}_l(\boldsymbol{\mathrm{C}}[t](0 : \tau_{\textbf{.}, max})))$, and from Lemma \ref{lemma-H-policy-supports-any-mean-arrival-rate-vector-inside-Lambda}, the $H$ policy supports any mean arrival rate vector $\boldsymbol{\lambda}$ that satisfies $(1+\epsilon)\boldsymbol{\lambda} \in \Lambda$ for any $\epsilon > 0$, but from Lemma \ref{lemma-Lambda-encompasses-all-supportable-arrival-rates}, $\Lambda$ is the region that encompasses all supportable arrival rates given the NSI structure in Sec. \ref{section-Structure-of-Heterogeneously-Delayed-NSI}, whence it follows that the $H$ policy is throughput optimal.\\
\ifdefined\ONECOLUMN
\begin{tabular}{p{17.2cm}p{2cm}}
& $\Box$
\end{tabular}
\else
\begin{tabular}{p{7.9cm}p{2cm}}
& $\blacksquare$
\end{tabular}
\fi

\begin{lemma}
\label{lemma-optimizing-solution-satisfies-threshold-property}
\raggedright
$F_l^{*}(\mathcal{P}_l(\boldsymbol{\mathrm{C}}[t](0 : \tau_{\textbf{.}, max}))) = $ $\idoperator_{\{C_l[t] \geq T_l^*(\mathcal{P}_l(\boldsymbol{\mathrm{C}}[t](0 : \tau_{\textbf{.}, max})))\}}$
\end{lemma}
\noindent
Proof:
The proof of this lemma is similar to the proof of Lemma 4.2 (part one) in \cite{Reddy_et_al_12}, and is therefore omitted.\\
\ifdefined\ONECOLUMN
\begin{tabular}{p{17.2cm}p{2cm}}
& $\Box$
\end{tabular}
\else
\begin{tabular}{p{7.9cm}p{2cm}}
& $\blacksquare$
\end{tabular}
\fi

\begin{lemma}
\label{lemma-optimizing-solution-depends-only-on-critical-set-NSI}
$T_l^{*}(\mathcal{P}_l(\boldsymbol{\mathrm{C}}[t](0 : \tau_{\textbf{.}, max}))) = T_l^{*}(\mathcal{CS}_l(\boldsymbol{\mathrm{C}}[t](0 : \tau_{\textbf{.}, max})))$
\end{lemma}
\noindent
Proof:
Even though a proof similar to the proof of Lemma 4.2 (part two) in \cite{Reddy_et_al_12} applies, we give an alternate more intuitive proof here. First, consider the NSI of link $l_1$ at time $t - \tau$, where $\tau < \tau_{l_1, min}, \; \tau_{l_1, min} = \min_{h \in \mathcal{L}}\tau_h(l_1)$. Since no transmitter has NSI of link $l_1$ at time $t - \tau$, none of the transmitters can make their thresholding decision based on $C_{l_1}[t - \tau]$, and therefore, $C_{l_1}[t - \tau]$ is absent from $CS_{l_i}(\boldsymbol{\mathrm{C}}[t](0 : \tau_{\textbf{.}, max}))$ for all $l_i \in \mathcal{L}$. Next, consider two links $l_i, l_j, i,j \neq 1$ such that $\tau_{l_j}(l_1) > \tau_{l_i}(l_1) + 1$ and $\tau_{l_j}(l_1)$ is the next larger delay value after $\tau_{l_i}(l_1)$ in the row corresponding to link $l_1$ in the table of delay values. Consider a delay value $\tau$ such that $\tau_{l_i}(l_1) < \tau < \tau_{l_j}(l_1)$, i.e. $\tau$ does not appear in the row corresponding to link $l_1$ in the table of delay values, or in other words, no transmitter has the NSI of link $l_1$ with delay $\tau$. Now, transmitter node of link $l_i$ does not need to make its thresholding decision based on $\tau$ since it has NSI of link $l_1$ with a delay smaller than $\tau$ (since $\tau_{l_i}(l_1) < \tau$) and the channel is Markovian. Therefore, $CS_{l_i}(\boldsymbol{\mathrm{C}}[t](0 : \tau_{\textbf{.}, max}))$ need not contain $C_{l_1}[t - \tau]$. Lastly, consider two links $l_i, l_j, i,j \neq 1$ as above, and a delay value $\tau$ such that $\tau_{l_i}(l_1) < \tau = \tau_{l_j}(l_1)$. While it may appear that $CS_{l_i}(.)$ need not contain $C_{l_1}[t - \tau]$ since $\tau_{l_i}(l_1) < \tau$ and the channel is Markovian, $C_{l_1}[t - \tau]$ is nonetheless required to be in $CS_{l_i}(.)$ since transmitter node of link $l_j$ makes its thresholding decision based on $\tau$, and $C_{l_1}[t - \tau]$ helps in arbitrating when comparing the thresholding functions $T_{l_i}(.)$ and $T_{l_j}(.)$. From the above arguments it follows that the optimal solution does not depend on any channel state information that is not critical NSI.\\
\ifdefined\ONECOLUMN
\begin{tabular}{p{17.2cm}p{2cm}}
& $\Box$
\end{tabular}
\else
\begin{tabular}{p{7.9cm}p{2cm}}
& $\blacksquare$
\end{tabular}
\fi

\begin{lemma}
\label{lemma-H-policy-supports-any-mean-arrival-rate-vector-inside-Lambda}
The $H$ policy supports any mean arrival rate vector $\boldsymbol{\lambda}$ that satisfies $(1+\epsilon)\boldsymbol{\lambda} \in \Lambda$ for any $\epsilon > 0$.
\end{lemma}
\noindent
Proof:
We first define a Lyapunov function of the queue lengths available in the system state $\boldsymbol{\mathrm{Y}}^{\mathcal{F}}[t]$ as follows:
{
\begin{equation}
\label{eqn-Lyapunov-function}
{\displaystyle V[t] := V[\boldsymbol{\mathrm{Y}}^{\mathcal{F}}[t]] := \sum_{l \in \mathcal{L}} Q_l^2[t] }
\end{equation}
}
\indent
For a given arrival rate vector $\boldsymbol{\mathrm{\lambda}}$ that satisfies $(1 + \epsilon) \boldsymbol{\mathrm{\lambda}} \in \Lambda$ such that $\epsilon > 0$, we show that the expected change, from one time slot to the next, in the sum of the squares of the queue lengths at all the links in the network, is negative for all but a finite number of states. This implies, from Foster's theorem (see \cite{Kumar12}), that the system state Markov chain $\boldsymbol{\mathrm{Y}}^{\mathcal{F}}[t]$ is positive recurrent, establishing that the network remains stochastically stable for this arrival rate vector.\\

\noindent
From Equation (\ref{eqn-Lyapunov-function}), we have
\ifdefined\ONECOLUMN
  {
  \begin{flalign}
     \Ex [\; V[t+1] & - V[t] \; | \; \boldsymbol{\mathrm{Q}}[t - \tau_{\textbf{.}, max}], \boldsymbol{\mathrm{C}}[t - \tau_{\textbf{.}, max}] \; ] \nonumber \\
     & = \Ex [\; \sum_{l \in \mathcal{L}} (Q_l[t+1] - Q_l[t])(Q_l[t+1] + Q_l[t]) \; | \;  \boldsymbol{\mathrm{Q}}[t - \tau_{\textbf{.}, max}], \boldsymbol{\mathrm{C}}[t - \tau_{\textbf{.}, max}] \; ],
  \end{flalign}
  }
where $\boldsymbol{\mathrm{Q}}[t - \tau_{\textbf{.}, max}] = \{Q_l[t - \tau_{l, max}]\}_{l \in \mathcal{L}}$ and $\boldsymbol{\mathrm{C}}[t - \tau_{\textbf{.}, max}]$ $= \{C_l[t - \tau_{l, max}]\}_{l \in \mathcal{L}}$.\\
\else
  {
  \begin{flalign}
     \textstyle \Ex [\; & V[t+1] - V[t] \; | \; \boldsymbol{\mathrm{Q}}[t - \tau_{\textbf{.}, max}], \boldsymbol{\mathrm{C}}[t - \tau_{\textbf{.}, max}] \; ] \nonumber &
  \end{flalign}
  }
  
  \vspace{-0.4in}
  \begin{equation*}
  \begin{split}
	\; = \textstyle \Ex [\; \sum_{l \in \mathcal{L}} (Q_l[t+1] - Q_l[t])&(Q_l[t+1] + Q_l[t]) \; | \;  \\ 
	& \boldsymbol{\mathrm{Q}}[t - \tau_{\textbf{.}, max}], \boldsymbol{\mathrm{C}}[t - \tau_{\textbf{.}, max}] \; ],
  \end{split}
  \end{equation*} 
where $\boldsymbol{\mathrm{Q}}[t - \tau_{\textbf{.}, max}] = \{Q_l[t - \tau_{l, max}]\}_{l \in \mathcal{L}}$ and $\boldsymbol{\mathrm{C}}[t - \tau_{\textbf{.}, max}]$ $= \{C_l[t - \tau_{l, max}]\}_{l \in \mathcal{L}}$.\\
\fi

\noindent
Since in each time slot, the arrivals into a queue and departures out of this queue are both bounded, we have
\ifdefined\ONECOLUMN
{
  \begin{flalign}
     \label{inequality-expected-drift-inequality1}
     \Ex [\; \sum_{l \in \mathcal{L}} (Q_l[&t+1] - Q_l[t])(Q_l[t+1] + Q_l[t])  \; | \; \boldsymbol{\mathrm{Q}}[t - \tau_{\textbf{.}, max}], \boldsymbol{\mathrm{C}}[t - \tau_{\textbf{.}, max}] \; ] \nonumber \\
	 & \; \leq K_1 + \Ex [\; \sum_{l \in \mathcal{L}} (Q_l[t+1] - Q_l[t])(2Q_l[t - \tau_{l, max}]) \; | \; \nonumber \boldsymbol{\mathrm{Q}}[t - \tau_{\textbf{.}, max}], \boldsymbol{\mathrm{C}}[t - \tau_{\textbf{.}, max}] \; ]
  \end{flalign}
}
\else
{
  \begin{flalign}
     \label{inequality-expected-drift-inequality1}
     \textstyle \Ex [\; \sum_{l \in \mathcal{L}} (Q_l[t+1] - Q_l[t])&(Q_l[t+1] + Q_l[t])  \; | \; \nonumber & \\
     & \boldsymbol{\mathrm{Q}}[t - \tau_{\textbf{.}, max}], \boldsymbol{\mathrm{C}}[t - \tau_{\textbf{.}, max}] \; ] \nonumber &
  \end{flalign}
}
{ \vspace{-0.2in}
  \begin{flalign}
	\; \leq K_1 + \textstyle \Ex [\; \sum_{l \in \mathcal{L}} (Q_l[t&+1] - Q_l[t])(2Q_l[t - \tau_{l, max}]) \; | \; \nonumber & \\
	& \boldsymbol{\mathrm{Q}}[t - \tau_{\textbf{.}, max}], \boldsymbol{\mathrm{C}}[t - \tau_{\textbf{.}, max}] \; ] & 
  \end{flalign}
}
\fi

\vspace{-0.in}
\noindent
Next, using the queue update equation we show that
\ifdefined\ONECOLUMN
{
  \begin{flalign}
	\Ex [\; \sum_{l \in \mathcal{L}} (Q_l[&t+1] - Q_l[t])(2Q_l[t - \tau_{l, max}]) \; | \; \boldsymbol{\mathrm{Q}}[t - \tau_{\textbf{.}, max}], \boldsymbol{\mathrm{C}}[t - \tau_{\textbf{.}, max}] \; ] \nonumber \\
	& \; \leq \Ex [\; \sum_{l \in \mathcal{L}} (R_{l, \tau_{\textbf{.}, max}}(\boldsymbol{\mathrm{T}}^*))(2Q_l[t - \tau_{l, max}])  \; | \; \nonumber \boldsymbol{\mathrm{Q}}[t - \tau_{\textbf{.}, max}], \boldsymbol{\mathrm{C}}[t - \tau_{\textbf{.}, max}] \; ] \nonumber
  \end{flalign}
}
\else
{
  \begin{flalign}
	\textstyle \Ex [\; \sum_{l \in \mathcal{L}} (Q_l[t+1] - Q_l[t])&(2Q_l[t - \tau_{l, max}]) \; | \;  & \nonumber \\
	& \boldsymbol{\mathrm{Q}}[t - \tau_{\textbf{.}, max}], \boldsymbol{\mathrm{C}}[t - \tau_{\textbf{.}, max}] \; ] \nonumber &
  \end{flalign}
}
{
 \vspace{-0.3in}
  \begin{flalign}
	 \; \leq \textstyle \Ex [\; \sum_{l \in \mathcal{L}} (R_{l, \tau_{\textbf{.}, max}}(\boldsymbol{\mathrm{T}}^*))&(2Q_l[t - \tau_{l, max}])  \; | \; \nonumber & \\ 
	 & \boldsymbol{\mathrm{Q}}[t - \tau_{\textbf{.}, max}], \boldsymbol{\mathrm{C}}[t - \tau_{\textbf{.}, max}] \; ] & \nonumber
  \end{flalign}
}\fi

\vspace{-0.1in}
\noindent
where $\boldsymbol{\mathrm{T}}^*$ is the optimal $\boldsymbol{\mathrm{T}}$ resulting from the optimization formulation in the scheduling policy.\\

\noindent
Using the queue update equation (Eq. (\ref{equation-queue-update-equation})), we have
\ifdefined\ONECOLUMN
  \begin{flalign}
	Q_l[t+1] - Q_l[t] \leq A_l[t] + S_l[t] \nonumber
  \end{flalign}
\else
\begin{tabular}{p{9.72cm}}
\\
$Q_l[t+1] - Q_l[t] \leq A_l[t] + S_l[t]$. 
\end{tabular} \\
\fi

\noindent
Multiplying both sides of this inequality by $2Q_l[t - \tau_{l, max}]$ and taking conditional expectation with respect to $\boldsymbol{\mathrm{Q}}[t - \tau_{\textbf{.}, max}]$ and $\boldsymbol{\mathrm{C}}[t - \tau_{\textbf{.}, max}]$, we have
\ifdefined\ONECOLUMN
{
  \begin{flalign}
	\Ex[\, &(Q_l[t+1] - Q_l[t])(2Q_l[t - \tau_{l, max}]) \; | \;  \boldsymbol{\mathrm{Q}}[t - \tau_{\textbf{.}, max}], \boldsymbol{\mathrm{C}}[t - \tau_{\textbf{.}, max}] \, ] \nonumber  \\
	& \leq \; \Ex[\, A_{max} \, 2Q_l[t - \tau_{l, max}] \; | \; \boldsymbol{\mathrm{Q}}[t - \tau_{\textbf{.}, max}], \boldsymbol{\mathrm{C}}[t - \tau_{\textbf{.}, max}] \, ] + \Ex[\, S_l[t] \, 2Q_l[t - \tau_{l, max}] \; | \; \boldsymbol{\mathrm{Q}}[t - \tau_{\textbf{.}, max}], \boldsymbol{\mathrm{C}}[t - \tau_{\textbf{.}, max}] \, ]  
  \end{flalign}
}
\else
{
  \begin{flalign}
	\Ex[\, (Q_l[t+1] - Q_l[t])(2Q_l[t - \tau_{l, max}]) \; | \; & \boldsymbol{\mathrm{Q}}[t - \tau_{\textbf{.}, max}], \nonumber & \\
	& \boldsymbol{\mathrm{C}}[t - \tau_{\textbf{.}, max}] \, ] \nonumber &
  \end{flalign}
}
{
 \vspace{-0.3in}
  \begin{flalign}
	\; \leq& \; \Ex[\, A_{max} \, 2Q_l[t - \tau_{l, max}] \; | \; \boldsymbol{\mathrm{Q}}[t - \tau_{\textbf{.}, max}], \boldsymbol{\mathrm{C}}[t - \tau_{\textbf{.}, max}] \, ] \nonumber & 
  \end{flalign}
}
{
 \vspace{-0.3in}
  \begin{flalign}
	\;\;\;\;+ \;\; \Ex[\, S_l[t] \, 2Q_l[t - \tau_{l, max}] \; | \; & \boldsymbol{\mathrm{Q}}[t - \tau_{\textbf{.}, max}], \boldsymbol{\mathrm{C}}[t - \tau_{\textbf{.}, max}] \, ] & 
  \end{flalign}
}
\fi

\vspace{-0.1in}
We note that the first term on the RHS is a constant, since the value of the term $Q_l[t - \tau_{l, max}]$ is given in $\boldsymbol{\mathrm{Q}}[t - \tau_{\textbf{.}, max}]$. We let $k_l$ represent this constant, and also note that for the $H$ policy, the factor $S_l[t]$ in the second term is the same as $R_{l, \tau_{\textbf{.}, max}}(\boldsymbol{\mathrm{T}}^*)$. Therefore, we have
\ifdefined\ONECOLUMN
{
  \begin{flalign}
	\Ex[\, (Q_l[&t+1] - Q_l[t])(2Q_l[t - \tau_{l, max}]) \; | \; \boldsymbol{\mathrm{Q}}[t - \tau_{\textbf{.}, max}], \boldsymbol{\mathrm{C}}[t - \tau_{\textbf{.}, max}] \, ] \nonumber \\
	& \;\; \leq k_l + \Ex[\, R_{l, \tau_{\textbf{.}, max}}(\boldsymbol{\mathrm{T}}^*) \, 2Q_l[t - \tau_{l, max}] \; | \;  \boldsymbol{\mathrm{Q}}[t - \tau_{\textbf{.}, max}], \boldsymbol{\mathrm{C}}[t - \tau_{\textbf{.}, max}] \, ] 
  \end{flalign}
}
\else
{
  \begin{flalign}
	\Ex[\, (Q_l[t+1] - Q_l[t])(2Q_l[t - \tau_{l, max}]) \; | \; &\boldsymbol{\mathrm{Q}}[t - \tau_{\textbf{.}, max}], \nonumber & \\
	&\boldsymbol{\mathrm{C}}[t - \tau_{\textbf{.}, max}] \, ] \nonumber &
  \end{flalign}
}
{
 \vspace{-0.3in}
  \begin{flalign}
	\;\; \leq k_l + \Ex[\, R_{l, \tau_{\textbf{.}, max}}&(\boldsymbol{\mathrm{T}}^*) \, 2Q_l[t - \tau_{l, max}] \; | \;  \nonumber & \\
	& \boldsymbol{\mathrm{Q}}[t - \tau_{\textbf{.}, max}], \boldsymbol{\mathrm{C}}[t - \tau_{\textbf{.}, max}] \, ] &
  \end{flalign}
}
\fi

\vspace{-0.1in}
\noindent
Summing LHS and RHS of this inequality over all $l$, and noting that expectation is a linear operator, we have
\ifdefined\ONECOLUMN
{
  \begin{flalign}
	\Ex[\, &\sum_{l \in \mathcal{L}} (Q_l[t+1] - Q_l[t])(2Q_l[t - \tau_{l, max}]) \; | \;  \boldsymbol{\mathrm{Q}}[t - \tau_{\textbf{.}, max}], \boldsymbol{\mathrm{C}}[t - \tau_{\textbf{.}, max}] \, ] \nonumber \\
    \label{inequality-bound-on-diference-of-queue-lengths}
	& \;\; \leq K_2 + \Ex[\, \sum_{l \in \mathcal{L}} R_{l, \tau_{\textbf{.}, max}}(\boldsymbol{\mathrm{T}}^*) \, 2Q_l[t - \tau_{l, max}] \; | \; \boldsymbol{\mathrm{Q}}[t - \tau_{\textbf{.}, max}], \boldsymbol{\mathrm{C}}[t - \tau_{\textbf{.}, max}] \, ]
  \end{flalign}
}
\else
{
  \begin{flalign}
	\textstyle \Ex[\, \sum_{l \in \mathcal{L}} (Q_l[t+1] - Q_l[t])(2Q_l[t - \tau_{l, max}]) \; | \; & \boldsymbol{\mathrm{Q}}[t - \tau_{\textbf{.}, max}], \nonumber & \\
	&\boldsymbol{\mathrm{C}}[t - \tau_{\textbf{.}, max}] \, ] \nonumber &
  \end{flalign}
}
{
 \vspace{-0.3in}
  \begin{flalign}
    \label{inequality-bound-on-diference-of-queue-lengths}
	\;\; \leq \textstyle K_2 + \Ex[\, \sum_{l \in \mathcal{L}} & R_{l, \tau_{\textbf{.}, max}}(\boldsymbol{\mathrm{T}}^*) \, 2Q_l[t - \tau_{l, max}] \; | \; \nonumber & \\ 
	&\quad\quad \boldsymbol{\mathrm{Q}}[t - \tau_{\textbf{.}, max}], \boldsymbol{\mathrm{C}}[t - \tau_{\textbf{.}, max}] \, ]&
  \end{flalign}
}
\fi

\vspace{-0.1in}
\noindent
From Inequalities (\ref{inequality-expected-drift-inequality1}) and (\ref{inequality-bound-on-diference-of-queue-lengths}), we have
\ifdefined\ONECOLUMN
{
  \begin{flalign}
	\Ex [\; &\sum_{l \in \mathcal{L}} (Q_l[t+1] - Q_l[t])(Q_l[t+1] + Q_l[t]) \; | \; \boldsymbol{\mathrm{Q}}[t - \tau_{\textbf{.}, max}], \boldsymbol{\mathrm{C}}[t - \tau_{\textbf{.}, max}] \; ] \nonumber \\
    \label{inequality-expected-drift-inequality3}
	& \;\; \leq K_3 + \Ex[\, \sum_{l \in \mathcal{L}} R_{l, \tau_{\textbf{.}, max}} (\boldsymbol{\mathrm{T}}^*) \, 2Q_l[t - \tau_{l, max}] \; | \; \boldsymbol{\mathrm{Q}}[t - \tau_{\textbf{.}, max}], \boldsymbol{\mathrm{C}}[t - \tau_{\textbf{.}, max}] \, ]
  \end{flalign}
}
\else
{
  \begin{flalign}
	\textstyle \Ex [\; \sum_{l \in \mathcal{L}} (Q_l[t+1] - Q_l[t])&(Q_l[t+1] + Q_l[t]) \; | \; \nonumber & \\ 
	& \boldsymbol{\mathrm{Q}}[t - \tau_{\textbf{.}, max}], \boldsymbol{\mathrm{C}}[t - \tau_{\textbf{.}, max}] \; ] \nonumber &
  \end{flalign}
}
{
 \vspace{-0.3in}
  \begin{flalign}
    \label{inequality-expected-drift-inequality3}
	\;\; \leq K_3 + \textstyle \Ex[\, \sum_{l \in \mathcal{L}} & R_{l, \tau_{\textbf{.}, max}} (\boldsymbol{\mathrm{T}}^*) \, 2Q_l[t - \tau_{l, max}] \; | \; \nonumber & \\ 
	& \quad\quad \boldsymbol{\mathrm{Q}}[t - \tau_{\textbf{.}, max}], \boldsymbol{\mathrm{C}}[t - \tau_{\textbf{.}, max}] \, ] &  
  \end{flalign}
}
\fi

\noindent
That is,
\ifdefined\ONECOLUMN
{
  \begin{flalign}
	\label{inequality-expected-drift-inequality2}
  	\Ex [\; &V[t+1] - V[t] \; | \; \boldsymbol{\mathrm{Q}}[t - \tau_{\textbf{.}, max}], \boldsymbol{\mathrm{C}}[t - \tau_{\textbf{.}, max}] \; ] \\
	& \;\;\leq K_3 + \Ex[\, \sum_{l \in \mathcal{L}} R_{l, \tau_{\textbf{.}, max}}(\boldsymbol{\mathrm{T}}^*) \, 2Q_l[t - \tau_{l, max}] \; | \; \nonumber \boldsymbol{\mathrm{Q}}[t - \tau_{\textbf{.}, max}], \boldsymbol{\mathrm{C}}[t - \tau_{\textbf{.}, max}] \, ]
  \end{flalign}
}
\else
{
  \begin{flalign}
  	\textstyle & \Ex \left[\; V[t+1] - V[t] \; | \; \boldsymbol{\mathrm{Q}}[t - \tau_{\textbf{.}, max}], \boldsymbol{\mathrm{C}}[t - \tau_{\textbf{.}, max}] \; \right] \nonumber &
  \end{flalign}
}
{
 \vspace{-0.2in}
  \begin{flalign}
	\label{inequality-expected-drift-inequality2}
	\;\;\leq K_3 + \textstyle \Ex[\, \sum_{l \in \mathcal{L}} & R_{l, \tau_{\textbf{.}, max}}(\boldsymbol{\mathrm{T}}^*) \, 2Q_l[t - \tau_{l, max}] \; | \; \nonumber & \\
	&\quad\quad \boldsymbol{\mathrm{Q}}[t - \tau_{\textbf{.}, max}], \boldsymbol{\mathrm{C}}[t - \tau_{\textbf{.}, max}] \, ] &
  \end{flalign}
}
\fi

\vspace{-0.1in}
\noindent
Since $(1 + \epsilon) \boldsymbol{\mathrm{\lambda}} \in \Lambda$, there exists $\{ \bar{\boldsymbol{\mathrm{\eta}}}(\boldsymbol{\mathrm{c}}) \}_{\boldsymbol{\mathrm{c}}}$ such that
{
\begin{equation}
  {\displaystyle \bar{\boldsymbol{\eta}} :=  \sum_{\boldsymbol{\mathrm{c}} \in \mathcal{C}^L} \pi (\boldsymbol{\mathrm{c}}) \bar{\boldsymbol{\eta}} (\boldsymbol{\mathrm{c}}) \in \Lambda} \,, \; \mbox{and} \quad
\end{equation}
}
{
\begin{equation}
  \label{inequality-involving-eta-bar-sub-l-of-c}
  {\displaystyle \sum_{\boldsymbol{\mathrm{c}} \in \mathcal{C}^L} \pi \left(\boldsymbol{\mathrm{c}} \right) \left( (1 + \epsilon) \lambda_l - \bar{\eta}_l \left(\boldsymbol{\mathrm{c}} \right) \right) \leq 0 \qquad \forall l \in \mathcal{L} }
\end{equation}
}  
  
\noindent
From the optimization formulation of the $H$ policy, we also have
\ifdefined\ONECOLUMN
{
  \begin{flalign}
	\sum_{l \in \mathcal{L}} (&\Ex [ \; R_{l, \tau_{\textbf{.}, max}}(\boldsymbol{{\mathrm{T}}^*}) \; | \; \boldsymbol{\mathrm{Q}}[t - \tau_{\textbf{.}, max}], \boldsymbol{\mathrm{C}}[t - \tau_{\textbf{.}, max}] = \boldsymbol{\mathrm{c}} \; ]) \, \nonumber \times ( 2Q_l[t - \tau_{l, max}] )  \nonumber \\
	\label{inequality-from-optimization-formulation}
	& \;\;\geq \sum_{l \in \mathcal{L}} \bar{\eta}_l(\boldsymbol{{\mathrm{c}}}) \, ( 2Q_l[t - \tau_{l, max}] )
  \end{flalign}
}
\else
{
  \begin{flalign}
	\textstyle \sum_{l \in \mathcal{L}} (\Ex [ \; R_{l, \tau_{\textbf{.}, max}}(\boldsymbol{{\mathrm{T}}^*}) \; | \; \boldsymbol{\mathrm{Q}}[t - \tau_{\textbf{.}, max}], &\boldsymbol{\mathrm{C}}[t - \tau_{\textbf{.}, max}] = \boldsymbol{\mathrm{c}} \; ]) \, \nonumber & \\
	& \times ( 2Q_l[t - \tau_{l, max}] )  \nonumber &
  \end{flalign}
}
{
 \vspace{-0.3in}
  \begin{flalign}
	\label{inequality-from-optimization-formulation}
	\;\;&\geq \textstyle \sum_{l \in \mathcal{L}} \bar{\eta}_l(\boldsymbol{{\mathrm{c}}}) \, ( 2Q_l[t - \tau_{l, max}] ) &  
  \end{flalign}
}
\fi

\vspace{-0.1in}
This is because, $\boldsymbol{\mathrm{T}}^*$ is an optimal threshold, and for a given $\boldsymbol{\mathrm{c}}$, the term  $\bar{\eta}_l(\boldsymbol{\mathrm{c}})$ in the RHS is the $l$th component of the point $\bar{\boldsymbol{\mathrm{\eta}}}(\boldsymbol{\mathrm{c}}) \in \mathbb{R}^L$, and this point $\bar{\boldsymbol{\mathrm{\eta}}}(\boldsymbol{\mathrm{c}})$ arises from a specific weighted combination of $\boldsymbol{\mathrm{T}}s$ (the weights being the fraction of times these $\boldsymbol{\mathrm{T}}s$ are chosen), where some of these $\boldsymbol{\mathrm{T}}s$ may not be optimal. The expectation term in the LHS results from giving a weight of $1$ to $\boldsymbol{\mathrm{T}}^*$. Also, given $\boldsymbol{\mathrm{T}}^*$, the term $\boldsymbol{\mathrm{Q}}[t - \tau_{\textbf{.}, max}]$ in the conditional expectation on the LHS is redundant.

Let $\tilde{K}$ be an upper bound on the LHS of Inequality (\ref{inequality-from-optimization-formulation})  such that this upper bound is at least twice as large as the value of the expression on the LHS. Then, the quantity $\tilde{K}$ minus the value of the expression on the RHS serves as an upper bound on the value of the expression on the LHS. That is,
\ifdefined\ONECOLUMN
{
  \begin{flalign}
	\sum_{l \in \mathcal{L}} (& \Ex [ \; R_{l, \tau_{\textbf{.}, max}}(\boldsymbol{{\mathrm{T}}^*}) \; | \; \boldsymbol{\mathrm{Q}}[t - \tau_{\textbf{.}, max}], \boldsymbol{\mathrm{C}}[t - \tau_{\textbf{.}, max}] = \boldsymbol{\mathrm{c}} \; ]) \, \nonumber \times ( 2Q_l[t - \tau_{l, max}] ) \nonumber \\
	\label{inequality-ingenious-bound}
	& \;\;\leq \; \tilde{K} - \sum_{l \in \mathcal{L}} \bar{\eta}_l(\boldsymbol{{\mathrm{c}}}) \, ( 2Q_l[t - \tau_{l, max}] ) 
  \end{flalign}
}
\else
{
  \begin{flalign}
	\textstyle \sum_{l \in \mathcal{L}} (\Ex [ \; R_{l, \tau_{\textbf{.}, max}}(\boldsymbol{{\mathrm{T}}^*}) \; | \; \boldsymbol{\mathrm{Q}}[t - \tau_{\textbf{.}, max}], & \boldsymbol{\mathrm{C}}[t - \tau_{\textbf{.}, max}] = \boldsymbol{\mathrm{c}} \; ]) \, \nonumber & \\ 
	& \times ( 2Q_l[t - \tau_{l, max}] ) \nonumber &
  \end{flalign}
}
{
 \vspace{-0.3in}
  \begin{flalign}
	\label{inequality-ingenious-bound}
	\;\;\leq \; & \textstyle \tilde{K} - \sum_{l \in \mathcal{L}} \bar{\eta}_l(\boldsymbol{{\mathrm{c}}}) \, ( 2Q_l[t - \tau_{l, max}] ) &
  \end{flalign}
}
\fi

\vspace{-0.1in}
\noindent
From Inequality (\ref{inequality-involving-eta-bar-sub-l-of-c}), we have
{
  \begin{flalign}
	\label{inequality-involving-eta-bar-sub-l-of-c-2}
	\displaystyle \sum_{\boldsymbol{\mathrm{c}} \in \mathcal{C}^L} \pi(\boldsymbol{\mathrm{c}}) \, \left( - \bar{\eta}_l(\boldsymbol{\mathrm{c}}) \right) \leq - (1 + \epsilon) \lambda_l &
  \end{flalign}
}

\noindent
Taking expectation on both sides of Inequality (\ref{inequality-ingenious-bound}) over $\boldsymbol{\mathrm{C}}[t - \tau_{\textbf{.}, max}]$, and noting that expectation is a linear operator, we get 
\ifdefined\ONECOLUMN
{
  \begin{flalign}
	\sum_{\boldsymbol{\mathrm{c}} \in \mathcal{C}^L} \mbox{\textit{\{}} (&\Ex [ \; \sum_{l \in \mathcal{L}} R_{l, \tau_{\textbf{.}, max}}(\boldsymbol{{\mathrm{T}}^*}) \; | \; \boldsymbol{\mathrm{Q}}[t - \tau_{\textbf{.}, max}], \nonumber \boldsymbol{\mathrm{C}}[t - \tau_{\textbf{.}, max}] = \boldsymbol{\mathrm{c}} \; ]) \, \times ( 2Q_l[t - \tau_{l, max}] ) \mbox{\textit{\}}} \; \pi(\boldsymbol{\mathrm{c}}) \nonumber \\
	& \;\;\leq \; \tilde{K} + \sum_{l \in \mathcal{L}} ( 2Q_l[t - \tau_{l, max}] )  \sum_{\boldsymbol{\mathrm{c}} \in \mathcal{C}^L} \pi(\boldsymbol{\mathrm{c}}) \, ( - \bar{\eta}_l(\boldsymbol{{\mathrm{c}}}) ) \nonumber \\
	& \;\;\leq \; \tilde{K} + \sum_{l \in \mathcal{L}} ( 2Q_l[t - \tau_{l, max}] ) \, (- (1 + \epsilon) \lambda_l ) \quad \mbox{(using  (\ref{inequality-involving-eta-bar-sub-l-of-c-2}))} \nonumber \\
	& \;\;\leq \; \tilde{K} - 2 \epsilon \sum_{l \in \mathcal{L}} \, ( Q_l[t - \tau_{l, max}] \, \lambda_l )  \nonumber
  \end{flalign}
}
\else
{
  \begin{flalign}
	\textstyle \sum_{\boldsymbol{\mathrm{c}} \in \mathcal{C}^L} \mbox{\textit{\{}} (\Ex [ \; & \textstyle \sum_{l \in \mathcal{L}} R_{l, \tau_{\textbf{.}, max}}(\boldsymbol{{\mathrm{T}}^*}) \; | \; \boldsymbol{\mathrm{Q}}[t - \tau_{\textbf{.}, max}], \nonumber & \\
	& \; \boldsymbol{\mathrm{C}}[t - \tau_{\textbf{.}, max}] = \boldsymbol{\mathrm{c}} \; ]) \, \times ( 2Q_l[t - \tau_{l, max}] ) \mbox{\textit{\}}} \; \pi(\boldsymbol{\mathrm{c}}) \nonumber &
  \end{flalign}
}
{
 \vspace{-0.25in}
  \begin{flalign}
	\;\;\leq \; & \textstyle \tilde{K} + \sum_{l \in \mathcal{L}} ( 2Q_l[t - \tau_{l, max}] )  \sum_{\boldsymbol{\mathrm{c}} \in \mathcal{C}^L} \pi(\boldsymbol{\mathrm{c}}) \, ( - \bar{\eta}_l(\boldsymbol{{\mathrm{c}}}) ) \nonumber &
  \end{flalign}
}
{
 \vspace{-0.25in}
  \begin{flalign}
	\;\;\leq \; & \textstyle \tilde{K} + \sum_{l \in \mathcal{L}} ( 2Q_l[t - \tau_{l, max}] ) \, (- (1 + \epsilon) \lambda_l ) \quad \mbox{(using  (\ref{inequality-involving-eta-bar-sub-l-of-c-2}))} \nonumber &
  \end{flalign}
}
{
 \vspace{-0.25in}
  \begin{flalign}
	\;\;\leq \; & \textstyle \tilde{K} - 2 \epsilon \sum_{l \in \mathcal{L}} \, ( Q_l[t - \tau_{l, max}] \, \lambda_l )  \nonumber & 
  \end{flalign}
}
\fi

\noindent
That is,
\ifdefined\ONECOLUMN
{
  \begin{flalign}
	\sum_{\boldsymbol{\mathrm{c}} \in \mathcal{C}^L} \mbox{\textit{\{}} & (\Ex [ \; \sum_{l \in \mathcal{L}} R_{l, \tau_{\textbf{.}, max}}(\boldsymbol{{\mathrm{T}}^*}) \; | \; \boldsymbol{\mathrm{Q}}[t - \tau_{\textbf{.}, max}], \nonumber \boldsymbol{\mathrm{C}}[t - \tau_{\textbf{.}, max}] = \boldsymbol{\mathrm{c}} \; ]) \times ( 2Q_l[t - \tau_{l, max}] ) \mbox{\}} \; \pi(\boldsymbol{\mathrm{c}}) \nonumber \\
  	\label{inequality-last-but-one}
	& \;\;\leq \; \tilde{K} - 2 \epsilon \sum_{l \in \mathcal{L}} \, ( Q_l[t - \tau_{l, max}] \, \lambda_l ) 
  \end{flalign}
}
\else
{
  \begin{flalign}
	\textstyle \sum_{\boldsymbol{\mathrm{c}} \in \mathcal{C}^L} \mbox{\textit{\{}} & (\Ex [ \; \textstyle \sum_{l \in \mathcal{L}} R_{l, \tau_{\textbf{.}, max}}(\boldsymbol{{\mathrm{T}}^*}) \; | \; \boldsymbol{\mathrm{Q}}[t - \tau_{\textbf{.}, max}], \nonumber & \\
	&\quad\quad \boldsymbol{\mathrm{C}}[t - \tau_{\textbf{.}, max}] = \boldsymbol{\mathrm{c}} \; ]) \times ( 2Q_l[t - \tau_{l, max}] ) \mbox{\}} \; \pi(\boldsymbol{\mathrm{c}}) \nonumber &
  \end{flalign}
}
{
 \vspace{-0.25in}
  \begin{flalign}
  \label{inequality-last-but-one}
	\;\;\leq \; & \textstyle \tilde{K} - 2 \epsilon \sum_{l \in \mathcal{L}} \, ( Q_l[t - \tau_{l, max}] \, \lambda_l ) &
  \end{flalign}
}
\fi

\noindent
From Inequalities (\ref{inequality-expected-drift-inequality2}) and (\ref{inequality-last-but-one}), we have
\ifdefined\ONECOLUMN
{
  \begin{flalign}
	\label{inequality-last-one}
	\Ex &\left[\; V[t+1] - V[t] \; | \; \boldsymbol{\mathrm{Q}}[t - \tau_{\textbf{.}, max}] \; \right] \leq K - 2 \epsilon \sum_{l \in \mathcal{L}} \, Q_l[t - \tau_{l, max}] \, \lambda_l 	
  \end{flalign}
}
\else
{
  \begin{flalign}
	\textstyle \Ex &\left[\; V[t+1] - V[t] \; | \; \boldsymbol{\mathrm{Q}}[t - \tau_{\textbf{.}, max}] \; \right] \nonumber &
  \end{flalign}
}
{
 \vspace{-0.35in}
  \begin{flalign}
	\label{inequality-last-one}
	\;\;\;\;\leq \;\; & K - 2 \epsilon \textstyle \sum_{l \in \mathcal{L}} \, Q_l[t - \tau_{l, max}] \, \lambda_l &
  \end{flalign}
}
\fi

\vspace{-0.4cm}
Since $K$ is a constant, the RHS in the above inequality is greater or equal to zero only for a finite number of states in the system state Markov chain $\boldsymbol{\mathrm{Y}}^{\mathcal{F}}[t]$, and lesser or equal to zero for the rest of the states in $\boldsymbol{\mathrm{Y}}^{\mathcal{F}}[t]$. It now follows from Foster's theorem (see \cite{Kumar12}), that the system state Markov chain $\boldsymbol{\mathrm{Y}}^{\mathcal{F}}[t]$ is positive recurrent. Hence the $H$ policy preserves the stochastic stability of the network for the given arrival rate vector  $\boldsymbol{\mathrm{\lambda}}$. Since $\boldsymbol{\mathrm{\lambda}}$ is any arbitrary vector in $\Lambda$ such that $(1 + \epsilon)\boldsymbol{\mathrm{\lambda}} \in \Lambda$, it follows that the $H$ policy preserves the stochastic stability of the network for all arrival rate vectors in the interior of $\Lambda$.\\
\ifdefined\ONECOLUMN
\begin{tabular}{p{17.2cm}p{2cm}}
& $\Box$
\end{tabular}
\else
\begin{tabular}{p{7.9cm}p{2cm}}
& $\blacksquare$
\end{tabular}
\fi

%% file: appendix-proof-run-time-complexity-of-LC1-and-LC2-policies.tex
We first count the number of comparison operations in steps \ref{step:recompute_tau_l_max}, \ref{step:H}, \ref{step:EC}, \ref{step:S}, and \ref{step-recompute-delays-and-expected-rates-one-last-time} when the number of links in \textit{ActiveSet} is $n$. We note that \textit{ActiveSet} has $L$ links in the first round, and that its size reduces by $1$ in each subsequent round. In step \ref{step:recompute_tau_l_max}, there are $n-1$ comparisons for each of the $n$ links in \textit{ActiveSet}. Therefore, the total number of comparisons in step \ref{step:recompute_tau_l_max} (for all rounds together) is $L(L-1) + (L-1)(L-2) + \ldots + 3^2$ $=\BigO(L^3)$. Step \ref{step:H} can be accomplished using $n-1$ comparisons yielding a total of $(L-1) + (L-2) + \ldots + 2$ $=\BigO(L^2)$ comparisons. In step \ref{step:EC}, after suppressing the row and column of any one of the $n-1$ links in $ActiveSet \setminus \{H\}$, we need $n-2$ comparisons for each of the remaining $n-1$ links in $ActiveSet \setminus \{K\}$ to compute the values $\{\tau_{l, max}\}_{l \in ActiveSet \setminus \{K\}}$, where $K$ is the link whose row and column we suppress. We further need another $n-1$ comparisons in the worst case to find whether eliminating this link reduces the delay for at least one channel, by comparing the $n-1$ newly computed delays with those computed earlier in step \ref{step:recompute_tau_l_max}. This gives us $(L-1)((L-1)(L-2)+(L-1)) + (L-2)((L-2)(L-3)+(L-2)) + \ldots + 3.(3.2+3) + 2.(2.1+2)$ $= \BigO(L^4)$ comparisons. In the \LCVARIANTONE policy, the set \textit{EC} has $n-1$ elements in the worst case, and hence step \ref{step:S} can be accomplished with $n-2$ comparisons for a total of $(L-2) + (L-3) + \ldots + 1$ $=\BigO(L^2)$ comparisons. In step \ref{step:S} of the \LCVARIANTTWO policy, the set \textit{EC} has $n-1$ elements in the worst case, and hence, as in step \ref{step:EC}, after suppressing the row and column for each link in \textit{EC}, we need $n-2$ comparisons for each of the remaining $n-1$ links in \textit{EC} to compute the values $\{\tau_{l, max}\}_{l \in ActiveSet \setminus \{K\}}$. We further need another $n-1$ comparisons to find the number of links for which the newly computed delays are lesser than the corresponding delays computed earlier in step \ref{step:recompute_tau_l_max}. This gives us $(L-1)((L-1)(L-2)+(L-1)) + (L-2)((L-2)(L-3)+(L-2)) + \ldots + 3.(3.2+3) + 2.(2.1+2)$ $= \BigO(L^4)$ comparisons. Step \ref{step-recompute-delays-and-expected-rates-one-last-time} contributes a constant to the runtime since there would only be two links left in \textit{ActiveSet}.\\
\indent
Next, we count the number of multiplications and additions required in steps \ref{step-compute-queue-length-weighted-conditional-expected-data-rates} and \ref{step-recompute-delays-and-expected-rates-one-last-time} in computing the queue-length weighted conditional expected values. Evaluation of the conditional expectation for each of the $n$ links in \textit{ActiveSet} requires $\mathcal{C}$ multiplications and $\mathcal{C}$ additions, and one additional multiplication is required for multiplying with queue-length. Hence, in all, $L(2\mathcal{C}+1) + (L-1)(2\mathcal{C}+1) + \ldots + 2(2\mathcal{C}+1)$ $ = \BigO(\mathcal{C}L^2)$ multiplications and additions are required. Therefore, the cost of comparisons, multiplications and additions together is $\BigO(\mathcal{C}L^2 + L^4)$.\\
\ifdefined\ONECOLUMN
\begin{tabular}{p{17.2cm}p{2cm}}
& $\Box$
\end{tabular}
\else
\begin{tabular}{p{7.9cm}p{2cm}}
& $\blacksquare$
\end{tabular}
\fi

%% file: appendix-proof-Exact-analytical-expression-for-the-saturated-system-throughput-of-LC-ELDR-policy.tex
\begin{figure*}[!t]
\begin{equation}
\label{defn-e-r}
e^{(r)} {}^{\dag} \coloneqq
\begin{dcases}
  \argminwithsuperscript_{\displaystyle k \in T \setminus \{ \mathop{\cup}_{1 \leq n \leq r-1}e^{(n)}, M^{(r)} \}}  & \Big\{ \OnePlus\big\{ \delta_{jk} > \max_{\displaystyle m \in T \setminus \{ j, k, \mathop{\cup}_{1 \leq n \leq r-1}e^{(n)} \}} \{ \delta_{jm} \}\big\}_{ j \in T \setminus \{ k, {\displaystyle \mathop{\cup}_{1 \leq n \leq r-1} } e^{(n)} \scriptsize \} } \\ 
  \quad & \times \; \Ex(k, \tau_k^{(r)}) \Big\}, \; \text{  if  } r < N-1, \text{ and } \OnePlus\{.\} \text{ above evaluates to 1 for at least} \\
  & \text{one } k \in T \setminus \{\mathop{\cup}_{1 \leq n \leq r-1}e^{(n)}, M^{(r)}\} \\
  & \\
  \hspace{14ex} 0 & \text{otherwise}
\end{dcases}
\end{equation} \\
{\footnotesize ${}^{\dag}$for the \LCVARIANTONE policy; $\;{}^{\S}$chooses the link with the largest $k$ if more than one link have the same minimum value} \vspace{-0.5cm}
\begin{flushleft}
\hrulefill
\end{flushleft}
\end{figure*}

First, we need to set up some notations. Let $N$ be the number of links in \textit{ActiveSet} and $T \coloneqq \{1, 2, \dots, N\}$ be the set of links in \textit{ActiveSet} before the call to Algorithm \ref{LC1Listing} (recall that in the case of a network with complete interference, and for the first call to Algorithm \ref{LC1Listing} in the case of a network with multiple interference sets, \textit{ActiveSet} will have all the links in the network). Let $\delta_{ij}$ for $1 \leq i, j \leq N$ be the delay value in row $i$ and column $j$ of the delay table (note that $\delta_{ii} = 0 \; \forall i$ in our model). Let $c_{k, \tau}$ be the realization of channel-state on link $l_k$ at time $t - \tau$. Further, let
\begin{align*}
\One \big\{ a \big\} & \coloneqq \left\{ \bfrac{1 \quad \mbox{if $a$ is true}}{\hspace{-0.14cm}0 \quad \mbox{otherwise}} \right. \\
\One {\big\{ {a_i}\big\} }_{i \in I} & \coloneqq \prod_{i \in I} \One\{a_i\} \\
\One_{+} {\big\{ {a_i}\big\} }_{i \in I} & \coloneqq \max{ \big\{ \One \{a_i\} \big\}}_{i \in I} \\
\Ex(k, \tau_k) & \coloneqq \Ex \Big[ C_{l_{k}}[t] \; \big\rvert \; C_{l_{k}}[t - \tau_k] = c_{k, \tau_{k}} \Big]
\end{align*}

We will call the ``\verb|while|'' loop body from line 3 to line 23, and also the computation in lines 24--29 in Algorithm \ref{LC1Listing} in Sec. \ref{section-low-complexity-scheduling-policies}, a ``round''. Thus, if the \LCVARIANTONE policy terminates at line 26 or at line 28, then it would have executed $N-1$ rounds (specifically, $N-2$ rounds in the body of the ``\verb|while|'' loop, and the last round [round $N-1$] in lines 24--29). 

Let us call the link that has not yet been eliminated from the set of contending links, an \textit{active} link. Let $M^{(r)}$ be the active link with the largest expected data rate in round $r$. That is,
$$ M^{(r)} \coloneqq \argmax_{\displaystyle k \in T \setminus \{ \mathop{\cup}_{1 \leq n \leq r-1}e^{(n)} \}} \Big\{\Ex(k, \tau_k^{(r)}) \Big\}, $$ where $e^{(r)}$ and $\tau_j^{(r)}$ are as defined below.

Let $e^{(r)}$ be the link other than $M^{(r)}$ that has the smallest expected data rate among all links that reduce the common delay value with which the channel state of at least one link can be accessed by the transmitter nodes of all the contending links, if this link is eliminated in round $r$ (i.e., among all links that reduce the maximum of the delay values in a row for at least one row [corresponding to an active link] in the table of delay values [after suppressing the rows and columns corresponding to links that have been eliminated up to round $r$], if this link is eliminated in round $r$). Equivalently, $e^{(r)}$ is the link that was eliminated in round $r$ (if $r$ is less than the current round number), or the link that will be eliminated in round $r$ (if $r$ is equal to the current round number). We can write $e^{(r)}$ as in Expr. (\ref{defn-e-r}). The computation of $e^{(r)}$ is illustrated in Figs. \ref{fig:Illustration-LC-ELDR-policy}(b) -- (e) for $r = 1$, and in Figs. \ref{fig:Illustration-LC-ELDR-policy}(f) -- (h) for $r = 2$, for the setting in Sec. \ref{subsec:dynamics-of-LC-ELDR-policy}. Note that the condition ``$e^{(r)} \neq 0$'' indicates that some link (other than the one with the largest expected data rate in round $r$) is eliminated in round $r$, and hence that the \LCVARIANTONE policy can proceed to round $r+1$ (i.e., round $r$ is not the last round).


We let $\tau_j^{(r)}$ (not to be confused with $\tau_l(h)$ defined in Sec. \ref{section-Structure-of-Heterogeneously-Delayed-NSI}) be equal to $\tau_{l_{j}, max}$ (see Sec. \ref{section-Structure-of-Heterogeneously-Delayed-NSI}) at the beginning of round $r$, where $\tau_{l_{j}, max}$ is calculated after masking the rows and columns pertaining to the links that have been eliminated in rounds 1 to $r-1$ (as illustrated in Figs. \ref{fig:Illustration-LC-ELDR-policy}(a), \ref{fig:Illustration-LC-ELDR-policy}(e), and \ref{fig:Illustration-LC-ELDR-policy}(h)). That is, 
$$
\tau_j^{(r)} \coloneqq \max_{\displaystyle k \in T \setminus \{j, \mathop{\cup}_{1 \leq n \leq r-1}e^{(n)} \}} \{ \delta_{jk} \}
$$
Note that $\tau_j^{(1)} \geq \tau_j^{(2)} \geq \dots \geq \tau_j^{(N-1)}$.\\
\indent
With these definitions in place, we are now ready to derive the required analytical expression for the expected saturated system throughput of the \LCVARIANTONE policy. Let $i \in T$ be the link chosen by the \LCVARIANTONE policy in a particular time slot, say $t$. We now consider the working of the \LCVARIANTONE policy when it is executed at the transmitter node of link $i$ in time slot $t$. From the listing of the \LCVARIANTONE policy in Sec. \ref{section-low-complexity-scheduling-policies}, we see that link $i$ can emerge as the ``winner'' (i.e., as the link chosen by the \LCVARIANTONE policy for carrying transmission) in time slot $t$ in two ways -- (i) from lines 25--26 (i.e., round $N-1$ is the last round),  or (ii) from lines 8--10,\nespace 14 (i.e., an intermediate round $r < N-1$ is the last round). Thus, link $i$ can emerge as the ``winner'' when the last round is any of 1 to $N-1$. Now, link $i$ will emerge as the ``winner'' when the last round is $r, \; 1 \leq r \leq N-1$, if it ``survives'' (i.e., if it not eliminated in) round 1, survives round 2, $\ldots,$ survives round $r$. \\
\indent
Let us first fix a particular last round $r, \; 1 \leq r \leq N-1$ (i.e, the \LCVARIANTONE policy runs from round 1 to round $r$ and terminates immediately after round $r$). When the last round is $r$, if link $i$ emerges as the ``winner'', its contribution to the total expected saturated system throughput is given by \\
\begin{equation*}
\begin{split}
\hspace{-4cm} \sum_{\substack{c_{j, \tau_{j}^{(q)}}\\ j \in T, \;\; q \in \{1, 2, \dots, r\}}} \kern-1em \Ex \bigg[ C_{l_{i}}[t] \; \times
\end{split}
\end{equation*}
\begin{equation*}
\begin{split}
\hspace{2cm} & \kern-5em \prod_{\tilde{r} = 1}^{r-1} \Big( \One\{\text{link } \text{$i$ is not eliminated in round $\tilde{r}$}\} \\
& \kern-2em \times \One\{\text{round $\tilde{r}$ is not the last round} \} \Big) \\
& \kern-2.5em \One\{\text{link } \text{$i$ is not eliminated in round $r$}\}\\
& \kern-2em \times \One\{\text{round $r$ is the last round} \} \\ 
& \kern-3.2em \; \Big\rvert \; C_{l_{i}}[t - \tau_{i}^{(r)}] = c_{i, \tau_{i}^{(r)}}, \\
& \kern-2.9em \big\{ C_{l_{i}}[t - \tau_{i}^{(q)}] = c_{i, \tau_{i}^{(q)}} \big\}_{\substack{q \in \{1, 2, \dots, r-1\}}}, \\
& \kern-3em \big\{ C_{l_{p}}[t - \tau_{p}^{(q)}] = c_{p, \tau_{p}^{(q)}} \big\}_{\substack{p \in T \setminus \{i\} \\ q \in \{1, 2, \dots, r\}}} \bigg] 
\end{split}
\end{equation*}
\begin{equation}
\label{eqn:high-level-conditions}
\kern-1em \times \Pr \big\{ C_{l_{1}}[t - \tau_1^{(1)}] = c_{1, \tau_1^{(1)}}, C_{l_{1}}[t - \tau_1^{(2)}] = c_{1, \tau_1^{(2)}}, 
$$
$$
\dots, C_{l_{N}}[t - \tau_N^{(r)}] = c_{N, \tau_N^{(r)}} \big\}
\end{equation}

\noindent
Now, consider a particular round $\tilde{r}$ of the \LCVARIANTONE policy. If link $i$ has survived (i.e., has not been eliminated in) the first $\tilde{r}-1$ rounds, then it will survive round $\tilde{r}$ if one of the following happens:
\begin{enumerate}
  \item link $i$ has the largest expected data rate in round $\tilde{r}$ (hence $i = H$ and therefore $i \notin EC$; recall that we set aside the link with the largest expected data rate in each round so that it will not be eliminated in that round [see steps 6 and 7 of Algorithm \ref{LC1Listing}]). We denote this condition as $s(i, \tilde{r}, 1)$.
  
  \item link $i$ does not have the largest expected data rate in round $\tilde{r}$ (hence $i \neq H$), and link $i$ does not reduce the maximum of the delay values in a row, for any row (corresponding to a link in \textit{ActiveSet}) in the table of delay values (after suppressing the rows and columns corresponding to links that have been eliminated up to round $\tilde{r}$) if it is eliminated in round $\tilde{r}$ (hence $i \notin EC$; see step 7 of Algorithm \ref{LC1Listing}), and some link (other than the one with the largest expected data rate in round $\tilde{r}$) reduces the maximum of the delay values in a row, for at least one row (i.e. link) if it is eliminated in round $\tilde{r}$ (i.e., $EC \neq \phi$). We denote this condition as $s(i, \tilde{r}, 2)$.
  
  \item link $i$ does not have the largest expected data rate in round $\tilde{r}$ (hence $i \neq H$), and link $i$ reduces the maximum of the delay values in a row, for at least one row (corresponding to a link in \textit{ActiveSet}) in the table of delay values (after suppressing the rows and columns corresponding to links that have been eliminated up to round $\tilde{r}$) if it is eliminated in round $\tilde{r}$ (hence $i \in EC$), and link $i$ does not have the smallest expected data rate among the links that reduce the maximum of the delay values in a row, for at least one row (i.e. link) if that link is eliminated in round $\tilde{r}$ (hence $i \neq S$; see step 16 of Algorithm \ref{LC1Listing}). We denote this condition as  $s(i, \tilde{r}, 3)$.
\end{enumerate}

\noindent
Now, $s(i, r, 1)$ can be written mathematically as:
\begin{equation*}
\begin{split}
\One& \Big\{ \Ex(i, \tau_i^{(r)}) > \max_{\substack{k \in T \setminus \{ {\displaystyle \mathop{\cup}_{1 \leq n \leq r-1} e^{(n)} } \scriptsize \} \\ \kern-3em k<i  }} \big\{\Ex(k, \tau_k^{(r)})\big\} \Big\} \\
\end{split}
\end{equation*}
\begin{equation}
\label{eqn:s-i-r-1}
\begin{split}
& \times \One \Big\{ \Ex(i, \tau_i^{(r)}) \geq \max_{\substack{k \in T \setminus \{ {\displaystyle \mathop{\cup}_{1 \leq n \leq r-1} e^{(n)} } \scriptsize \} \\ \kern-3em k>i  }} \big\{\Ex(k, \tau_k^{(r)})\big\} \Big\}
\end{split}
\end{equation}
In the expression for $s(i, r, 1)$ above, we note that splitting the comparison of the expected data rates into two -- namely, (i) comparison with (the maximum of the expected data rates of) transmitters that are less than $i$, and (ii) comparison with transmitters that are greater than $i$, creates a lexicographic ordering among the transmitters. This is required to consistently resolve the ``winner'' in case there is a tie in the expected data rates of multiple transmitters -- we always resolve in favor of the smallest numbered transmitter (as a convention) in case of a tie.

Next, $s(i, r, 2)$ can be written mathematically as:\\
\begin{equation*}
\begin{split}
  \quad \OnePlus& \Big\{ \Ex(i, \tau_i^{(r)}) \leq \max_{\substack{k \in T \setminus \{ {\displaystyle \mathop{\cup}_{1 \leq n \leq r-1} e^{(n)} } \scriptsize \} \\ \kern-3em k<i  }} \big\{\Ex(k, \tau_k^{(r)}) \big\}, \\
\end{split}
\end{equation*}
\begin{equation}
\label{eqn:s-i-r-2}
\begin{split}
  & \;\;\, \Ex(i, \tau_i^{(r)}) < \max_{\substack{k \in T \setminus \{ {\displaystyle \mathop{\cup}_{1 \leq n \leq r-1} e^{(n)} } \scriptsize \} \\ \kern-3em k>i  }} \big\{\Ex(k, \tau_k^{(r)}) \big\} \Big\} \\
  & \kern-1em \times \One \Big\{ \delta_{ji} \leq \max_{m \in T \setminus \{i, j, {\displaystyle \mathop{\cup}_{1 \leq n \leq r-1} e^{(n)} } \scriptsize \}} \{ \delta_{jm} \} \Big\}_{ \substack{ j \in T \setminus \{i, \\ \kern-1em {\displaystyle \mathop{\cup}_{1 \leq n \leq r-1} \kern-1em e^{(n)} } \scriptsize \} } } \\
  & \kern-1em \times \One \big\{ e^{(r)} \neq 0 \big\}
\end{split}
\end{equation}

Finally, $s(i, r, 3)$ can be written mathematically as:\\
\begin{equation*}
\begin{split}
  \OnePlus& \Big\{ \Ex(i, \tau_i^{(r)}) \leq \max_{\substack{k \in T \setminus \{ { \displaystyle \mathop{\cup}_{1 \leq n \leq r-1} e^{(n)} } \scriptsize \} \\ \kern-3em k<i  }} \big\{\Ex(k, \tau_k^{(r)}) \big\}, 
\end{split}
\end{equation*}
\begin{equation*}
\begin{split}
  \;\quad \Ex(i, \tau_i^{(r)}) < \max_{\substack{k \in T \setminus \{ {\displaystyle \mathop{\cup}_{1 \leq n \leq r-1} e^{(n)} } \scriptsize \} \\ \kern-3em k>i  }} \big\{\Ex(k, \tau_k^{(r)}) \big\} \Big\} \\
\end{split}
\end{equation*}
\begin{equation*}
\begin{split}
   & \kern-2em \times \OnePlus \Big\{ \delta_{ji} > \max_{m \in T \setminus \{i, j, {\displaystyle \mathop{\cup}_{1 \leq n \leq r-1} e^{(n)} } \scriptsize \}} \{ \delta_{jm} \} \Big\}_{ \substack{j \in T \setminus \{i, \\ \kern-1em {\displaystyle \mathop{\cup}_{1 \leq n \leq r-1} \kern-1em e^{(n)} } \scriptsize \}} }\\
\end{split}
\end{equation*}
\begin{equation}
\label{eqn:s-i-r-3}
\begin{split}
  \kern-12em \; \times \, \One \big\{ e^{(r)} \neq i \big\}
\end{split}
\end{equation}
\noindent
From (\ref{eqn:high-level-conditions}), (\ref{eqn:s-i-r-1}), (\ref{eqn:s-i-r-2}) and (\ref{eqn:s-i-r-3}), the expression for the contribution of link $i$ to the total expected saturated system throughput when link $i$ emerges as the ``winner'' when round $r, \; 1 \leq r \leq N-1$, is the last round becomes
\begin{equation*}
\begin{split}
\kern-3em
\sum_{\substack{c_{j, \tau_{j}^{(q)}}\\ j \in T \\ q \in \{1, 2, \dots, r\}}} \kern-1em \Ex \Bigg[ &  C_{l_i}[t] \; \prod_{\tilde{r}=1}^{r-1} \Big(\OnePlus \big\{s(i, \tilde{r}, n)\big\}_{n \in \{ 1, 2, 3 \}} \\
\end{split}
\end{equation*}
\begin{equation*}
\begin{split}
\hspace{1cm} & \kern+3em \times \One\big\{e^{(\tilde{r})} \neq 0 \big\} \Big) \One \big\{s(i,r,1) \big\} \One\big\{e^{(r)} = 0\big\} \\ 
& \kern+4em \bigg\rvert \; C_{l_i} \big[ t - \tau_i^{(r)} \big] = c_{i, \tau_i^{(r)}},  \\
& \kern+4.5em \Big\{C_{l_i} \big[ t - \tau_i^{(q)} \big] = c_{i, \tau_i^{(q)}} \Big\}_{q \in \{ 1, 2, \dots, r-1 \}}, \\
& \kern+4.5em \Big\{C_{l_p} \big[ t - \tau_p^{(q)} \big] = c_{p, \tau_p^{(q)}} \Big\}_{\substack{ p \in T \setminus \{i\} \\ q \in \{ 1, 2, \dots, r \}}} \Bigg] \\
& \kern+1em \times \prod_{m = 1}^{N} \left( \pi_{m} \big( c_{m, \tau_{m}^{(1)}} \big) \prod_{n=1}^{r-1} p_{{}_{{}_{m,}} \; c_{m, \tau_{m}^{(n)}} \; c_{m, \tau_{m}^{(n+1)}}}^{(\tau_m^{(n)} - \tau_m^{(n+1)})} \right)
\end{split}
\end{equation*}
where $\pi_{m}(c)$ is the steady-state probability of being in state $c$ in the channel-state Markov chain of link $m$, and $p_{m, \, ij}^{(k)}$ is the probability of reaching state $j$ from state $i$ in $k$ steps in the channel-state Markov chain of link $m$.\footnote{$\pi_{m}(c) = \pi(c) \text{ and } p_{m, \, ij}^{(k)} = p_{ij}^{(k)} \; \forall m \in \mathcal{L}$ in our model; see Sec. \ref{section-Network-Model}}

Now, summing over all possible last rounds $r = 1 \text{ to } N-1$, and over all links $i = 1 \text{ to } N$, we get the expression for the expected saturated system throughput of the \LCVARIANTONE policy to be
\begin{equation*}
\begin{split}
\sum_{i=1}^{N} \sum_{r=1}^{N-1} \sum_{\substack{c_{j, \tau_{j}^{(q)}}\\ j \in T \\ q \in \{1, 2, \dots, r\}}} \kern-1em \Ex \Bigg[ &  C_{l_i}[t] \; \prod_{\tilde{r}=1}^{r-1} \Big(\OnePlus \big\{s(i, \tilde{r}, n)\big\}_{n \in \{ 1, 2, 3 \}} \\
& \kern-3em \times \One\big\{e^{(\tilde{r})} \neq 0 \big\} \Big) \One \big\{s(i,r,1) \big\} \One\big\{e^{(r)} = 0\big\} \\ 
\end{split}
\end{equation*}
\begin{equation*}
\begin{split}
& \bigg\rvert \; C_{l_i} \big[ t - \tau_i^{(r)} \big] = c_{i, \tau_i^{(r)}},  \\
\end{split}
\end{equation*}
\begin{equation*}
\begin{split}
& \kern+7.5em \Big\{C_{l_i} \big[ t - \tau_i^{(q)} \big] = c_{i, \tau_i^{(q)}} \Big\}_{q \in \{ 1, 2, \dots, r-1 \}}, \\
\end{split}
\end{equation*}
\begin{equation*}
\begin{split}
& \kern+7em \Big\{C_{l_p} \big[ t - \tau_p^{(q)} \big] = c_{p, \tau_p^{(q)}} \Big\}_{\substack{ p \in T \setminus \{i\} \\ q \in \{ 1, 2, \dots, r \}}} \Bigg] \\
\end{split}
\end{equation*}
\begin{equation*}
\begin{split}
& \kern+4em \times \prod_{m = 1}^{N} \left( \pi_{m} \big( c_{m, \tau_{m}^{(1)}} \big) \prod_{n=1}^{r-1} p_{{}_{{}_{m,}} \; c_{m, \tau_{m}^{(n)}} \; c_{m, \tau_{m}^{(n+1)}}}^{(\tau_m^{(n)} - \tau_m^{(n+1)})} \right)
\end{split}
\end{equation*}
as required.\\
\ifdefined\ONECOLUMN
\begin{tabular}{p{17.2cm}p{2cm}}
& $\Box$
\end{tabular}
\else
\begin{tabular}{p{7.9cm}p{2cm}}
& $\blacksquare$
\end{tabular}
\fi

%% file: appendix-proof-Exact-analytical-expression-for-the-saturated-system-throughput-of-LC-ERDMC-policy.tex
The expression for the expected saturated system throughput of the \LCVARIANTTWO policy is the same as the expression in Proposition \ref{lemma-saturated-system-throughput-of-LC1-policy}, except that we need to redefine $e^{(r)}$. For that, in addition to carrying over all the notations we established in Appendix \ref{appendix-proof-saturated-system-throughput-of-LC1-policy}, we need the following notations:
\begin{align*}
\OneInfty \big\{ a \big\} & \coloneqq \left\{ \bfrac{1 \quad \mbox{if $a$ is true}}{\hspace{-0.14cm}\infty \quad \mbox{otherwise}} \right. \\
s_{+} {\big\{ {a_i}\big\} }_{i \in I} & \coloneqq \sum_{i \in I} { \One \big\{ a_i \big\} }
\end{align*}

\begin{align*}
s_+(&k, r) \coloneqq s_+\big\{ \delta_{jk} > \\ 
& \max_{\displaystyle m \in T \setminus \{ j, k, \mathop{\cup}_{1 \leq n \leq r-1}e^{(n)} \}} \{ \delta_{jm} \}\big\}_{ \substack{\kern-2em j \in T \setminus \{ k, \\ 
  {\displaystyle \mathop{\cup}_{1 \leq n \leq r-1} } e^{(n)} \scriptsize \} } }
\end{align*}\\
For the \LCVARIANTTWO policy, we redefine $e^{(r)}$ to be the link other than $M^{(r)}$ that has the smallest expected data rate among all links that reduce the common delay value with which the channel state of a link can be accessed by the transmitters corresponding to all the contending links, for the largest number of links, if this link is eliminated in round $r$ (i.e., among all links that reduce the maximum of the delay values in a row for the maximum number of rows [corresponding to active links] in the table of delay values [after suppressing the rows and columns corresponding to links that have been eliminated up to round $r$], if this link is eliminated in round $r$). We can write redefined $e^{(r)}$ as in Expr. (\ref{defn-e-r-LC-ERDMC}). The required expression for the expected saturated system throughput of the \LCVARIANTTWO policy is then obtained by following the rest of the arguments as in the proof of Proposition \ref{lemma-saturated-system-throughput-of-LC1-policy} in Appendix \ref{appendix-proof-saturated-system-throughput-of-LC1-policy}.
\ifdefined\ONECOLUMN
\begin{tabular}{p{17.2cm}p{2cm}}
& $\Box$
\end{tabular}
\else
\begin{tabular}{p{7.9cm}p{2cm}}
& $\blacksquare$
\end{tabular}
\fi
\newpage
\begin{minipage}{1.0\textwidth}
	\begin{equation}
	\label{defn-e-r-LC-ERDMC}
	e^{(r)} {}^{\ddag} \coloneqq
	\begin{dcases}
	\argminwithsuperscript_{\displaystyle k \in T \setminus \{ \mathop{\cup}_{1 \leq n \leq r-1}e^{(n)}, M^{(r)} \}} & \Big\{ \OneInfty \big\{ s_+(k,r) = \max_{\displaystyle k' \in T \setminus \{ \mathop{\cup}_{1 \leq n \leq r-1}e^{(n)}, M^{(r)} \}} s_+(k', r) \big\} \\ 
	\quad & \times \; \Ex(k, \tau_k^{(r)}) \Big\}, \; \text{  if  } r < N-1, \text{ and } \OneInfty\{.\} \text{ above evaluates to 1 for at least} \\
	& \text{one } k \in T \setminus \{\mathop{\cup}_{1 \leq n \leq r-1}e^{(n)}, M^{(r)}\} \\ 
	& \\
	\hspace{14ex} 0 & \text{otherwise}
	\end{dcases}
	\end{equation} \\
	{\footnotesize ${}^{\ddag}$for the \LCVARIANTTWO policy; $\;{}^{\S}$chooses the link with the largest $k$ if more than one link have the same minimum value} \vspace{-0.5cm}
	\begin{flushleft}
		\hrulefill
	\end{flushleft}
\end{minipage}

%% file: Throughput_Optimal_and_Fast_Near-Optimal_Scheduling_with_Heterogeneously_Delayed_Network-State_Information__Extended_Version__v2.bbl
\begin{thebibliography}{99}
%
%
\bibitem{Reddy_et_al_12}
{A. Reddy, S. Banerjee, A. Gopalan, S. Shakkottai, and L. Ying}, {``On Distributed Scheduling with Heterogeneously Delayed Network-State Information,''} {\textit{Queueing Systems}}, {vol. 72}, {pp. 193 -- 218}, {2012} 

\ifExtendedVersion
\else
\bibitem{SupplementaryMaterial}
{S. Narasimha, and J. Kuri}, {``Throughput Optimal and Fast Near-Optimal Scheduling with Heterogeneously Delayed Network-State Information (Appendices),'' [Document submitted along with, and as supplementary material to, the manuscript]} 
\bibitem{ExtendedVersion}
{S. Narasimha, and J. Kuri}, {``Throughput Optimal and Fast Near-Optimal Scheduling with Heterogeneously Delayed Network-State Information (Extended Version),'' {arXiv preprint}, 2015 [Temporary submission identifier is: submit/1237827. Allotment of official arXiv identifier is pending as of the date of this submission. This document can be located on arxiv.org through a search using the document title. Alternatively, the first/corresponding author can be contacted for a copy of this document and/or the official arXiv identifier.]}
\fi

\bibitem{Tassiulas_Ephremides_92}
{L. Tassiulas, and A. Ephremides}, {``Stability Properties of Constrained Queueing Systems and Scheduling Policies for Maximum Throughput in Multihop Radio Networks,''} {\textit{IEEE Transactions on Automatic Control}}, {vol. 37, no. 12}, {pp. 1936 -- 1948}, {1992}


\bibitem{Moharir_Shakkottai_13}
{S. Moharir, and S. Shakkottai}, {``MaxWeight vs. BackPressure: Routing and Scheduling in Multi-Channel Relay Networks,''} {\textit{IEEE INFOCOM 2013}}, {pp. 1537 -- 1545}, {2013}

\bibitem{Tassiulas_Ephremides_93}
{L. Tassiulas, and A. Ephremides}, {``Dynamic Server Allocation to Parallel Queues with Randomly Varying Connectivity,''} {\textit{IEEE Transactions on Information Theory}}, {vol. 39, no. 2}, {pp. 466 -- 478}, {1993}

\bibitem{Shakkottai_Stolyar_00}
{S. Shakkottai, and A. Stolyar}, {``Scheduling for Multiple Flows Sharing a Time-Varying Channel: The Exponential Rule,''} {\textit{American Mathematical Society Translations, Series}}, {vol. 2}, {pp. 2002}, {2000}

\bibitem{Joo_Shroff_06}
{C. Joo, and N. Shroff}, {``Performance of Random Access Scheduling Schemes in Multi-hop Wireless Networks,''} {\textit{Fortieth Asilomar Conference on Signals, Systems and Computers (ACSSC '06)}}, {pp. 1937 -- 1941}, {2006}

\bibitem{Rajagopalan09}
{S. Rajagopalan, D. Shah, and J. Shin}, {``Network Adiabatic Theorem: An Efficient Randomized Protocol for Contention Resolution,''} {\textit{SIGMETRICS/Performance}}, {pp. 133 -- 144}, {2009}

\bibitem{Bordenave08}
{C. Bordenave, D. McDonald, and A. Proutiere}, {``Performance of Random Medium Access Control An Asymptotic Approach,''} {\textit{SIGMETRICS Perform. Eval. Rev.}}, {vol. 36, no. 1}, {pp. 1 -- 12}, {2008}

\bibitem{Marbach207}
{P. Marbach, A. Eryilmaz, and A. Ozdaglar}, {``Achievable Rate Region of CSMA Schedulers in Wireless Networks with Primary Interference Constraints,''} {\textit{IEEE Conference
on Decision and Control}}, {pp. 1156 -- 1161}, {2007}

\bibitem{Gupta06}
{P. Gupta, and A. L. Stolyar}, {``Optimal Throughput Allocation in General Random-Access Networks,''} {\textit{40th Annual Conference on Information Sciences and Systems}}, {pp. 1254 -- 1259}, {2006}

\bibitem{Liu07}
{J.Liu, and A. L. Stolyar}, {``Distributed Queue-Length based Algorithms for Optimal End-to-End Throughput Allocation and Stability in Multi-hop Random Access Networks,''} {\textit{45th Allerton Conference on Communication, Control, and Computing, Urbana-Champaign, IL}}, {2007}

\bibitem{Stolyar08}
{A. L. Stolyar}, {``Dynamic Distributed Scheduling in Random Access Networks,''} {\textit{Journal of Applied Probability}}, {vol. 45, no. 2}, {pp. 297 -- 313}, {2008}

\bibitem{Jiang08}
{L. Jiang, and J. Walrand}, {``A Distributed CSMA Algorithm for Throughput and Utility Maximization in Wireless Networks,''} {\textit{IEEE/ACM Transactions on Networking}}, {vol. 18, no. 3}, {pp. 960 -- 972}, {2008}

\bibitem{Modiano06}
{E. Modiano, D. Shah, and G. Zussman}, {``Maximizing Throughput in Wireless Networks via Gossiping,''} {\textit{ACM
SIGMETRICS/Performance}}, {pp. 27 -- 38}, {2008}

\bibitem{Sanghavi07}
{S. Sanghavi, L. Bui, and R. Srikant}, {``Distributed Link Scheduling with Constant Overhead,''} {\textit{ACM SIGMETRICS Perform. Eval. Rev.}}, {vol. 35, no. 1}, {pp. 313 -- 324}, {2007}


\bibitem{Pantelidou07}
{A. Pantelidou, A. Ephremides, and A. Tits}, {``Joint Scheduling and Routing for Ad-hoc Networks Under Channel State Uncertainty,''} {\textit{WiOpt}}, {pp. 1 -- 8}, {2007}

\bibitem{Ying08}
{L. Ying, and S. Shakkottai}, {``On Throughput Optimality with Delayed Network-State Information,''} {\textit{Technical  Report}}, {2008}

\bibitem{Kumar12}
{A. Kumar}, {2012}, {``Discrete Event Stochastic Processes: Lecture Notes for an Engineering Curriculum,''} {Retrieved from \url{http://www.ece.iisc.ernet.in/~anurag/books/anurag/spqt.pdf} }


\end{thebibliography}
